\documentstyle[11pt,aaspp4,flushrt,amssym]{article}

\begin{document}

\slugcomment{Astrophysical Journal, in press}

\title{The CNOC2 Field Galaxy Luminosity Function I:\\
       A Description of Luminosity Function Evolution}

\author{Huan Lin \altaffilmark{1,2,3,8}, 
        H. K. C. Yee \altaffilmark{1,8}, 
        R. G. Carlberg \altaffilmark{1,8}, 
        Simon L. Morris \altaffilmark{4,8},\\ 
        Marcin Sawicki \altaffilmark{1,5,8},
        David R. Patton \altaffilmark{6,8}, 
        Greg Wirth \altaffilmark{6,7,8},  
        \& Charles W. Shepherd \altaffilmark{1}}

\altaffiltext{1}{Department of Astronomy, University of
  Toronto, 60 St.\ George Street, Toronto, ON M5S 3H8, Canada, 
  lin, hyee, carlberg, sawicki, shepherd@astro.utoronto.ca}

\altaffiltext{2}{Now at Steward Observatory, University of
  Arizona, 933 N.\ Cherry Avenue, Tucson, AZ 85721, USA,
  hlin@as.arizona.edu}

\altaffiltext{3}{Hubble Fellow}

\altaffiltext{4}{Dominion Astrophysical Observatory, Herzberg Institute of
  Astrophysics, Victoria, BC V8X 4M6, Canada,
  Simon.Morris@hia.nrc.ca}

\altaffiltext{5}{Now at California Institute of Technology, 320-47,
  Pasadena, CA 91125, USA, sawicki@mop.caltech.edu}

\altaffiltext{6}{Department of Physics \& Astronomy, University of
  Victoria, P.O.\ Box 3055, Victoria, BC V8W 3P6, Canada,
  patton@uvastro.phys.uvic.ca}

\altaffiltext{7}{Now at W. M. Keck Observatory, Kamuela, HI 96743,
  USA, wirth@keck.hawaii.edu}

\altaffiltext{8}{Visiting Astronomer, Canada-France-Hawaii
  Telescope, which is operated by the National Research Council of
  Canada, the Centre Nationale de la Recherche Scientifique of
  France, and the University of Hawaii.}

\begin{abstract}

We examine the evolution of the galaxy luminosity function (LF) using
a sample of over 2000 galaxies, with $0.12 < z < 0.55$ and 
$17.0 < R_c < 21.5$, drawn from the Canadian Network for Observational
Cosmology Field Galaxy Redshift
Survey (CNOC2), at present the largest such sample at intermediate redshifts.
We use $U \! B V \! R_c I_c$ photometry and the spectral energy
distributions (SED's) of Coleman, Wu, \& Weedman (1980) to classify our
galaxies into early, intermediate, and late types, for which we compute
luminosity functions in the rest-frame $B$, $R_c$, and $U$ bandpasses.
In particular, we adopt a convenient parameterization of LF evolution
including luminosity and number density evolution, and take care to
quantify correlations among our LF evolution parameters.
We also carefully measure and account for sample selection effects 
as functions of galaxy magnitude and color.

Our principal result is a clear quantitative separation of luminosity and
density evolution for different galaxy populations, and the finding
that the character of the LF evolution is strongly dependent on
galaxy type.
Specifically, we find that the early- and intermediate-type LF's
show primarily brightening at higher redshifts
and only modest density evolution, while the late-type LF is best
fit by strong number density increases at higher $z$, with little
luminosity evolution.
We also confirm the trend seen in previous smaller $z \lesssim 1$ samples
of the contrast between the strongly increasing luminosity density of
late-type galaxies and the relatively constant luminosity density of
early-type objects.
Specific comparisons against the Canada-France and Autofib redshift
surveys show general agreement among our LF evolution results,
although there remain some detailed discrepancies.
In addition, we use our number count and color distribution data to further 
confirm the validity of our LF evolution models to $z \sim 0.75$, and 
we also show that our results are not significantly affected by
potential systematic effects, such as surface brightness selection, 
photometric errors, or redshift incompleteness.

\end{abstract}


\section{Introduction}

The luminosity function (LF) is a basic and fundamentally important
statistic used to study galaxy populations and
their evolution.
In particular, measurement of the field galaxy luminosity function at
different redshifts provides a simple means of describing the
global changes seen in the galaxy population with lookback time.
These LF data, together with complementary information from galaxy
number counts and color distributions, among others, supply some of the key
observations that help shape our picture of how galaxies evolve (e.g., see
reviews by \cite{koo92}; \cite{ell97}).
It remains an important problem and a difficult theoretical challenge to properly
interpret the variety of galaxy evolution data, including the LF,
in terms of models of galaxy formation and evolution that
self-consistently incorporate relevant physical processes
such as star formation, feedback, and gravity (e.g., \cite{bau98}; 
\cite{col94}; \cite{kau97}).

Recent observational progress in measuring the field galaxy LF
has spanned a very wide range of redshifts, including improved estimates
using larger local ($z \sim 0$) and intermediate-redshift ($z \lesssim
1$) samples, as well as the first observations for
high-redshift ($z \sim 3$) galaxies. 
In the local $z \lesssim 0.2$ regime, the galaxy LF is now
determined routinely using redshift survey samples containing thousands 
of galaxies, and the LF is most commonly measured 
for the rest-frame optical $B$ (\cite{lov92};
\cite{mar94b}; \cite{daC94}; \cite{zuc97}; \cite{rat98}; \cite{col98})
and $R$ bands (\cite{lin96lum}; \cite{gel97}).
However, despite the very large sample sizes, there remains
controversy in the determination of the local LF, with regard to both its
normalization and shape.
The controversy results from potential systematic effects which 
may adversely affect LF measurements in local surveys at bright magnitudes; 
these include surface brightness selection effects,
systematic photometric errors,
possible local galaxy underdensities, and the small volumes over which
intrinsically faint galaxies are visible (see discussion and review in
\cite{ell97}).
The uncertainties in the local LF normalization and shape
make it more difficult to use the local results as low-redshift 
anchors for models of galaxy evolution.
Nonetheless, a robust result from local LF measurements appears to be
the dependence of the LF on galaxy type, in the sense that
the faint end of the LF is consistently dominated by galaxies of later
morphology (\cite{mar94a}; \cite{mar98}), 
later spectral type (\cite{bro98}; \cite{col98}),
stronger line emission (\cite{lin96lum}; \cite{zuc97}), 
or bluer color (\cite{mar97}; \cite{met98}).

The LF situation is somewhat less controversial at intermediate redshifts 
($0.2 \lesssim z \lesssim 1$).
Recent deep redshift surveys, with samples of typically hundreds of
galaxies, have consistently found similar trends in the evolution of
different types of galaxies (\cite{lil95b}; \cite{ell96}; \cite{cow96};
\cite{lin97}; \cite{hey97}; \cite{sma97}; \cite{liu98}; \cite{hog98};
\cite{deL97}).
Namely, there exists a distinct contrast between the rapid
evolution seen in the LF of late-type, blue, starforming galaxies and
the relatively mild changes observed in the LF of early-type, red,
quiescent objects.
Also, type-dependent LF differences similar to those seen at low
redshifts are also observed at intermediate $z$.
The depth of the photometry required for 
intermediate-redshift surveys renders them less susceptible than their 
local counterparts to surface brightness selection effects, and
multicolor data are also typically available for moderate-redshift samples,
permitting more accurate galaxy classification.
On the other hand, the sample sizes are smaller 
and the redshift completeness is not as high as for low-$z$ surveys, 
so that the random errors on the luminosity function are in general larger
than those for local LF's.

A much larger intermediate-redshift sample will, however, be provided by
the Canadian Network for Observational Cosmology (CNOC) Field Galaxy
Redshift Survey, hereafter denoted CNOC2. 
The CNOC2 survey has as its primary goal the study
of the evolution of galaxy clustering and galaxy populations at
intermediate redshifts $0.1 \lesssim z \lesssim 0.7$.
In order to accomplish its objectives, CNOC2 will acquire some 5000 galaxy
redshifts at $R_c < 21.5$, thus making a dramatic improvement over
other intermediate-$z$ surveys in terms of sample size.
In addition, nearly all CNOC2 galaxies have multicolor $U \! B V \! R_c I_c$
photometry, which is more extensive color coverage
than is available for the vast majority of other redshift surveys.
This multicolor information permits galaxy classifications using fits
to broadband colors computed from model galaxy spectral energy
distributions (SED's).
Consequently, luminosity functions may be calculated and studied
for different galaxy populations, as well as for
a number of different rest-frame bandpasses such as $B$, $R_c$, and $U$.
In addition, the multicolor photometry permits more accurate computation of
$k$-corrections and related quantities and also allows detailed
checks of the survey's redshift completeness as functions of galaxy color and
type.

In this paper we will examine the evolution of the luminosity function
for different galaxy populations, using an interim but
statistically complete sample of over 2000 CNOC2 galaxies with $R_c < 21.5$.
Though this is only half of the ultimate CNOC2 survey sample, our interim
data set nevertheless comprises 
the largest intermediate-$z$ redshift survey at present.
The combination of large sample size, multicolor data, and careful
control of redshift selection effects should allow CNOC2 to give the
best quantitative LF constraints thus far at intermediate redshifts.

The outline of the paper is as follows.
In \S~\ref{data} we describe details of the CNOC2 data sample, and we
will also discuss redshift success rates, use of statistical weights to
correct for incompleteness, and surface brightness selection effects.
Then, in \S~\ref{methods} we detail our methods for galaxy
classifications and for fitting the luminosity function and its associated
evolution parameters.
Our LF evolution results are described in \S~\ref{results}, where
we focus in particular on quantifying luminosity and
density evolution in the LF's of different galaxy populations.
In addition, we will also compute galaxy SED type distributions, 
number counts, and color distributions, as well as
examine the impact that various potential systematic effects may have
on our LF evolution results.
In \S~\ref{comparisons}, we compare our LF's to those derived
from a number of previous intermediate-redshift galaxy surveys, in
particular the two next largest samples, the Canada-France Redshift Survey
(Lilly et al.\ 1995a,b) and the composite Autofib Redshift Survey
(\cite{ell96}; \cite{hey97}; plus references therein).
Finally, we summarize our conclusions in \S~\ref{conclusions}.

CNOC2 is now complete with respect to data acquisition, 
and LF studies using the full CNOC2 sample will be 
forthcoming once all the data have been reduced.
All the {\em presently} available, fully reduced, and statistically complete 
CNOC2 data are contained in the interim sample defined below in
\S~\ref{data}, and 
this is the first in a series of papers studying LF evolution in
the CNOC2 data set.
The second paper (\cite{lin99}; hereafter Paper II) will address LF
evolution in the context of physically-motivated galaxy evolution
models.
In contrast, in the present paper we focus on a {\em description} of
the redshift-dependent changes in the luminosity function, and will
not attempt to {\em explain} those changes in terms of physical
processes; that is the province of Paper II.
We emphasize that our use of ``luminosity evolution'' and 
``density evolution'' should strictly be construed to describe the
apparent changes in the LF, and may or may not correspond to true
physical changes in individual galaxies.
Not too surprisingly, it turns out that the apparent LF evolution will
be sensitive to the details of the galaxy classification scheme (and choice
of SED's as a function of $z$) that one adopts; this will be
elaborated further in \S~\ref{description} and in Paper II.
Nonetheless, the descriptive approach we take in this paper is a
simple and effective way of characterizing and quantifying 
the changes in the LF's of different intermediate-redshift galaxy
populations, and it is also in essence the typical approach taken in 
previous LF studies.

For the Hubble constant $H_0 = 100 \ h $~km
s$^{-1}$ Mpc$^{-1}$, $h = 1$ should be assumed in this paper if the $h$ 
dependence is not explicitly shown.
We also adopt a deceleration parameter $q_0 = 0.5$ throughout except
where otherwise specified (in particular $q_0 = 0.1$ will be used 
on occasion). 

\section{The CNOC2 Survey Data} \label{data}

A detailed description of the CNOC2 Field Galaxy Redshift Survey will be 
given in Yee et al.\ (1999; see also \cite{yee96}, 1997). 
We summarize the relevant points here. 

The survey covers four widely separated areas, hereafter denoted
as ``patches'', on the sky. 
In this paper we use data from the two
CNOC2 patches 0223+00 and 0920+37 (named by RA and Dec).
Observations for these two patches were
obtained during six observing runs at the Canada-France-Hawaii Telescope
(CFHT) over the period February 1995 to August 1997. 
Both photometry and spectroscopy were done using the CFHT Multi-Object
Spectrograph (MOS). 
Each of these two patches is a mosaic of 19 MOS fields 
(each $\sim 9' \times 8'$) covering nearly 1400 arcmin$^2$
(but see next paragraph).
Each patch is roughly in the shape of the letter L, with dimensions
approximately $54'$ EW and $80'$ NS (see \cite{yee97}).

Photometry was obtained using MOS in imaging mode for five bands
(approximate $5\sigma$ limits in parentheses): Kron-Cousins $R_c$ (24.0)
and $I_c$ (23.0), and Johnson $B$ (24.6), $V$ (24.0), and $U$ (23.0).
We restrict our sample to the 34 MOS fields (of the 38 total in the
0223 and 0920 patches) which have full $U \! BV \! R_cI_c$ coverage. 
The sky coverage of our sample is then 2490 arcmin$^2$, and the
comoving sample volume is $1.13 (1.50) \times 10^5 \ h^{-3}$ Mpc$^3$
for $q_0 = 0.5 (0.1)$.
Photometric reductions (object detection, star-galaxy classification, and 
photometry) were done using an improved version of the Picture
Processing Package (PPP; \cite{yee91}; \cite{yee96}). 
Objects were selected in the $R_c$ band for the survey's spectroscopic sample
and we adopt $R_c = 21.5$ as the nominal spectroscopic
completeness limit (see \S~\ref{weights} below).

Multislit spectroscopy was carried out on CFHT MOS.
We used a band-limiting filter to
restrict the wavelength coverage to 4400-6300\AA, in order to
increase the multiplexing efficiency such that 
typically 90-100 objects may be observed per slit mask.
The band-limiting filter does however restrict the redshift range over
which spectroscopic features important for redshift
measurement may be seen.
We adopt the nominal redshift completeness range $0.12 < z < 0.55$,
based on the observability of the Ca~II H+K (3968, 3933\AA)
absorption feature important for early-type galaxies.
Over this same redshift range, the [OII] $\lambda$3727 emission feature
important for late-type galaxies is also observable, except for $z
\lesssim 0.2$, where we expect [OIII] $\lambda\lambda$5007, 4959 and
H$\beta$ (4861\AA)
emission to substitute for the unobservable [OII] $\lambda$3727.
Spectroscopic reductions and redshift measurements 
were carried out using custom-written programs and 
standard IRAF routines.
The rms error in the velocity measurements is about 100 km s$^{-1}$,
as determined empirically from redundant spectroscopic observations.

We correct our photometry for extinction from the Milky Way using the 
dust maps of Schlegel et al.\ (1998).
We convert Schlegel et al.'s $E(B-V)$ values
to magnitudes of extinction in the $U \! BV \! R_cI_c$ bands using the
procedure described in their Appendix B, adopting the Milky Way
extinction curves of O'Donnell (1994) and Cardelli et al.\ (1989).
The extinction variation within each of the 0223 and 0920 patches is
small, so we simply apply a single correction for each patch as a whole:
\begin{equation}
\left\{
\begin{array}{lrrr}
         & 0223 & 0920 & \\
\Delta U & -0.171 & -0.059 & \\
\Delta B & -0.140 & -0.048 & \\
\Delta V & -0.108 & -0.037 & \\
\Delta R_c & -0.083 & -0.029 & \\
\Delta I_c & -0.064 & -0.022 & \\
\end{array}
\right. .
\end{equation}

\subsection{Redshift Success Rates and Statistical Weights} \label{weights}

\begin{figure}
\plotone{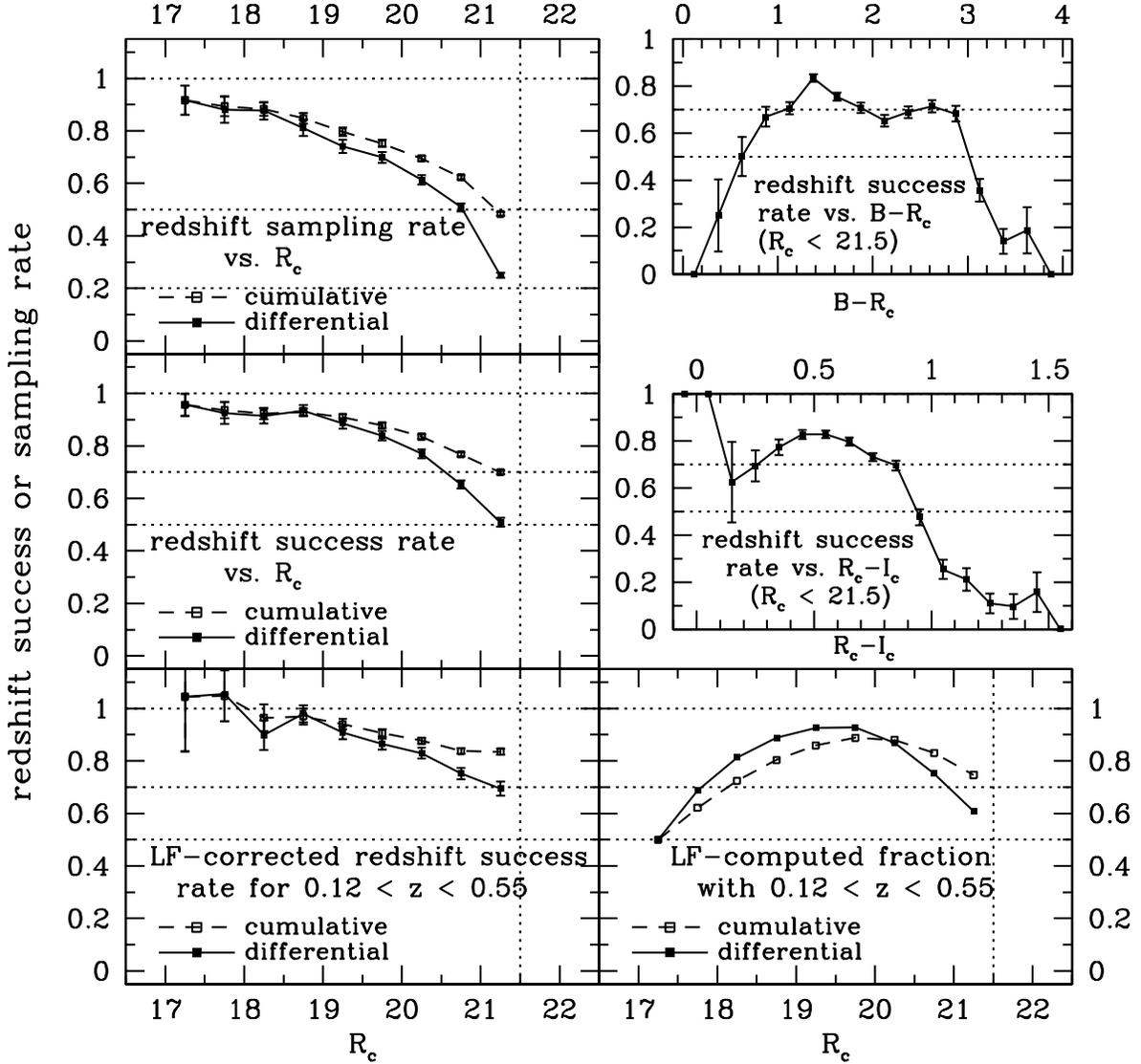}
\caption{
 ({\em Top left}) Redshift sampling rate (fraction of {\em all}
  galaxies with redshifts) as a function of apparent magnitude $R_c$;
 ({\em middle left}) redshift success rate (fraction of {\em
   spectroscopically observed} galaxies with redshifts) vs.\ $R_c$;
 ({\em top 2 panels on right}) redshift success rate vs.\ $B-R_c$ and
  $R_c-I_c$ colors; 
 ({\em bottom right}) fraction of all galaxies with $0.12 < z < 0.55$
  as a function of $R_c$, computed
  from the best-fit evolving 
  $B_{AB}$ luminosity function derived in \S~\ref{Bresults};
 ({\em bottom left}) luminosity-function corrected
  redshift success rate vs.\ $R_c$, appropriate 
  for galaxies within the nominal $0.12 < z <
  0.55$ completeness limits (see discussion in text).
 All uncertainties are calculated assuming simple $\protect\sqrt{N}$ errors.
 }
\label{figsf}
\end{figure}

For reasons of observational efficiency, we cannot target for
spectroscopy all galaxies in our fields, and like
the majority of other redshift surveys, we do not successfully measure
a redshift from every spectrum.
We thus need to derive a set of statistical weights so that we can 
account for incompleteness in the CNOC2 redshift sample for
our luminosity function and other analyses.
Figure~\ref{figsf} (top left panel) shows our redshift sampling rate
as a function of apparent magnitude $R_c$, where the redshift sampling rate
is defined as the fraction of galaxies with redshifts
among all galaxies in our photometric catalog.
The differential redshift sampling rate is about 20\%
at the nominal spectroscopic completeness limit $R_c = 21.5$, 
and the cumulative sampling rate is about 50\% for 
$17.0 < R_c < 21.5$.
Since we do not put a spectroscopic slit on every object, the redshift
sampling rate is different from the redshift success rate,
defined as the fraction of {\em spectroscopically
  observed} galaxies with redshifts.
Our redshift success rate is also plotted in 
Figure~\ref{figsf} (middle left panel).
As expected, the redshift success rate declines with fainter apparent
magnitude and hence decreasing signal-to-noise ratio in our spectra. 
The {\em raw} success rate ranges from over 90\% for $R_c < 19$
to about 50\% at $R_c = 21.5$, with an overall cumulative success rate of 70\%
for $17.0 < R_c < 21.5$.
However, this raw success rate is biased low by galaxies with 
redshifts outside the nominal CNOC2 $0.12 < z < 0.55$ 
completeness range.
As we will show at the end of this section, we can correct for this
bias and estimate a {\em corrected} redshift success
rate solely for $0.12 < z < 0.55$ galaxies; this improves the 
success rate to 70\% at $R_c = 21.5$, and to about
85\% cumulatively for $17.0 < R_c < 21.5$.

The simplest way to derive a statistical weight is just to use the
inverse of the redshift sampling rate. 
This will be correct if, at each value of $R_c$, 
the spectroscopically failed objects constitute the same population
as the spectroscopically successful ones,
in terms of the distribution of both their spectral types and redshifts.
However, this will not be true in general, as our ability to measure a
redshift will be a function of the spectral type and redshift of a
particular galaxy.
For example, given an early- and a late-type galaxy with the same
apparent magnitude $R_c$,
the early-type galaxy will yield a lower signal-to-noise CNOC2
spectrum (4400-6300\AA) because of its redder spectral energy distribution.
Moreover, our finite spectral window means that our redshift failures
will be biased toward objects outside of our nominal $0.12 < z < 0.55$
completeness range.
We illustrate these points by plotting
in Figure~\ref{figsf} 
the redshift success rate as a function of $B-R_c$ 
(top right panel) and $R_c-I_c$ (middle right) colors,
showing that there are indeed some obvious color and hence galaxy type 
dependences in our success rate
(but note that the same observed color can result from galaxies 
of somewhat different types spread out across a range of redshifts).
Note in particular that the steep decline seen in the success rate for 
$B-R_c \gtrsim 3$ or $R_c-I_c \gtrsim 1$ is caused by higher-redshift 
($z \gtrsim 0.55$) early-type galaxies whose Ca~II H+K features have
shifted out of our spectral window
(see \S~\ref{classification}).
Also, the success rate drops for $B-R_c \lesssim 1$ because
of lower-redshift ($z \lesssim 0.2$) late-type galaxies for which the 
[OII] $\lambda$3727 emission line is likewise outside the spectral
window.

One way to proceed is to apply photometric redshift methods (e.g., \cite{saw97}; 
\cite{con95}; \cite{koo85}) on our $U \! BV \! R_c I_c$ data
to obtain approximate redshifts and spectral types for all galaxies
in our photometric catalog (with or without spectroscopic redshifts),
and subsequently derive an
estimate of our spectroscopic redshift sampling rate as a function of
intrinsic galaxy type and redshift, as well as of $R_c$.
However, this procedure will be postponed to a future paper analyzing the full
4-patch CNOC2 data set.
For the present paper we adopt a simpler but sufficient procedure, and
will just estimate our redshift sampling rates in joint bins of
$R_c$, $B-R_c$, and $R_c-I_c$.
Specifically, for a galaxy $i$ which has a redshift, 
we will define its statistical weight $W_i$ (the inverse of the
redshift sampling rate) by
\begin{equation} \label{eqwt}
W_i \equiv \frac{N[R_{c,i} , \ (B-R_c)_i , \ (R_c-I_c)_i]}
                {N_z[R_{c,i} , \ (B-R_c)_i , \ (R_c-I_c)_i]} \ ,
\end{equation}
that is, just the ratio of all galaxies $N$ to those galaxies with 
redshifts $N_z$, but where both $N$ and $N_z$ include only those galaxies
$j$ that lie within the following magnitude and color bounds
relative to galaxy $i$:
\begin{equation} \label{eqwtbds}
\left\{
\begin{array}{l}
 \vert R_{c,i} - R_{c,j} \vert \leq 0.25 \\
 \vert (B-R_c)_i - (B-R_c)_j \vert \leq 0.25 \\ 
 \vert (R_c-I_c)_i - (R_c-I_c)_j \vert \leq 0.1 
\end{array}
\right. . 
\end{equation}
This particular magnitude- and color-dependent weighting
scheme does in fact account for our somewhat complicated redshift 
selection effects; that we obtain sensible results will be shown below in
\S~\ref{counts} where we compare our LF-computed number counts and color
distributions with the observations. 
Note also that we are using the sample as a whole to calculate our weights.
We are thus ignoring some real field-to-field variations in our redshift
success rate, due primarily to observational factors, in particular
seeing.
These variations need to be accounted for in galaxy clustering
analyses, but should not be important for the LF analysis of this
paper.
A more detailed discussion of our selection effects will be found in 
Yee et al.\ (1999).

Next, we show in Figure~\ref{figsf} (bottom right panel) the fraction
of galaxies, as a function of $R_c$, 
within our $0.12 < z < 0.55$ redshift completeness range,
computed using the best-fit evolving $B$-band LF
that we will obtain below in \S~\ref{Bresults}.
This fraction peaks at about 90\% at $R_c \approx 19.5$, but declines
to about 60\% at $R_c \approx 17.5$ and $R_c \approx 21.5$;
overall, the cumulative fraction is about 75\% for $17.0 < R_c < 21.5$.
We can then compute, as a function of $R_c$,
a corrected redshift success rate $f_{\rm corrected}$
for $0.12 < z < 0.55$ galaxies using
\begin{equation}
f_{\rm corrected}(R_c) = \frac{N_z(R_c)}{N_{obs}(R_c) \times F_z(R_c)} \ ,
\end{equation}
where $N_z$ is the number of galaxies with redshifts in the
range $0.12 < z < 0.55$, $N_{obs}$ is the total number of
spectroscopically observed galaxies, and $F_z$ is the LF-derived
fraction of galaxies with $0.12 < z < 0.55$.
This more relevant corrected success rate $f_{\rm corrected}$ 
is also plotted in Figure~\ref{figsf} 
(bottom left panel), where we see that the differential rate improves to about
70\% at $R_c = 21.5$ and the cumulative rate increases to about
85\% for $17.0 < R_c < 21.5$.
These success rates are comparable to those obtained in other large
intermediate-$z$ surveys (e.g., \cite{cra95}; \cite{ell96}).
Also, in \S~\ref{zincomplete} below, we 
discuss the impact of any potential residual $z$-dependent
incompleteness on our results.
Finally, note that $f_{\rm corrected}$ is computed for illustrative
purposes only; it is {\em not} used to weight the data 
in any of our analyses.

\subsection{Surface Brightness Selection Effects} \label{SB}

\begin{figure}
\plotone{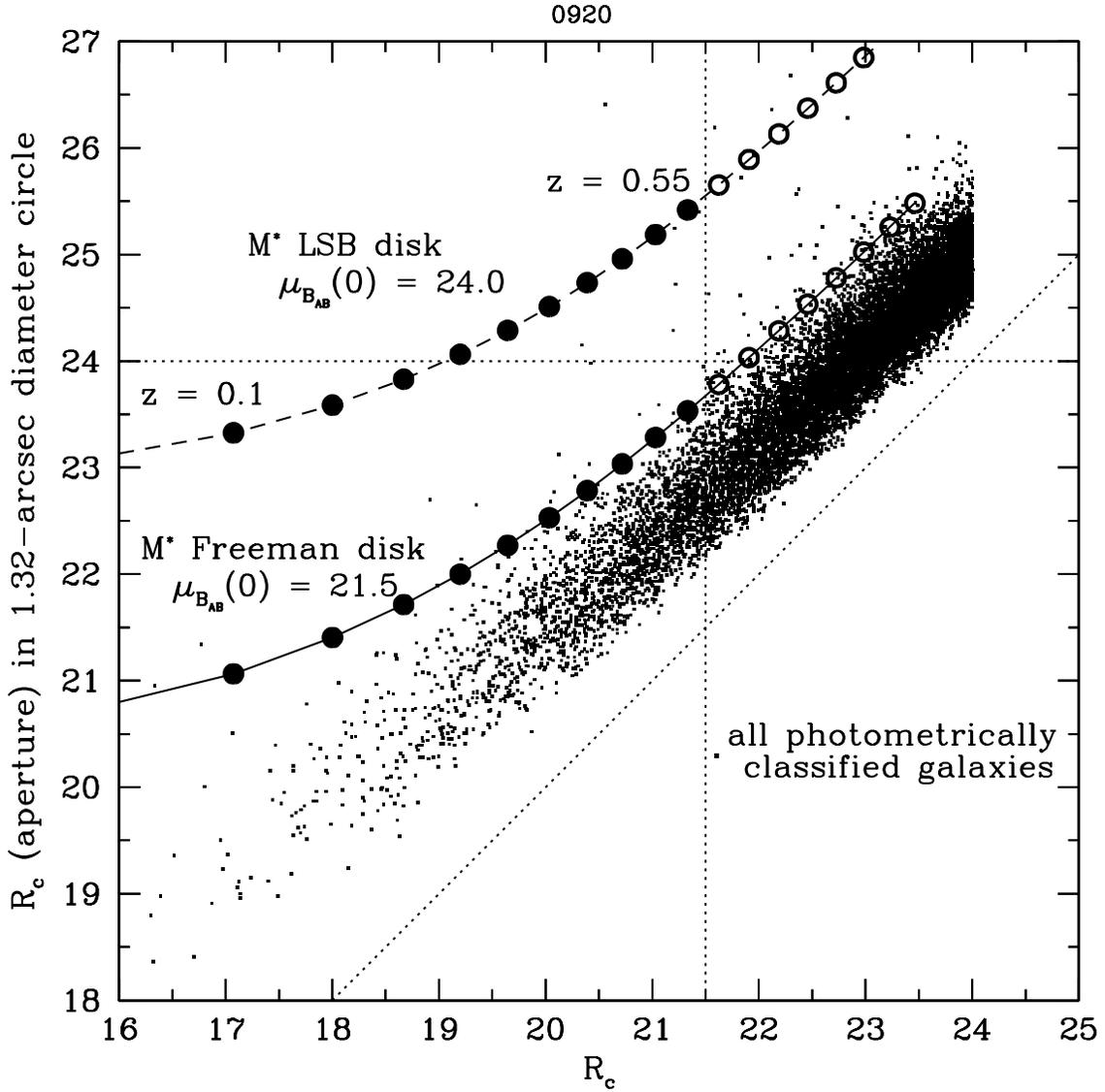}
\caption{
 Central aperture magnitude $R_c$(aperture), defined in a
  1.32$\arcsec$-diameter circle, vs.\ apparent magnitude $R_c$, 
  plotted for objects
  photometrically classified as galaxies or probable galaxies in the
  CNOC2 0920 patch. 
 The dotted diagonal line is just $R_c = R_c$(aperture), the 
  dotted vertical line indicates the $R_c =
  21.5$ nominal spectroscopic limit, and the dotted horizontal line is
  the approximate central aperture magnitude limit $R_c$(aperture) $=
  24.0$ (see discussion in text). 
 Also plotted are the redshift tracks
  for two face-on disk galaxies, one a Freeman disk ({\em lower solid curve})
  with central surface brightness $\mu_{B_{AB}}(0) = 21.5$ mag arcsec$^{-2}$, 
  and the other a
  low surface brightness (LSB) disk ({\em upper dashed curve})
  with $\mu_{B_{AB}}(0) = 24.0$ mag arcsec$^{-2}$. 
 Both galaxies have absolute magnitude 
  $M_{B_{AB}} = -19.5 + 5 \log \ h \approx M^*$. 
 The tracks are calculated for 1$\arcsec$ seeing and 
  using $k$-corrections for an Sbc galaxy (CWW).
 The circles indicate redshifts at intervals
  $\Delta z = 0.05$, starting at $z = 0.1$ on the left. 
 The filled circles denote the nominal CNOC2 redshift completeness range
  $0.10 \lesssim z < 0.55$.
 }
\label{figsb}
\end{figure}

Unaccounted surface brightness selection effects may seriously 
bias calculation of the luminosity function,
especially for low-redshift samples with relatively
shallow photometry and bright limiting isophotes
(e.g., \cite{fer95}; \cite{dal98}).
Although surface brightness selection effects are less problematic for
intermediate-$z$ surveys like CNOC2 with deeper imaging, it is 
nonetheless important to quantify the survey's effective surface brightness
limits.
We do so in Figure~\ref{figsb}, where we plot apparent magnitude $R_c$ vs.\
a central aperture magnitude $R_c$(aperture), 
for CNOC2 objects classified as galaxies or probable galaxies by PPP.
$R_c$(aperture) will serve as our measure of the central surface
brightness, and the aperture used is a circle with 
diameter 1.32$\arcsec$ (corresponding to 3
pixels for our largest-pixel-size STIS2 CCD).
The vertical line indicates the nominal $R_c = 21.5$ spectroscopic
limit, and the horizontal line is our estimate of the central surface
brightness completeness limit, $R_c$(aperture) $= 24.0$ (or 24.3 $R_c$
mag arcsec$^{-2}$).
The latter limit is conservatively 
estimated as 0.5 mag brighter than the turnover in
the number count histogram for $R_c$(aperture).
Also plotted is the track, as a function of redshift, for a fiducial
face-on exponential disk galaxy with the Freeman (1970) central surface
brightness value $\mu_{B_{AB}}(0) = 21.5$ mag arcsec$^{-2}$ and an absolute
magnitude $M_{B_{AB}} = -19.5 + 5 \log h \approx M^*_B$.
The assumed seeing is a Moffat profile with 1$\arcsec$ FWHM, nearly
the CNOC2 average (0.9$\arcsec$).
We calculate $k$-corrections using an Sbc galaxy spectral energy
distribution (\cite{cww}); we have checked that using an E or Im SED
instead makes little difference for our conclusions.

The vast majority of our galaxies lie brightwards in central surface brightness
relative to the $M^*$ Freeman disk track, even though we should be
sensitive to lower surface brightness objects;
this is very similar
to what Lilly et al.\ (1995a) found in a completely analogous plot for the 
Canada-France Redshift Survey.
Note that our Freeman disk model is a pure exponential disk only, so
that the addition of a bulge component needed for a more realistic
galaxy would result immediately in a higher central surface brightness.
Likewise, a sub-$M^*$ Freeman disk or an inclined $M^*$ Freeman disk 
will also have tracks that are everywhere brighter in $R_c$(aperture) (vertically
below in the plot) compared to the face-on $M^*$ Freeman disk track shown.
These other tracks pass more centrally through the observed galaxy distribution,
but the $M^*$ Freeman disk track serves as a useful central surface
brightness lower bound for the vast majority of galaxies.

Given the redshift track of a particular type of
galaxy, our survey will be flux-limited with respect to that galaxy
type {\em if} the track first crosses the $R_c = 21.5$ vertical boundary instead
of the $R_c$(aperture) $= 24.0$ horizontal boundary;
otherwise we will need to consider the surface brightness limit
explicitly in our analyses.
Figure~\ref{figsb} shows that we are indeed flux-limited with respect
to the $M^*$ Freeman track across the entire CNOC2 nominal redshift 
range $0.12 < z < 0.55$, and since that Freeman track is basically
a lower surface brightness bound for the bulk of our galaxies, 
we may conclude that the $R_c < 21.5$ CNOC2 photometric sample is
essentially free of central surface brightness selection effects.

Also shown in Figure~\ref{figsb} is the track for an $M^*$ low surface
brightness (LSB) disk galaxy with $\mu_{B_{AB}}(0) = 24.0$ mag
arcsec$^{-2}$, ten times fainter than that of a Freeman disk.
This LSB disk crosses our central surface brightness boundary and 
exits our sample by $z \approx 0.25$.
We are thus not complete in surface brightness to this type of LSB galaxy
over the full CNOC2 redshift range. 
Nevertheless, as Figure~\ref{figsb} shows,
we should still be sensitive to galaxies with somewhat higher
surface brightnesses 
(but still fainter than the $M^*$ Freeman track) over a fairly broad
range of redshift and apparent magnitude. 
However, very few of these LSB galaxies 
faintwards of the $M^*$ Freeman track are detected within our sample.
Hence, the number of these LSB galaxies 
is apparently quite small compared to that of the more ``normal''
objects to which the survey is complete.
We have not attempted to check whether the number density of our ``LSB'' galaxies
is quantitatively consistent with recent results at low redshifts (e.g.,
\cite{spr97}; see review by \cite{ib97}), as that would take us too
far afield, requiring us to examine detailed issues of 
surface brightness measurements, LSB galaxy definitions, surface
brightness evolution from intermediate to low redshifts, etc.
We will however explore these issues in future analyses of the galaxy surface
brightness distributions in the CNOC2 sample.

\section{Methods} \label{methods}

\subsection{Galaxy Classification} \label{classification}

\begin{figure}
\plotone{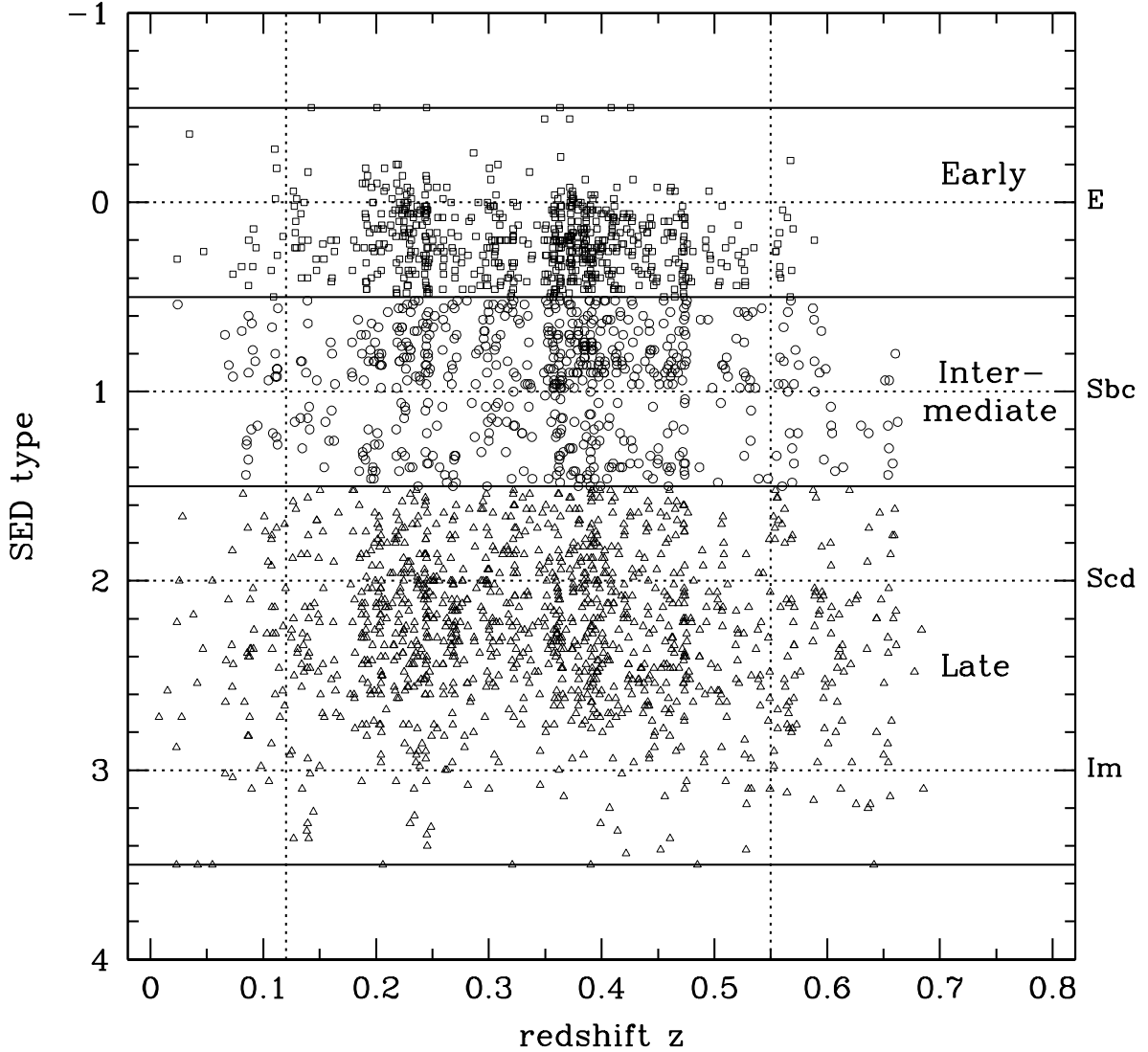}
\caption{
 Spectral energy distribution (SED) type vs.\ redshift for
  CNOC2 galaxies with $R_c < 21.5$.
 The SED types are determined by least-squares fits of 
  $U \! B V \! R_cI_c$ magnitudes to those computed from the SED's of
  Coleman, Wu, \& Weedman (1980; CWW), as described in the text. 
 Numerical SED types for the four original CWW SED's are assigned as
  indicated by the dotted horizontal lines,
  while the boundaries defining the ``Early'' ({\em squares}), 
  ``Intermediate'' ({\em circles}), and ``Late'' ({\em triangles})
  CNOC2 galaxy categories are indicated by the solid
  horizontal lines.
 The dotted vertical lines indicate the CNOC2 nominal redshift
  completeness range $0.12 < z < 0.55$.
 }
\label{figsed}
\end{figure}

We classify CNOC2 galaxies using least-squares fits of our
$U \! B V \! R_c I_c$ colors to those computed from the galaxy spectral energy
distributions (SED's) of Coleman, Wu, \& Weedman (1980; hereafter CWW). 
The CWW colors are computed using filter transmission curves taken
from Buser \& Kurucz (1978) for $U \! B V$, and from Bessel (1990) for
$R_c I_c$.
As shown in Figure~\ref{figsed}, we assign numerical 
values to the four CWW SED's as follows: 0 = E (average of the M31
bulge and M81 bulge SED's), 1 = Sbc, 2 = Scd, 3 = Im. We linearly
interpolate between neighboring SED's using 50 equal steps in the
computed broadband magnitudes,
and also allow linear extrapolations to SED types $-0.5$ and $+3.5$.
The best-fitting SED type is plotted against redshift in
Figure~\ref{figsed} for CNOC2 galaxies with $R_c < 21.5$. 
The galaxies are assigned to three categories, ``Early,'' ``Intermediate,''
and ``Late'' according to:
\begin{equation}
\begin{array}{lrcr}
{\rm Early} & -0.50 & \leq {\rm SED \ type} < & 0.50 \nonumber \\
{\rm Intermediate} & 0.50 & \leq {\rm SED \ type} < & 1.50  \\
{\rm Late} & 1.50 & \leq {\rm SED \ type} \leq & 3.50 \nonumber 
\end{array} .
\end{equation}
Before fitting, we also add
$-0.05$ mag to the $I_c$ magnitudes computed 
from the E, Sbc, and Scd SED's (but not the Im SED), in order to
empirically match the observed $R_c-I_c$ colors of CNOC2 galaxies, 
but otherwise we make no further adjustments to the CWW SED's.
This is an ad hoc procedure and may be symptomatic of a general 
limitation of the CWW SED set, specifically that they are based 
on a very small number of observed local galaxies which should not be
expected to represent the full galaxy population in
every exacting detail.
However, lacking a better SED set, we nonetheless choose the CWW
set for simplicity, and note that aside from the above exception, 
the CWW galaxy colors do give a reasonable match to the observed CNOC2 
galaxy colors.
In \S~\ref{description}, we will discuss the general implications 
of the particular choice of SED set and classification scheme on 
our galaxy evolution results.
We also do not attempt to fit for dust extinction in the CNOC2 galaxies
themselves (though we do correct for Milky Way extinction as described
in \S~\ref{data}); this will be addressed instead in Paper II.

Note that our SED types are ``stellar population'' types derived from
broadband galaxy colors, and would be more closely related to 
classifications derived from galaxy spectra than from galaxy 
morphologies.
{\em We stress that our galaxy SED types are not and should not be
interpreted as morphological types}.
Of course there are correlations between galaxy types derived
separately from colors, spectra, and morphologies; we postpone an
examination of the similarities and differences among these various 
classification schemes to future CNOC2 papers.

\begin{figure}
\plotone{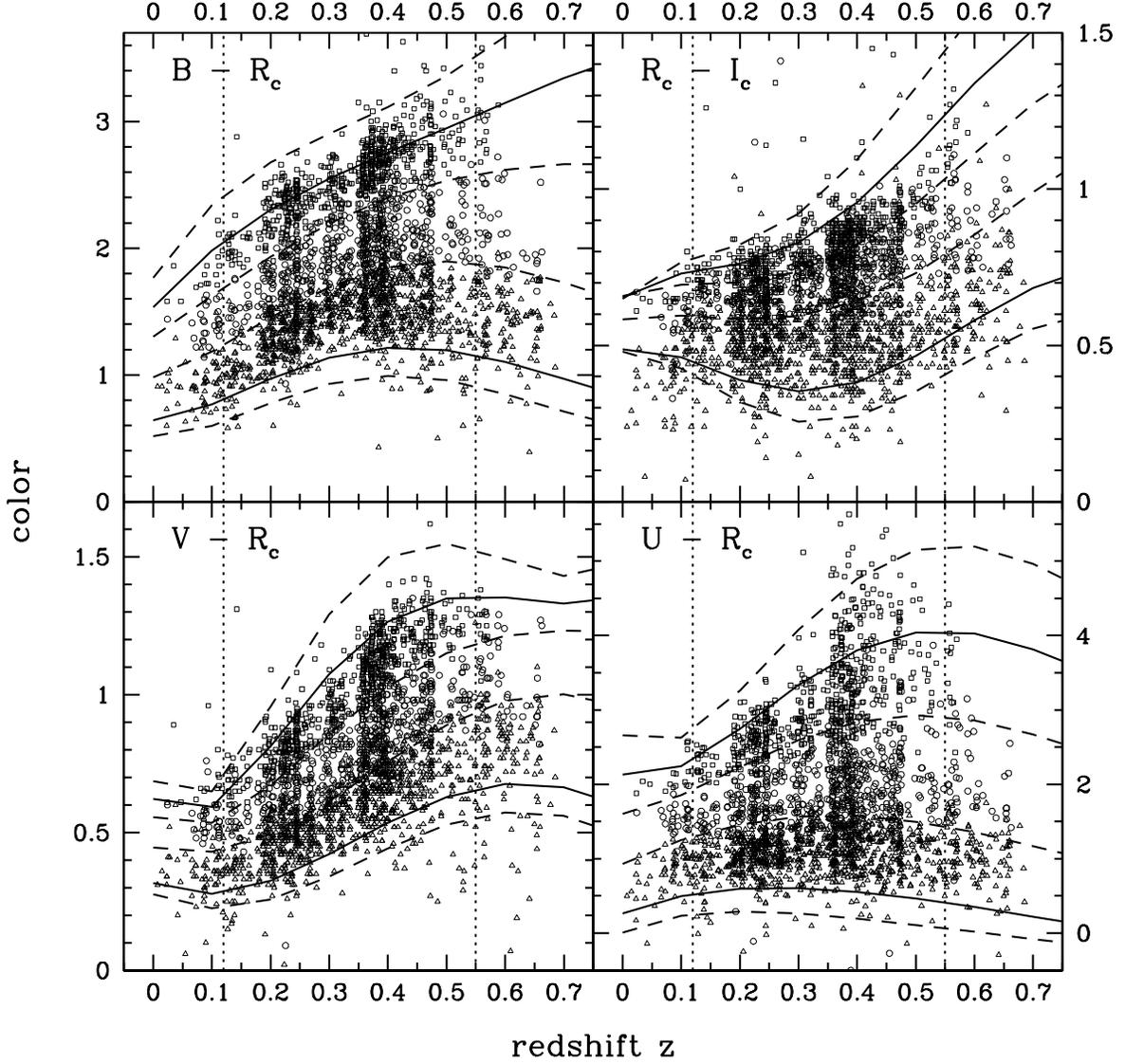}
\caption{
 Plot of various observed colors vs.\ redshift for CNOC2 galaxies 
  with $R_c < 21.5$: $B-R_c$ ({\em top left}), $R_c-I_c$ ({\em top
  right}), $V-R_c$ ({\em bottom left}), and $U-R_c$ ({\em bottom
  right}).
 The {\em upper and lower solid curves} show the colors for the
 original CWW E and Im SED's, respectively.
 The {\em dashed curves} (corresponding to the solid lines 
  in Figure~\protect\ref{figsed})
  show the colors for those interpolated and
  extrapolated CWW SED types which
  define the boundaries of early ({\em squares}), 
  intermediate ({\em circles}) and late ({\em triangles}) CNOC2 galaxies
  (the same classifications as in Figure~\protect\ref{figsed}).
 The dotted vertical lines indicate the CNOC2 nominal redshift
  completeness range $0.12 < z < 0.55$.
 }
\label{figcol4}
\end{figure}

The visual impression from Figure~\ref{figsed} is that there are no
obvious type- or redshift-dependent incompletenesses, except for the
lack of galaxies earlier than Sbc at redshifts $z \gtrsim 0.6$ (Ca~II
H+K redshifts out of our spectral window 4400-6300\AA) and the
dearth of intermediate-type galaxies at $z \lesssim 0.05$; 
both cases are outside our redshift completeness range $0.12 < z < 0.55$.
Also, there are no redshifts at $z > 0.7$ 
because [OII] $\lambda$3727 redshifts beyond the red end of our spectra.
(The exceptions are a handful of higher-$z$ AGN's/QSO's that are excluded from
our analysis and are not plotted.)
Figure~\ref{figcol4} compares the $B-R_c$, $R_c-I_c$, $V-R_c$, and $U-R_c$
colors observed for CNOC2 galaxies with those computed from the CWW
SED's, showing that the SED's do indeed span the range of actual
galaxy colors.

We will compute absolute magnitudes, $k$-corrections and other needed
quantities using the best-fitting (interpolated or extrapolated)
CWW SED for each individual galaxy.
Note in particular that the absolute magnitudes we use (in $B$,
$R_c$, or $U$) are calculated directly from the best-fitting SED
(thus making full use of the available $U \! B V \! R_c I_c$ data),
rather than from any single apparent magnitude.
For example, the $U$ absolute
magnitude would {\em not} be derived by adding $-25 - 5 \log d_L$ and a
$k$-correction to the $U$ apparent magnitude; 
instead, we would calculate the $U$ absolute magnitude by direct integration
of the best-fit SED convolved with the $U$-band filter response function.

\subsection{Computing the LF} \label{computeLF}

We compute the luminosity function using standard maximum-likelihood
methods (Sandage, Tammann, \& Yahil 1979; Efstathiou, Ellis, \&
Peterson 1988, hereafter EEP) which are unbiased by density
inhomogeneities in the galaxy distribution. 
Our procedure essentially follows that given in 
Lin et al.\ (1996a, 1997), and is only summarized briefly here,
except that we will describe in more detail our present methods for
parameterizing and fitting the {\em evolution} of the luminosity function.

Given a survey of $N$ galaxies at redshifts $z_i$, 
we form the likelihood $\cal L$ for those galaxies to possess their observed
absolute magnitudes $M_i$:
\begin{equation}
{\ln \cal L} \equiv \ln p(M_1, \ldots , M_N \vert z_1, \ldots , z_N)
            = \sum_{i=1}^{N} W_i \ln p_i + {\rm constant} \ .
\end{equation}
Here $W_i$ is the weight described previously in \S~\ref{weights}, and
$p_i$ is the individual conditional probability
\begin{equation}
p_i \equiv p(M_i \vert z_i) \propto
    \phi(M_i) \left/ \int^{\min[M_{\rm max}(z), M_2]}_{\max[M_{\rm min}(z), M_1]}
                     \phi(M) dM \right. \ ,
\end{equation}
where $M_1$ and $M_2$ are the global absolute magnitude limits we
impose on the sample ($M_1 < M_i < M_2$),
$M_{\rm min}$ and $M_{\rm max}$ are the
absolute magnitude limits at $z_i$ that correspond to the 
survey's apparent magnitude limits, and $\phi(M)$ is the differential
luminosity function whose parameters we determine by maximizing $\ln
\cal L$.

For the form of $\phi(M)$, we adopt the usual Schechter (1976)
parameterization,
\begin{equation} \label{eqschech}
\phi(M) = (0.4 \ln 10) \ \phi^* \ [10^{0.4(M^*-M)}]^{1+\alpha} \
                         \exp [-10^{0.4(M^*-M)}] \ ,
\end{equation}
with characteristic magnitude $M^*$, faint-end slope $\alpha$, and
normalization $\phi^*$. 
We also use the nonparametric ``steps'' function of EEP,
\begin{equation}
\phi(M) = \phi_k \ , \ \ M_k - \Delta M / 2 < M < M_k + \Delta M / 2 \ , \ \
                     k = 1, \ \ldots , \ N_p \ ,
\end{equation}
which we refer to hereafter as the SWML (stepwise maximum
likelihood) LF. The details for computing $\phi(M)$ via maximum
likelihood and for estimating errors are as given in EEP and Lin et
al.\ (1996a, 1997).

To parameterize evolution in the luminosity function, we adopt the
following simple model for the redshift dependence of the Schechter parameters:
\begin{eqnarray}
M^*(z) & = & M^*(0) - Q z \nonumber \\ 
       & = & M^*(z=0.3) - Q (z-0.3) \nonumber \\ 
\alpha(z) & = & \alpha(0) \label{eqmodel} \\
\rho(z) & = & \rho(0) 10^{0.4 P z} \nonumber \ .
\end{eqnarray} 
We thus take $M^*$ to vary linearly with redshift, at a rate
quantified by $Q$, which we call 
the $M^*$ or luminosity evolution parameter. 
Note that we will fit for $M^*(z=0.3)$ since this is a better 
constrained quantity than $M^*(0)$, 
given the mean redshift $z \approx 0.3$ for CNOC2 galaxies.
We also make the null assumption that $\alpha$ does not change with
redshift, so that the shape of the LF stays the same. 
Since $\alpha$ is fixed, the normalization parameter $\phi^*$
and the total galaxy number density $\rho = \int \phi(M) dM$ are
essentially equivalent.
We then take $\rho$ to vary with
$z$ as determined by the density evolution parameter $P$ defined above.
The expression for $\rho$ in equation~(\ref{eqmodel}) is chosen for
convenience, so that the luminosity density $\rho_L = \int L \phi(M) dM$
(where $L \propto 10^{-0.4 M}$) may be written as:
\begin{equation}
\rho_L(z) = \rho_L(0) 10^{0.4 (P+Q) z} \ ,
\end{equation}
where $P+Q$ then measures the linear rate of evolution of $\rho_L$ with redshift.
Also, note that $\rho$ as defined above may be approximated by 
$\rho(z) \approx \rho(0) (1+Pz)$, so that
$P$ is merely the coefficient of the linear term in the expansion of
$\rho$ in powers of $z$.

We first estimate $M^*(z=0.3)$, $\alpha$, and $Q$ together using the usual
maximum likelihood method, and by design this is independent of
density fluctuations or density evolution, so that both $\phi^*(0)$
and $P$ have to be determined separately, beginning with $P$.
In the case of a non-evolving LF, it is possible to derive 
maximum-likelihood estimates of 
$\rho(z)$ without prior knowledge of the luminosity function
(\cite{sau90}; \cite{lov92}; \cite{fis92}). 
This is completely analogous to the case above where we may estimate
$M^*$ and $\alpha$ independently of galaxy density variations. 
For the more general evolving LF defined above, we may still determine
$P$ without knowing $M^*(z=0.3)$ or $\alpha$, but not without first knowing 
$Q$. 
Given a value of $Q$, we may convert an observed absolute
magnitude $M_i$ at $z = z_i$ to an evolution-corrected absolute
magnitude at some fiducial redshift, say $z = 0$ (the actual redshift
does not matter): $M_i(0) \equiv M_i(z_i) + Q z_i$. 
Then, with any given $Q$ and the resulting set of $M_i(0)$, we compute the
likelihood that those galaxies will have their observed
redshifts $z_i$:
\begin{equation}
\ln {\cal L}^\prime \equiv 
          \ln p^\prime (z_1, \ldots , z_N \vert M_1(0), \ldots , M_N(0), Q)
            = \sum_{i=1}^{N} W_i \ln p^\prime_i + {\rm constant} \ .
\end{equation}
Here the individual conditional probabilities are (cf. \cite{sau90};
\cite{fis92}) 
\begin{eqnarray}
p^\prime_i & \equiv & p(z_i \vert M_i(0), Q) \nonumber \\
           & \propto &
  \phi^\prime(M_i(0), z=0) \ \rho(z_i) 
  \left/ \int^{\min[z_{\rm max}(M_i(0)), z_2]}_{\max[z_{\rm min}(M_i(0)), z_1]}
         \phi^\prime(M_i(0), z=0) \ \rho(z) \ \frac{dV}{dz} \ dz \right.
         \nonumber \\
           & = &
  \rho(z_i) 
  \left/ \int^{\min[z_{\rm max}(M_i(0)), z_2]}_{\max[z_{\rm min}(M_i(0)), z_1]}
         \rho(z) \ \frac{dV}{dz} \ dz \right. \nonumber \\
           & = &
  10^{0.4 P z_i}
  \left/ \int^{\min[z_{\rm max}(M_i(0)), z_2]}_{\max[z_{\rm min}(M_i(0)), z_1]}
         10^{0.4 P z} \ \frac{dV}{dz} dz \right. \ ,
\end{eqnarray}
where $\phi^\prime$ is $\phi$ with $\phi^*$ set to unity
($\phi^\prime$ has units of mag$^{-1}$),
$z_1$ and $z_2$ are the global redshift limits we
impose on the sample, and
$z_{\rm min}$ and $z_{\rm max}$ are the redshift limits over which
galaxy $i$ may be observed, given the
survey's apparent magnitude limits and the assumed rate of evolution
specified by $Q$. 
$P$ may then be readily determined using maximum likelihood,
once given the previously found best-fit value for $Q$. The fifth and final
parameter $\phi^*(0)$ is then computed via straightforward summation
(cf.\ Lin et al.\ 1996a, 1997)
\begin{equation}
\phi^*(0) = \frac{1}{V} \sum_i \frac{W_i}{S(z_i) 10^{0.4 P z_i}}
          \left/
            \int_{M_1}^{M_2} \phi^\prime(M,z=0) \ dM
          \right. \ ,
\end{equation}
where $V$ is the survey volume and 
$S(z)$ is the selection function, defined by
\begin{equation} \label{eqselfunc}
S(z) \equiv \int^{\min[M_{\rm max}(z), M_2]}_{\max[M_{\rm min}(z), M_1]}
            \phi(M,z) \ dM
     \left/ 
       \int^{M_2}_{M_1} \phi(M,z) \ dM
     \right. \ .
\end{equation}

Once we have fit for all the LF parameters, we will
calculate luminosity densities $\rho_L$ as a function of redshift using
\begin{equation} \label{eqrhoL}
\rho_L(z_a < z < z_b) = \frac{1}{V(z_a < z < z_b)} \sum_{z_a < z_i < z_b} 
     W_i \ 10^{-0.4 M_i} / S_L(z_i) \ ,
\end{equation}
where
\begin{equation} \label{eqsl}
S_L(z) = \int^{\min[M_{\rm max}(z), M_2]}_{\max[M_{\rm min}(z), M_1]}
            10^{-0.4 M} \phi(M,z) \ dM
     \left/ 
       \int 10^{-0.4 M} \phi(M,z) \ dM
     \right. \ .
\end{equation} 
That is, we sum over the luminosities of our observed galaxies, but
weighted by the factor $S_L(z)$, which uses the luminosity function 
$\phi$ to extrapolate
for the luminosity of unobserved galaxies lying outside the accessible
survey flux limits. 
Also, we will express $\rho_L$ in units of $h$ W Hz$^{-1}$ Mpc$^{-3}$
using the conversion given in Lilly et al.\ (1996); specifically,
one $M_{B_{AB}} = -19.5 + 5 \log h$ galaxy per Mpc$^3$ produces a luminosity
density of $2.85 \times 10^{21} \ h^{-2}$ W Hz$^{-1}$ Mpc$^{-3}$.

Finally, we estimate uncertainties in $\rho_L$ and $\phi^*(0)$ using
both bootstrap resampling and an estimate of the uncertainty contributed by
galaxy density fluctuations.
We apply bootstrap resampling (e.g., \cite{bar84}) to the
{\em full photometric} sample (not just to those galaxies with redshifts),
re-calculate statistical weights $W_i$ (as in \S~\ref{weights}) anew 
for galaxies with redshifts in each bootstrap resample,
and then re-fit our LF evolution model and re-compute  
luminosity densities.
This process should account for the uncertainties in $\phi^*(0)$ and
$\rho_L(z)$ contributed by sampling and weighting fluctuations and 
by our fitting procedure.
This does not account for the additional uncertainty arising from galaxy 
density fluctuations, which we estimate instead using an integral
over the galaxy clustering power spectrum $P(k)$; 
see \S~3 of Lin et al.\ (1997) for details.
For $P(k)$ we adopt the local result from the Las Campanas Redshift
Survey (\cite{lin96ps}, eqs.\ [23,24]), but adjusted (only) for the {\em linear}
clustering evolution at the higher redshifts sampled in CNOC2; 
this is done as appropriate for both the $q_0 = 0.5$ and 0.1 cosmologies we 
consider.
We then take the overall error on $\phi^*(0)$ and $\rho_L$ to be
the quadrature sum of the bootstrap resampling and density fluctuation
error contributions.

\section{Results} \label{results}

\subsection{Evolution of the $B_{AB}$-band LF} \label{Bresults}

We apply the LF fitting methods and evolution model of
\S~\ref{computeLF} to our nominally complete 
$17.0 < R_c < 21.5$ sample, subdivided into early,
intermediate, and late galaxy types as described in \S~\ref{classification}.
The sample details and fit parameters are given in Table~\ref{tablfB}.
We first concentrate on our LF results in the $B$ band, shown in
Figure~\ref{figlfsB} for
the three galaxy types and for each of three redshift bins in the
range $0.12 < z < 0.55$.
For ease of comparison against previous surveys, we will
report our $B$-band LF results in the $AB$ system (\cite{oke72})
using the transformation $B_{AB} = B - 0.14$ (\cite{fsi95}).
The points in the figure show the nonparametric SWML LF estimates in each
individual type-redshift bin, while the solid lines indicate the
results of our 5-parameter LF evolution model ($M^*(z=0.3)$, $\alpha$,
$\phi^*(0)$, $P$, and $Q$), fit to the full redshift completeness
range $0.12 < z < 0.55$ for each of the three galaxy types. 
Figure~\ref{figlfsB} should allow us to judge how well our parametric
LF model matches the nonparametric LF estimates, which we obtained without
making any assumptions about the form that the LF evolution takes.

\subsubsection{Some Technical Considerations}

\begin{figure}
\plotone{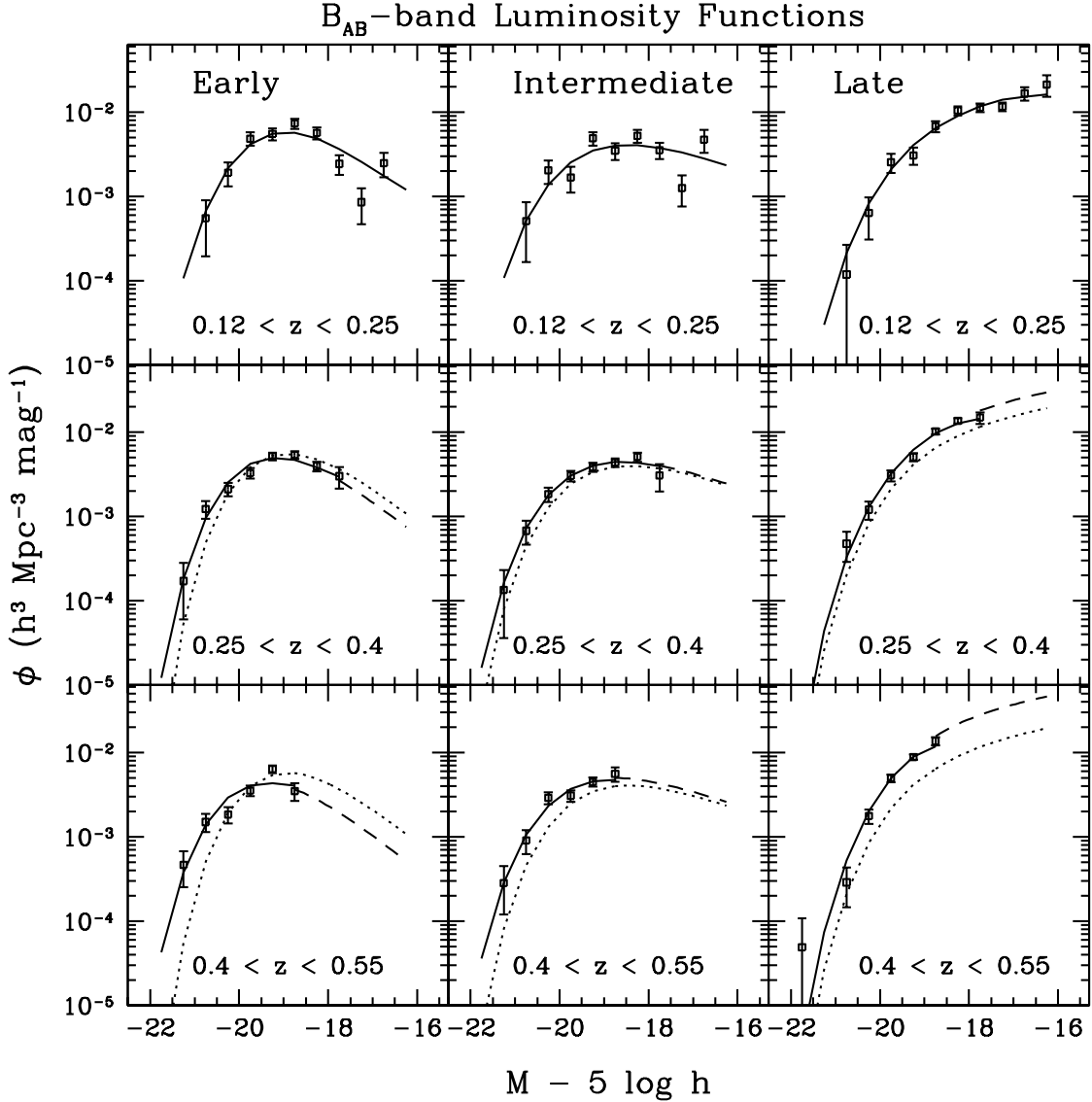}
\caption{
 $B_{AB}$-band luminosity functions for early-, intermediate-, and
  late-type ({\em left to right}) 
  CNOC2 galaxies, plotted for the three indicated redshift bins ({\em z
  increases from top to bottom}) in the
  range $0.12 < z < 0.55$.
 We show both our best-fit parametric evolving LF models ({\em solid
  curves}) as well as our nonparametric SWML LF estimates ({\em points
  with $1\sigma$ errors}).
 Also shown are fiducial LF's ({\em dotted curves}) 
  from the lowest-redshift bin for each galaxy type, and 
  extrapolations ({\em dashed curves}) 
  of our best-fit parametric LF to absolute magnitudes
  fainter than those accessible by the survey in each redshift bin.
 Results shown are for $q_0 = 0.5$.
 Please see text for a detailed discussion of the presentation of the
  data, as there are a number of subtleties involved.
 }
\label{figlfsB}
\end{figure}

There are, however, a number of subtleties involved in comparing the 
SWML LF estimates $\phi_k$ to the parametric evolving LF estimate
$\phi(M,z)$. 
The nonparametric $\phi_k$ are binned in both $M$ and $z$, while the
parametric $\phi(M,z)$ is not, and it may be unclear at what redshift
we should evaluate $\phi(M,z)$ in order to compare against the $\phi_k$.
For example, a simple procedure such as
plotting the parametric LF models evaluated
at the average redshift of each redshift bin in Figure~\ref{figlfsB} will
actually result in noticeable discrepancies (at the bright and faint
ends of the LF) when compared against the
$\phi_k$, even when there should be none.
The proper thing to do is actually to calculate a weighted
average of $\phi(M,z)$ over the appropriate intervals in $M$ and $z$.
Specifically, we follow a procedure given by EEP, but modified for our sample.
We note first that in general $\phi_k \neq \phi(M = M_k)$, even in the
absence of LF evolution.
As shown by EEP in their equation (2.15), 
the $\phi_k$ are actually related to the parametric $\phi(M)$ by
a weighted integral over $\phi(M)$, 
where the weights are just the expected number $N(M)$ of galaxies of 
absolute magnitude $M$ observable by the survey.
In the limit that the bin size $\Delta M \rightarrow 0$, and with no evolution,
$\phi_k$ would indeed converge to $\phi(M = M_k)$.
For CNOC2 we need to modify EEP's original equation (2.15) to
account for our use of an evolving LF, as well as 
for cosmological and $k$-correction effects important for our
intermediate-$z$ sample.
Specifically, for an absolute magnitude bin
$M_k - \Delta M/2 < M < M_k + \Delta M/2$ and a redshift bin 
$z_1 < z < z_2$, we may define the quantity
\begin{eqnarray}
\phi_{{\rm parametric},k}(z_1 < z < z_2)
  & \equiv & 
    \int^{M_k + \Delta M/2}_{M_k - \Delta M/2} 
     \int^{\min[z_{\rm max}(M),z2]}_{\max[z_{\rm min}(M),z1]} 
       \phi(M,z) \frac{dN}{dz dM}(M,z) \ dz \ dM
     \nonumber \\
  & & \left/
    \int^{M_k + \Delta M/2}_{M_k - \Delta M/2} 
     \int^{\min[z_{\rm max}(M),z2]}_{\max[z_{\rm min}(M),z1]} 
     \frac{dN}{dz dM}(M,z) \ dz \ dM 
    \right. \\ 
  & = &
    \int^{M_k + \Delta M/2}_{M_k - \Delta M/2} 
     \int^{\min[z_{\rm max}(M),z2]}_{\max[z_{\rm min}(M),z1]} 
     \phi^2(M,z) \frac{dV}{dz} \ dz \ dM \nonumber \\
  & &  \left/ 
    \int^{M_k + \Delta M/2}_{M_k - \Delta M/2} 
     \int^{\min[z_{\rm max}(M),z2]}_{\max[z_{\rm min}(M),z1]} 
     \phi(M,z) \frac{dV}{dz} \ dz \ dM 
    \right. \ , \label{eqphiint} 
\end{eqnarray}
where $\frac{dN}{dM}(M,z) = \phi(M,z) dV$ is the expected number of galaxies per
unit magnitude at redshift $z$, 
and $z_{\rm min}(M)$ and $z_{\rm max}(M)$ are the minimum
and maximum redshift, respectively, at which a galaxy of absolute
magnitude $M$ may be seen, given our survey's apparent magnitude
limits and cosmological and $k$-correction effects.
The parametric LF estimates we plot in Figure~\ref{figlfsB} and
elsewhere are those given by Equation~(\ref{eqphiint}); this is the
right way to compare our parametric LF fits against the 
directly-computed nonparametric SWML estimates.

In addition, in Figure~\ref{figlfsB} we also show low-redshift
fiducial LF's (dotted curves) to facilitate comparison from one redshift
bin to another.
For this purpose we use $\phi(M, z=0.175)$ (i.e., $\phi$ 
evaluated at nearly the average redshift of the lowest-$z$ bin), but
appropriately averaged using Equation~(\ref{eqphiint}) over the
higher-redshift intervals.
Also to facilitate bin-to-bin comparisons, 
we show extrapolations (dashed curves) of the parametric LF's 
faintwards of the faintest absolute magnitude accessible in each redshift bin.
For this purpose we simply choose $\phi(M, z=(z1+z2)/2)$ 
(i.e., $\phi$ evaluated at the average redshift of the bin), 
since $\phi_{{\rm parametric},k}$ defined above is zero at these
magnitudes (no galaxies observable there!).
Note the slight disconnections between the solid and dashed curves for
the late-type parametric LF's in the two highest-$z$ bins in
Figure~\ref{figlfsB}; these are artificial and are examples of the ``noticeable
discrepancies'' mentioned above.

\begin{figure}
\plotone{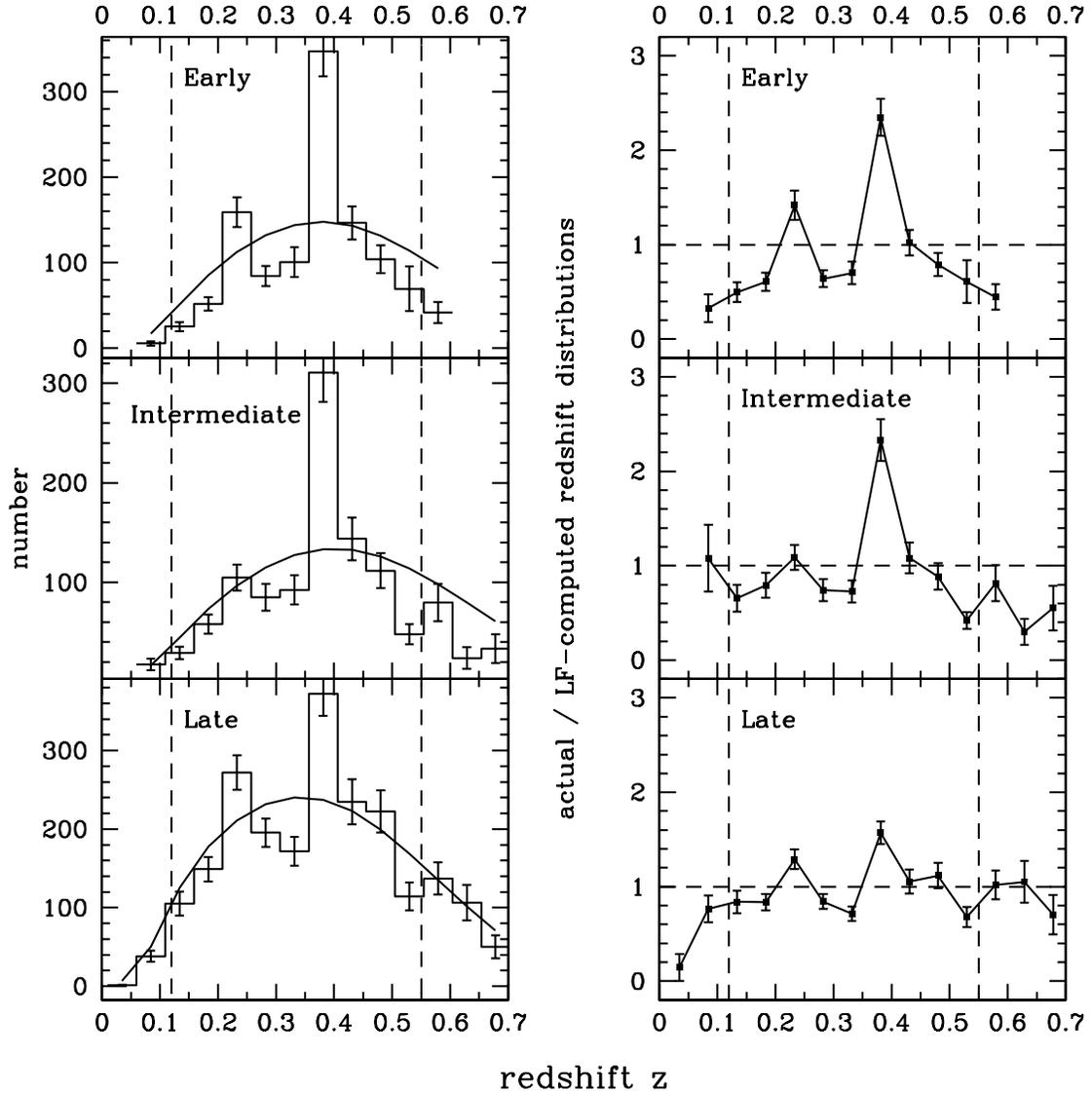}
\caption{
 ({\em Left 3 panels}) Weighted redshift histograms for early,
  intermediate, and late CNOC2 galaxies with
  apparent magnitudes $17.0 < R_c < 21.5$
  and absolute magnitudes $-22 < M_{B_{AB}} - 5 \log h < -16$ ($q_0 = 0.5$).
 The galaxies have been weighted by the factors $W_i$ defined in
  equation (\ref{eqwt}) to correct for incompleteness. 
 The smooth curves are the redshift distributions computed from the
  best-fit evolving $B_{AB}$ LF model derived in \S~\ref{Bresults}.
 ({\em Right 3 panels}) The ratio of the actual weighted redshift
 histogram to the LF-computed redshift histogram.
 The dashed vertical lines indicate the CNOC2 nominal redshift
  completeness range $0.12 < z < 0.55$.
 All uncertainties are computed assuming simple $\protect\sqrt{N}$ errors.
 }
\label{fighist}
\end{figure}

For clarity in seeing the evolution
trends, we have purposefully matched the normalizations of the 
SWML and parametric LF fits in each redshift bin of Figure~\ref{figlfsB}.
Specifically, we set
\begin{equation} \label{eqnorm}
\sum_{k = 1}^{N_p} \phi_k V(M_k) = 
\sum_{k = 1}^{N_p} \phi_{{\rm parametric},k} V(M_k) \ ,
\end{equation}
where $V(M_k)$ is the volume (within the redshift limits of each
bin) over which a galaxy of absolute magnitude $M_k$ may be seen in
our survey.
We do this to take out the
effects of strong density fluctuations present in the survey,
clearly seen in the (weighted) redshift histograms shown in Figure~\ref{fighist}
(left panels), particularly for early- and intermediate-type galaxies
in the $0.25 < z < 0.4$ bin.
Note from Figure~\ref{fighist} (right panels) that the
ratio of actual to LF-computed redshift distributions are reasonably centered
on unity and do not show conspicuous systematic trends with redshift,
indicating that the normalization $\phi^*(0)$ and the 
number density evolution parameter $P$ in our fits
are indeed good matches to the data.
(To construct the galaxy redshift histograms, we have first weighted 
each galaxy by $W_i$ to correct for redshift incompleteness; this
makes construction of the corresponding LF-computed redshift histogram
much simpler, since we are then freed from modeling the somewhat 
complicated magnitude-
and color-dependent selection effects of our survey.)

\subsubsection{Description of the LF Evolution} \label{description}

Setting the above technical considerations aside and 
returning to Figure~\ref{figlfsB}, 
we may note that our LF model does indeed appear to be a reasonable 
description of the data, as seen in the good agreement between the
parametric and nonparametric SWML fits
(but recall we have matched their normalizations, so that we are
really only assessing the validity of the $M^*$, $\alpha$ and $Q$ parameters).
The comparison also shows that our simple assumption of a fixed
$\alpha$ is quite reasonable, though of course at higher redshifts it
becomes increasingly difficult to constrain the faint-end slope of the
luminosity function. 
Note also that our fixed-$\alpha$ result differs from that found in
the Autofib Redshift Survey (\cite{ell96}; \cite{hey97}); 
see \S~\ref{compautofib} below.

\begin{figure}
\plotone{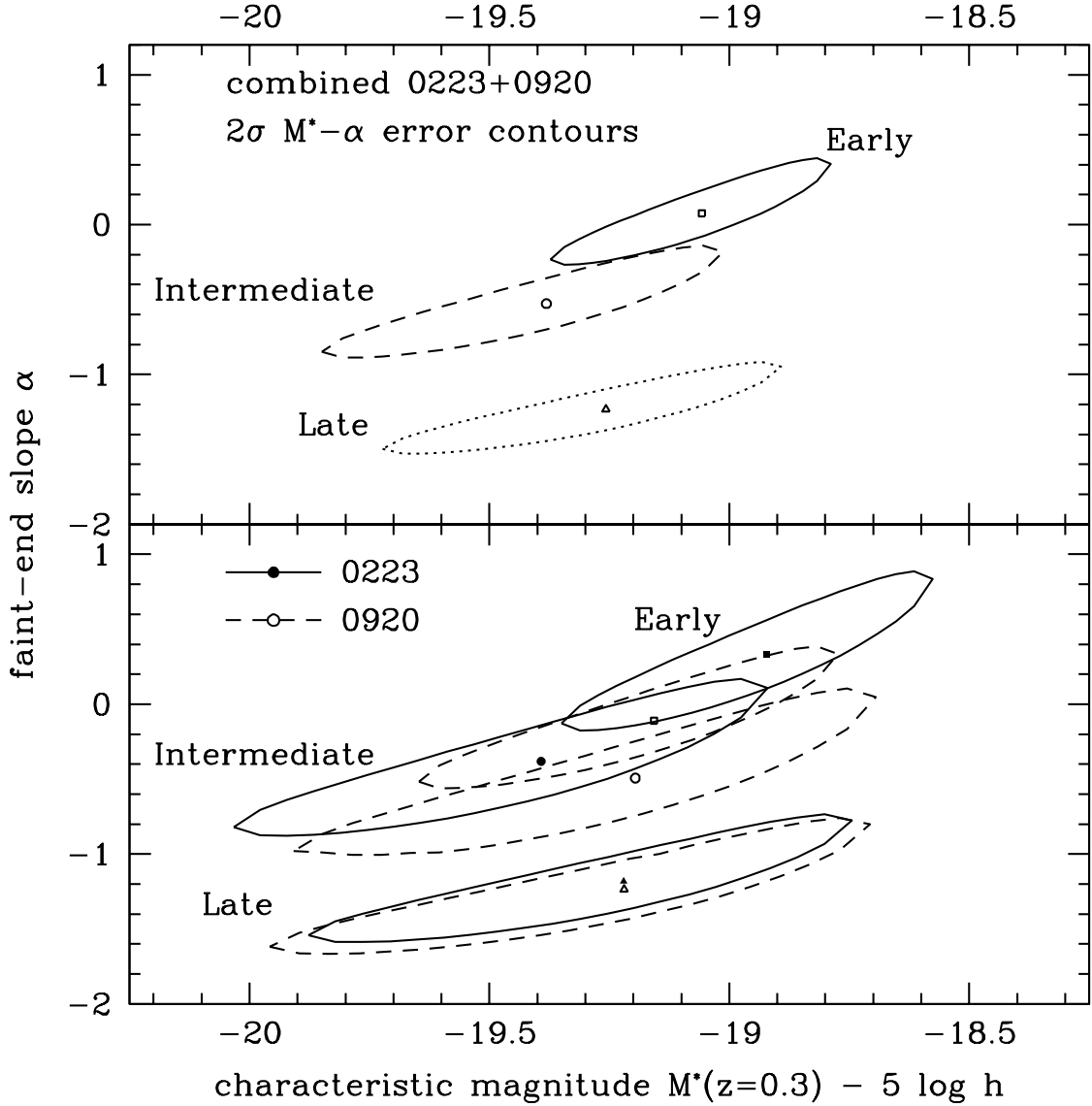}
\caption{
 $2\sigma$ error contours in $M^*(z=0.3)$ vs.\ $\alpha$ 
  for the $B_{AB}$ luminosity functions of various CNOC2 samples.
 ({\em Top}) Error contours for early, intermediate, and late types for
  the full 0223+0920 sample.
 ({\em Bottom}) Comparison of error contours for early, intermediate,
  and late types for the 0223 ({\em solid contours and filled points})
  and 0920 ({\em dashed contours and open points}) patches.
 }
\label{figmaB}
\end{figure}

Figure~\ref{figlfsB} also shows that the three galaxy types have
conspicuously different LF's, with faint-end slopes $\alpha$ steepening
from $\alpha = +0.1$ for early types to $\alpha = -1.2$ for late
types (Table~\ref{tablfB}).
These clear LF differences are indeed significant, 
as shown in Figure~\ref{figmaB}
(top panel), where we see non-overlapping or barely-touching 
2$\sigma$ error contours in 
$M^*(z=0.3)$ and $\alpha$ for the three galaxy types.

\begin{figure}
\plotone{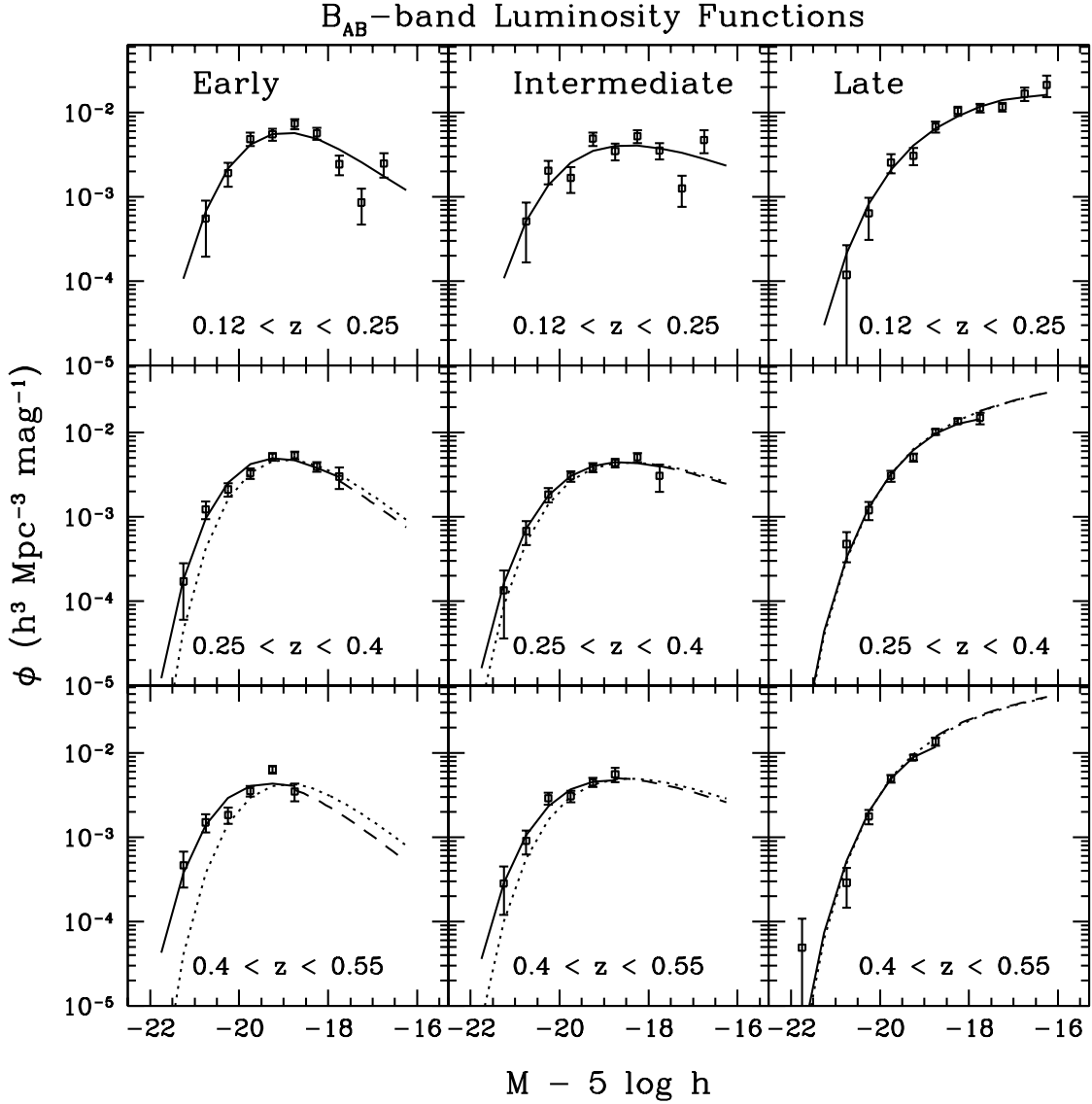}
\caption{
 Same as Figure~\protect\ref{figlfsB} but with the fiducial
  low-redshift LF ({\em dotted curves}) rescaled to take out the
  effects of density evolution; see text for details.
 }
\label{figlfsBa}
\end{figure}

Figure~\ref{figlfsB} demonstrates that the LF's for all 
three galaxy types do indeed evolve.
The impression is that the early- and intermediate-type LF's are not
changing much in number density, but are rather brightening in $M^*$
at higher redshifts.
For the late-type LF, it is harder to discern visually (because of the
steepness of the LF) whether the definite changes seen result from
increasing number density, brightening $M^*$, or a combination of the
two.
We can isolate the luminosity evolution component
of the LF's by rescaling the dotted fiducial $z = 0.175$ LF in each
panel by the factor $10^{\{ 0.4 P [(z_1+z_2)/2 - 0.175)] \} }$, to
explicitly take out the effect of the number density evolution
parameter $P$.
We do this in Figure~\ref{figlfsBa}, where we confirm our earlier impression
that the early- and intermediate-type LF's are evolving primarily in $M^*$.
In contrast, the rescaled low-$z$ fiducial late-type LF is a
good match to the results in the two higher-redshift bins, indicating
that the observed late-type LF evolution seen before in
Figure~\ref{figlfsB} is driven primarily by number density changes.
(Note that an apparent change in number density does not necessarily
imply mergers; changes in the star formation duty cycle for late-type
galaxies may also mimic the effect of true mergers.) 

Now, an important consideration 
mentioned in the Introduction needs to be kept in mind, namely the 
sensitivity of the LF evolution 
results to the precise {\em choice} of SED's used to classify galaxies. 
In particular, the present choice of {\em non-evolving} CWW SED's 
obviously does not account for the evolution of the colors of galaxies
with time, so that during the course of its evolution, a particular
galaxy may actually cross the type boundaries we have defined.
It is thus better to use more physically motivated 
{\em evolving} galaxy SED's (e.g., produced by models such as those of
\cite{bc96}) to properly track the paths different galaxies may take in
the space of redshift vs.\ color. 
Not surprisingly, the resulting galaxy
classifications will in general differ from the ones we make based on
the CWW SED's, and importantly, the conclusions we draw on the
rates of luminosity and density evolution for different galaxy
populations will also be different in general.
However, there are a myriad of possible evolving SED's
that one may choose by varying 
parameters such as star formation history, stellar initial mass
function, epoch of galaxy formation, metallicity, dust content, and
others, so that there is no unique set of SED's that one should
obviously pick a priori.
In LF Paper II, we will examine evolution in the CNOC2 sample using
these physically motivated evolving galaxy SED models.
In the present paper, though, we will use only the non-evolving CWW SED's
for galaxy classification.
Thus, the LF evolution constraints we derive here
should strictly be considered as
{\it descriptions} of galaxy evolution within the
framework of non-evolving SED's, rather than as
{\it explanations} of galaxy evolution in terms of more physically
motivated processes.
In other words, we are using the terms ``luminosity evolution'' and
``density evolution'' purely to describe the changes in the LF's of
the galaxy populations we have defined, and those terms {\em
may} not correspond to the true evolutionary processes those
galaxies are actually undergoing.
(The latter is not precluded, though.
As we will find in Paper II, the luminosity evolution we see
in early and intermediate galaxies still holds true when we
use physically-motivated evolving SED's.)

Also, we caution that the LF constraints will be weaker and the
errors larger when our LF models are extrapolated outside the
nominal CNOC2 redshift limits.
For example, the errors are approximately doubled for $M^*(z=0)$ 
compared to $M^*(z=0.3)$; specifically $M^*(z=0) = -18.58 \pm 0.23,
-19.11 \pm 0.34$, and $-19.20 \pm 0.35$ for early-, intermediate-, and 
late-type galaxies, respectively. 
There are only about 200 galaxies in our LF sample with 
$0.12 < z < 0.2$ to constrain the lowest-$z$ behavior of our LF evolution
model.
These galaxies alone do in fact give an overall best fit 
$M_{B_{AB}}^\ast = -19.5$ and $\alpha = -0.9$, in reasonable agreement with 
results from much larger local redshift samples (e.g., \cite{lov92}),
and our $M^*(z=0)$ and $\alpha$ values for intermediate- and 
late-type galaxies are also in good agreement with local results 
(e.g., Figure~8 of \cite{col98}).
However, our early-type LF may have a fainter
$M^*(z=0)$ and shallower $\alpha$ compared to local values 
(e.g., \cite{col98}, but cf.\ also \cite{bro98}).
In future work we will compare in more careful detail our results with
those of large local surveys, in order to further check the validity of 
extrapolations of our LF models (see also \S\S~\ref{counts} 
and \ref{compcfrs}).

\begin{figure}
\plotone{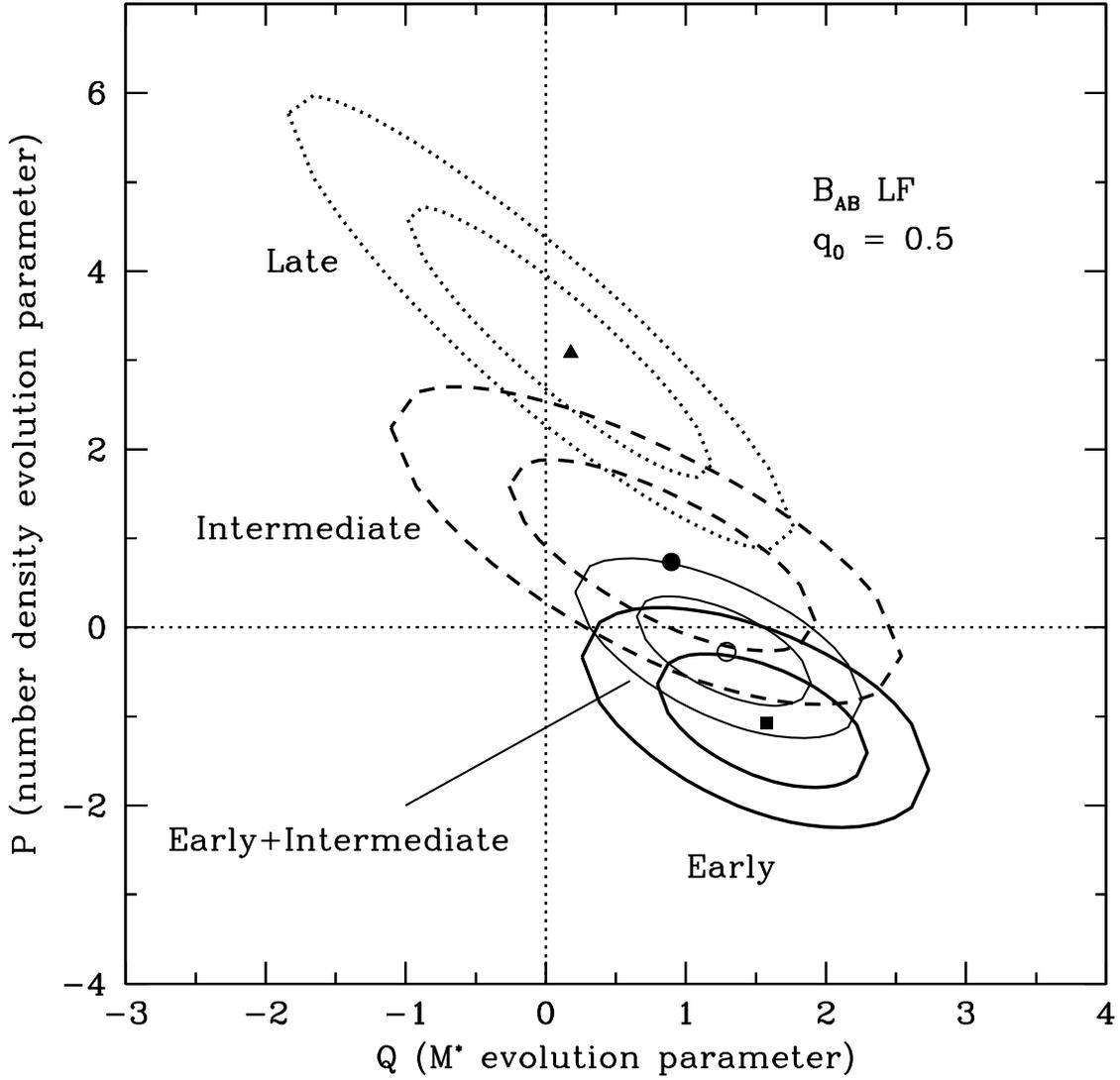}
\caption{
 $1\sigma$ and $2\sigma$ error contours in $P$ 
  (number density evolution parameter)
  vs.\ $Q$ ($M^*$ evolution parameter) for the $B_{AB}$ luminosity
  functions of early ({\em solid contours and filled square}), 
  intermediate ({\em dashed contours and filled circle}), 
  late ({\em dotted contours and filled triangle}), 
  and early+intermediate ({\em light solid contours and open circle}) CNOC2
  samples.
 Results shown are for $q_0 = 0.5$.
 The intersection of the horizontal and vertical dotted lines indicates
 no-evolution, $P = Q = 0$.
 }
\label{figpqB}
\end{figure}

Keeping the above caveats in mind, 
our impression so far is that evolution in early- and 
intermediate-type galaxies is
dominated by brightening in $M^*$ at higher $z$, while the evolution
in late-type galaxies is caused by increasing number densities at
higher redshifts. 
This impression is borne out in the $P$ vs.\ $Q$ error
contours shown in Figure~\ref{figpqB}: early and intermediate types show
positive luminosity evolution, with a combined $Q = 1.3$, but little density
evolution, $P = -0.3$, while late types show strong positive
density evolution, $P = 3.1$, but little $M^*$ evolution, with $Q
= 0.2$ (Table~\ref{tablfB}). 
However, Figure~\ref{figpqB} also shows that we need to be somewhat
cautious, and keep the correlated nature and fairly large size of the
$P$-$Q$ error contours in mind. 
Not surprisingly, our ability to decouple density and luminosity
evolution depends on the shape of the luminosity function: for early
and intermediate types, the LF has a shallow $\alpha$ and a
conspicuous ``knee'' near $M^*$, while for late types the LF is steep
and it becomes correspondingly harder to measure subtle changes in
$M^*$ with redshift.
Thus, although the late-type sample is the largest among the three types, 
the late-type $P$-$Q$ contour is the most elongated and
the most difficult one for which to separately constrain density and
luminosity evolution.
Moreover, even for the two earlier-type samples, no-evolution 
($P = 0, \ Q = 0$) is ruled out at only somewhat better than the 
2$\sigma$ level.
Nonetheless, it does appear to be a fairly robust conclusion from 
Figure~\ref{figpqB} that late types occupy a different region of
$P$-$Q$ parameter space than early and intermediate types, 
so that the form of the LF evolution of late-types is
distinct from that of early- and intermediate-type galaxies,

\subsubsection{Evolution of the Luminosity Density} \label{evolution_lumden}

\begin{figure}
\plotone{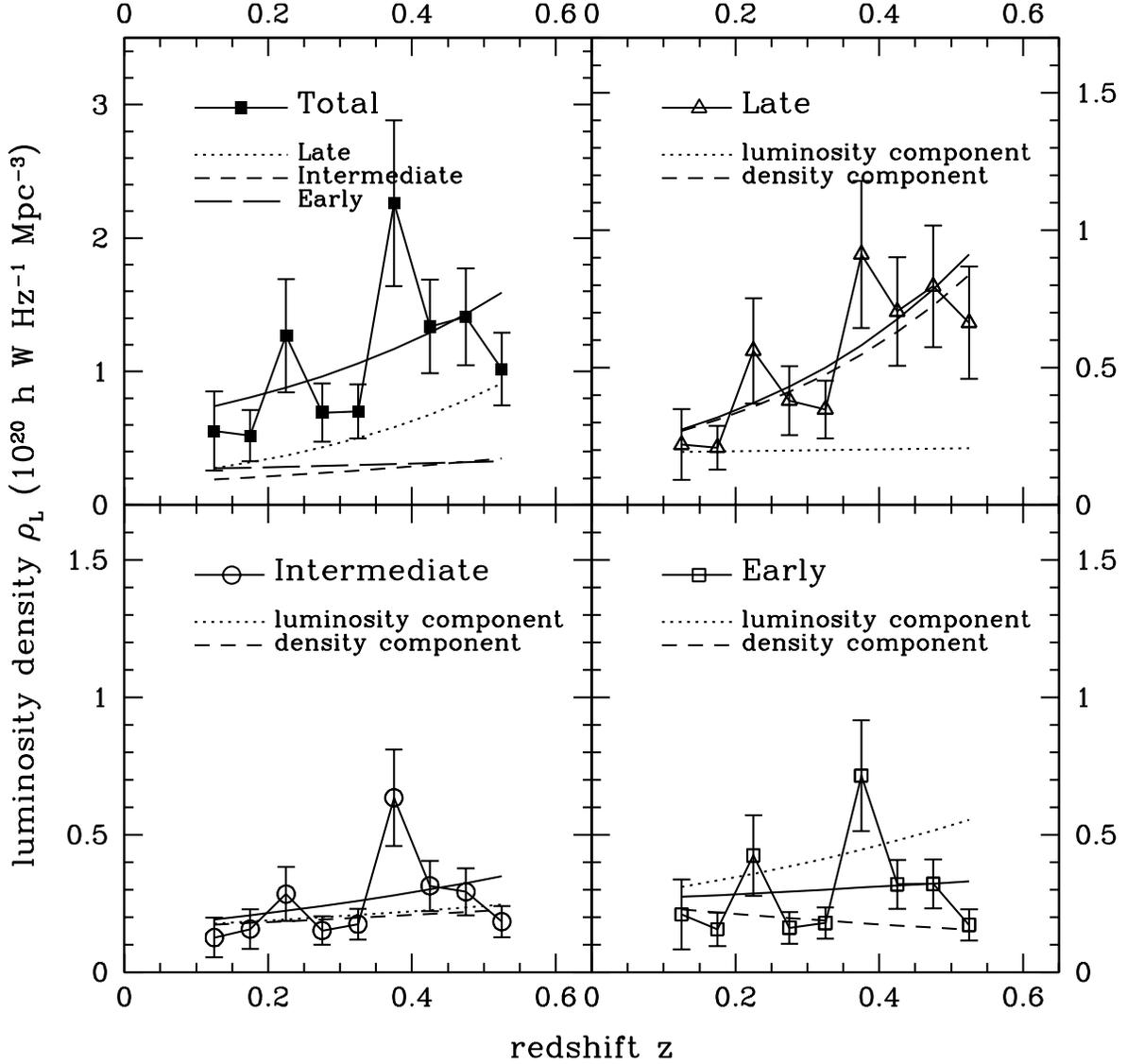}
\caption{
 Redshift evolution of the CNOC2 rest-frame $B_{AB}$ luminosity density
  $\rho_L(z)$, shown for the early, intermediate, late, and total
  galaxy samples.
 We plot both the directly-observed but LF-weighted ({\em
  points}) as well as the LF-computed ({\em solid lines})
  luminosity densities.
 We also show the separate luminosity-evolution ({\em dotted
  curves}) and density-evolution ({\em dashed curves}) components of
  the overall LF-computed luminosity density evolution curves.
 Results shown are for $q_0 = 0.5$.
 }
\label{figldB}
\end{figure}

Despite the difficulties in decoupling $P$ and $Q$, 
we can nevertheless robustly constrain the sum $P+Q$, which measures
the rate of evolution of the rest-frame luminosity density $\rho_L(z)$.
The error contours in Figure~\ref{figpqB} are elongated roughly along
lines of constant $P+Q$ (and thus compressed in the orthogonal
direction), so that the contours are actually most effective for 
constraining the sum as opposed to $P$ and $Q$ separately.
The luminosity density of late types evolves at a significantly
more rapid rate ($P+Q = 3.3$) than that of early and intermediate
types ($P+Q = 0.5$ and 1.6, respectively).
This is shown in more detail in Figure~\ref{figldB}, 
where we plot $\rho_L(z)$ for
the 3 galaxy types individually, as well as summed together.
We also tabulate our luminosity density results 
in Table~\ref{tabrhoL}.
Clearly, the late-type population shows the most strongly increasing
$\rho_L(z)$, while the early and intermediate types show much milder
increases at higher redshift.
Note that the luminosity densities for the three types are 
roughly equal at $z \approx 0.1$, but by $z \approx 0.55$ 
late-type galaxies
predominate and account for over half of the total luminosity
density. 
Also shown in Figure~\ref{figldB} are the separate 
contributions to $\rho_L(z)$ from the luminosity and number density evolution
components for each of the three galaxy types.
Compared to the $P$-$Q$ plot or even the LF plots, the curves for 
these individual components most clearly illustrate 
the different LF evolution trends we discussed earlier (but keeping
the caveats in mind).
The late-type galaxy LF is dominated by strong density evolution, with
nearly no luminosity evolution. The intermediate-type LF shows 
positive luminosity evolution plus weak positive density
evolution, resulting in mild positive evolution in $\rho_L$.
The early-type LF shows positive luminosity evolution which is
nearly compensated by negative density evolution, yielding
a very weak positive evolution in the luminosity density.

\begin{figure}
\plotone{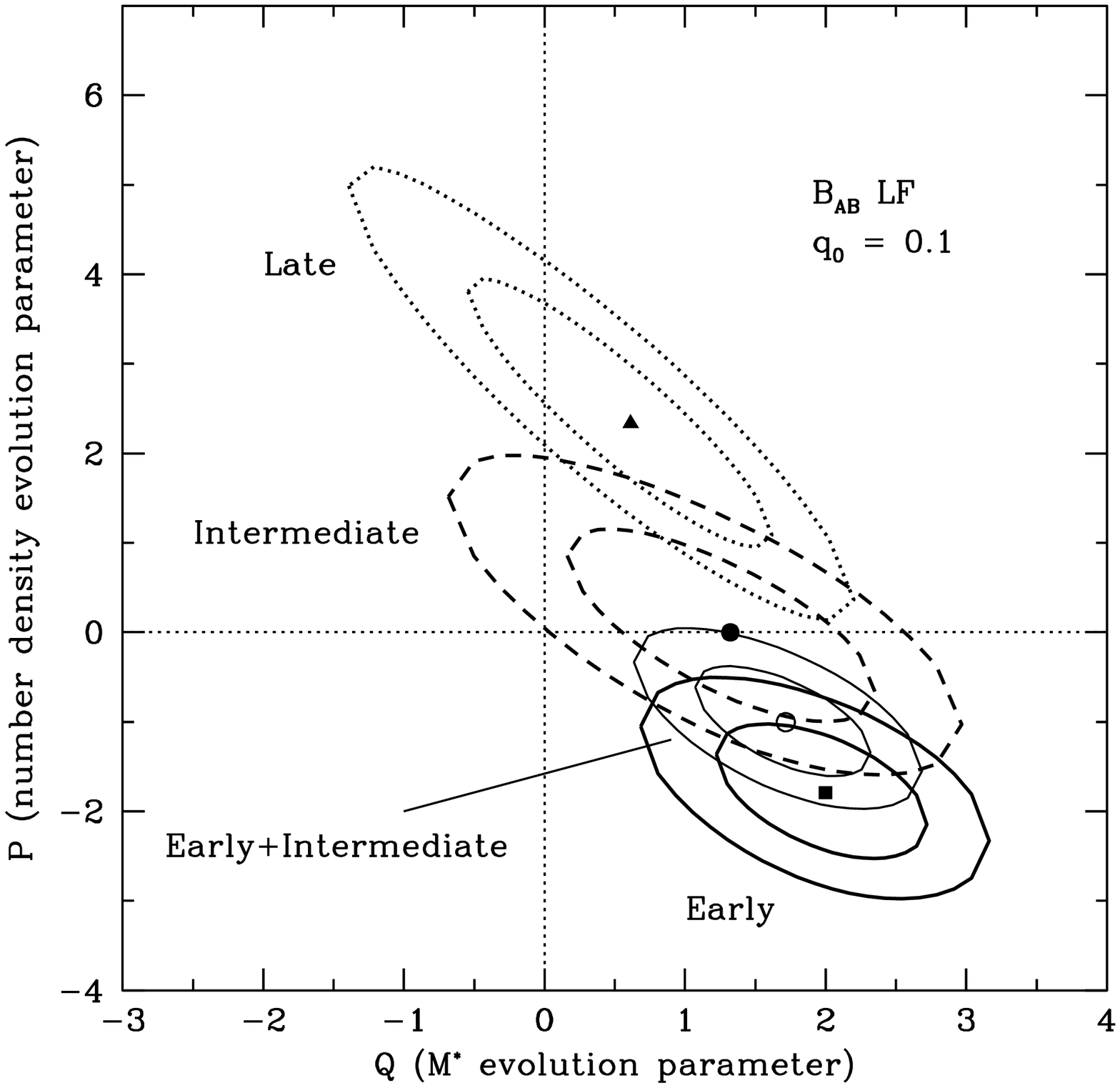}
\caption{
 Same as Figure~\protect\ref{figpqB} but for $q_0 = 0.1$.
 }
\label{figpqBqo0.1}
\end{figure}

\begin{figure}
\plotone{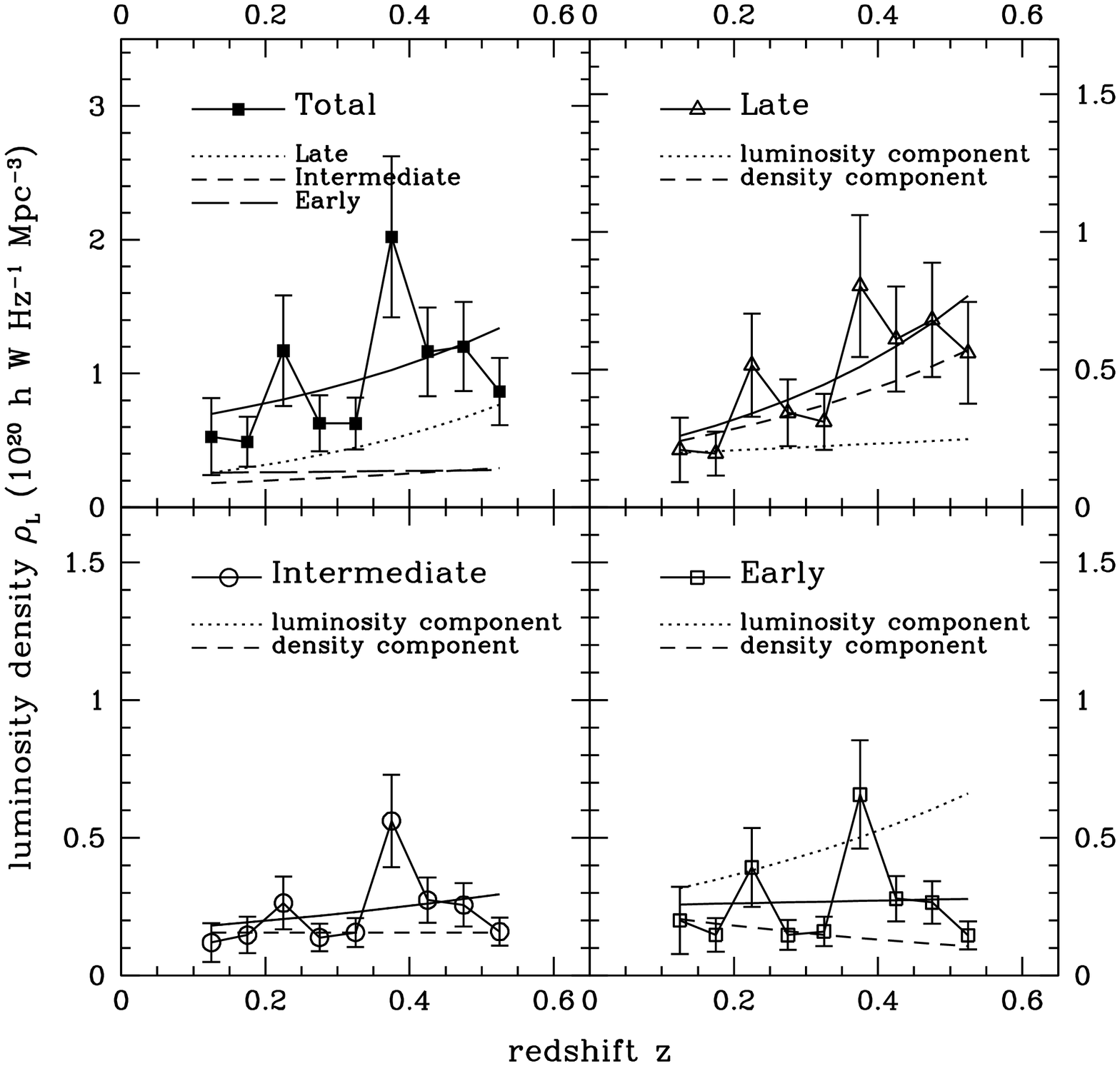}
\caption{
 Same as Figure~\protect\ref{figldB} but for $q_0 = 0.1$.
 }
\label{figldBqo0.1}
\end{figure}

These general conclusions are not altered much by adopting a $q_0 =
0.1$ instead of a $q_0 = 0.5$ cosmology.
Our $q_0 = 0.1$ results are also tabulated in Tables~\ref{tablfB} and
\ref{tabrhoL},
and the corresponding $P$-$Q$ contours and $\rho_L(z)$ plots are shown
in Figures~\ref{figpqBqo0.1} and \ref{figldBqo0.1}, respectively.
To first order, absolute magnitudes change with $q_0$ as
$M(z,q_0) \approx M(z,q_0 = 0.5) + (q_0 - 0.5)z$, and the differential
volume element varies as
$\frac{dV}{dz}(z,q_0) \approx 
 \frac{dV}{dz}(z,q_0 = 0.5) [1 - 2(q_0 - 0.5)z]$.
We thus expect $\Delta Q \approx +0.4$, $\Delta P \approx -0.8$, and
$\Delta (P+Q) \approx -0.4$ in
going from $q_0 = 0.5$ to $q_0 = 0.1$, and indeed that is what we
approximately find quantitatively.
Qualitatively, this means more positive luminosity evolution, but
more negative number density and luminosity density evolution.
In particular, no-evolution ($P = Q = 0$) for the early 
and intermediate types combined may be ruled out at higher 
significance than was possible for the $q_0 = 0.5$ case.
Otherwise, though, the general LF evolution trends follow those for
the $q_0 = 0.5$ cosmology.

\begin{figure}
\plotone{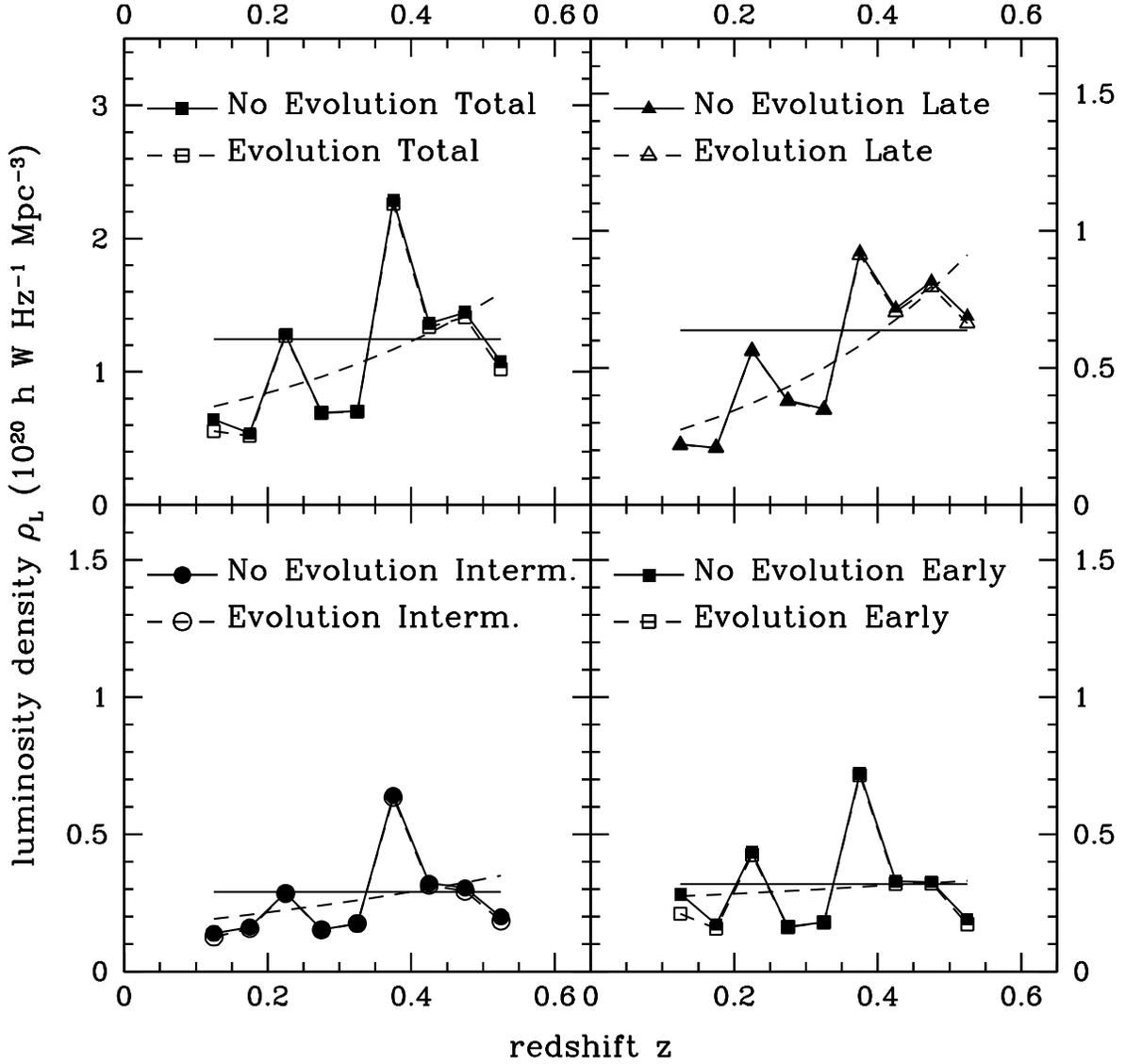}
\caption{
 Redshift evolution of the CNOC2 $B_{AB}$ luminosity density
  $\rho_L(z)$, shown for the early, intermediate, late, and total
  galaxy samples.
 We plot both the directly-observed but LF-weighted ({\em
  points}) and the LF-computed ({\em lines})
  luminosity densities, where the LF's have been fit using either an 
  evolving model ({\em dashed curves and open points}), or a
  non-evolving model ({\em solid horizontal lines and filled points}) 
  with $P = Q = 0$.
 Results shown are for $q_0 = 0.5$.
 }
\label{figldBnoev}
\end{figure}

Finally in this subsection, we make the most minimal assumptions and
fit the LF for the three galaxy types with non-evolving Schechter
functions, and compare the resulting trends of luminosity density vs.\
redshift against those obtained from the evolving models above; this
is done in Figure~\ref{figldBnoev} (for $q_0 = 0.5$ only). 
Reassuringly, we find that the
actual trend of $\rho_L$ with redshift (the {\em points} in the
figure, not the curves) is insensitive to whether we 
use non-evolving or evolving LF's 
(in eqs.~[\ref{eqrhoL}] and [\ref{eqsl}]). 
The late-type $\rho_L$ always rises sharply compared to the weak
increases observed for the earlier types.
However, as seen in the top two panels of
Figure~\ref{figldBnoev}, a constant $\rho_L$, as required by a 
non-evolving LF, is clearly a bad description of the late-type $\rho_L(z)$,
and consequently for the total $\rho_L(z)$ as well, since late types
make up the greatest contribution. 
On the other hand, a non-evolving constant $\rho_L(z)$ does seem to be
reasonable for the two earlier types. 
This is not surprising, as Figure~\ref{figpqB} shows that 
although an evolving LF is preferred by the data,
a non-evolving LF is ruled out at only about 2$\sigma$ for
early and intermediate galaxies.
A larger data set will be needed in order to make
a stronger statement regarding no-evolution vs.\ evolution, and the
doubled size of the final CNOC2 sample should allow significantly
improved constraints on these early- and intermediate-type galaxies.
In addition, we are also calibrating 
photometric redshifts using the multicolor data for our 
spectroscopic-redshift sample.
Application of photometric redshifts to those CNOC2 galaxies without
spectroscopic redshifts should provide another factor of two
increase in the number of $R_c < 21.5$ galaxies that may be used in 
our LF studies, thereby allowing further improvements 
in our evolution constraints.

\subsection{The $R_c$- and $U$-band LF's} \label{RUresults}

The availability of $U \! B V \! R_c I_c$ colors in conjunction with the
CWW SED types makes it a
straightforward matter to calculate the appropriate $k$-corrections
and derive the LF in bands other than $B$, the most typical choice.
We do so for the $R_c$ and $U$ bands.
No extrapolations are required of our color data to derive rest-frame
$U$ magnitudes.
Although extrapolations are required to obtain rest-frame $R_c$ magnitudes,
the needed $k$-corrections are not large ($\lesssim 1$ mag) and are
well-constrained by the SED classifications.
Note that here we are always using a $17.0 < R_c < 21.5$ sample; we are
{\em not} varying the band used for galaxy {\em selection} (not until
\S\S~\ref{Bselect} and \ref{compautofib} below)

The best-fit $R_c$ and $U$ LF and evolution parameters are given in 
Table~\ref{tablfRU}, and the luminosity densities are given
in Table~\ref{tabrhoL}.
The $R_c$- and $U$-band LF's themselves will not be plotted since those
figures would look very similar to Figure~\ref{figlfsB} for the
$B_{AB}$ LF's.
Essentially the same trends observed earlier for the $B_{AB}$ LF's
are seen for $R_c$ and $U$ as well, and our earlier discussion
applies.
We also find that the best-fit faint-end slopes $\alpha$ are
independent of band. 
For all three galaxy types, the full range in best-fit $\alpha$ values
is only about 0.2 
among the three bands $B_{AB}$, $R_c$, and $U$.
Thus, to convert the LF results from one band to another, it is a good
approximation to keep $\alpha$ fixed and just apply 
an appropriate offset in $M^*$ based on
the mean rest-frame color for that galaxy type.
Moreover, it also turns out that the evolution parameters $Q$ and $P$
agree well from band to band. 
In retrospect this is not surprising.
We expect that $Q$ should stay the same because our galaxy
classification scheme is based on selecting galaxies of similar
rest-frame colors at different redshifts.
For example, however much the average $M_B$ changes with redshift for our 
early-type galaxies, $M_R$ for those same galaxies should change by
about the same amount, since by definition,
the rest-frame color $M_B-M_R$ of our early-type galaxies needs to
stay about constant with redshift.
Thus, for populations of similar rest-frame color, 
$Q$ and subsequently $P$ will be approximately 
independent of which band is chosen for the LF.

\begin{figure}
\plotone{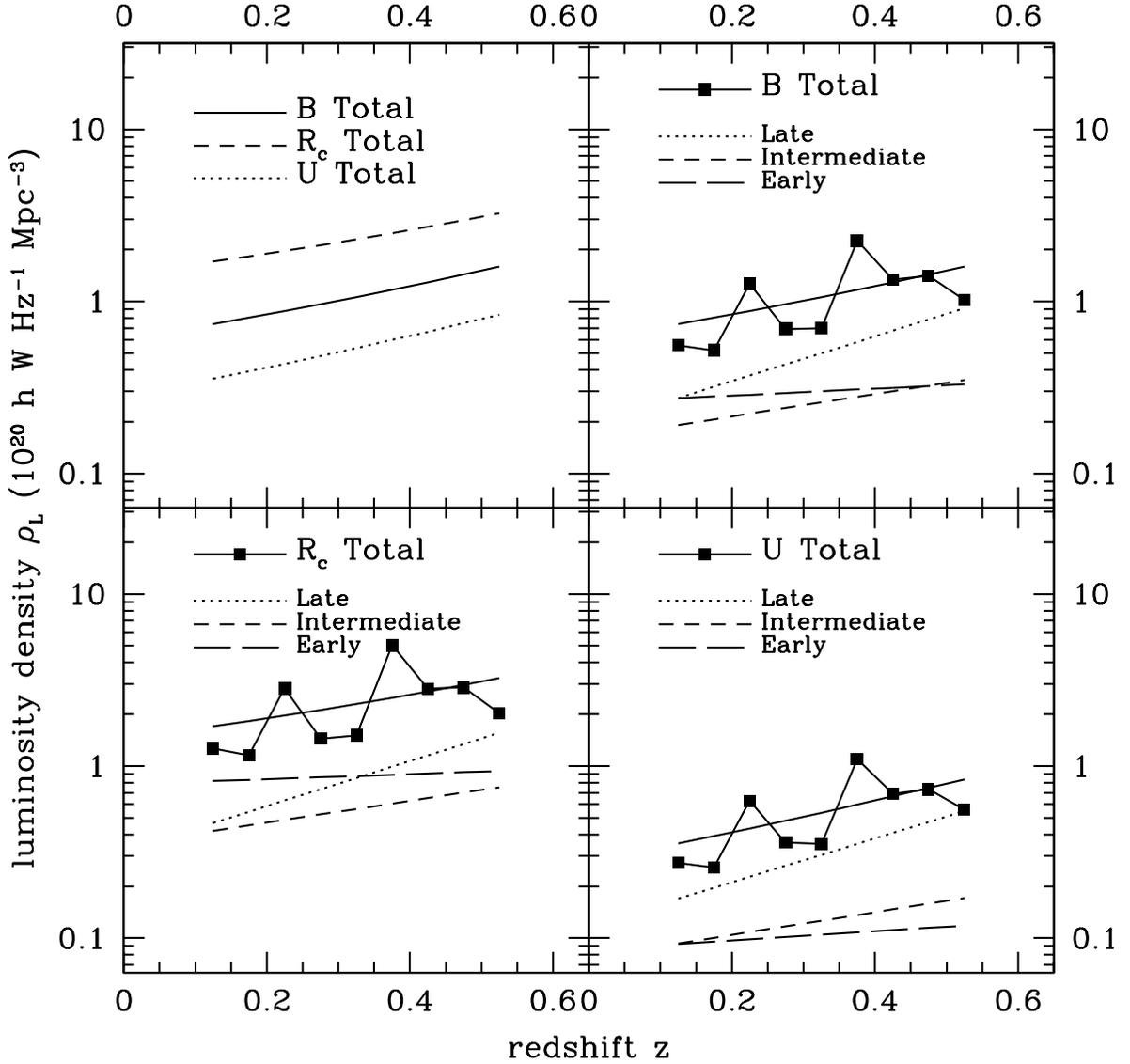}
\caption{
 Redshift evolution of the CNOC2 rest-frame luminosity density
  $\rho_L(z)$, shown for the $B$, $R_c$, and $U$ bands.
 The top left panel compares the total $\rho_L(z)$ for the three
  bands, while the other three panels break down each band into
  results by galaxy type.
 Note that unlike in previous figures, $\rho_L$ is plotted here on a
  logarithmic scale to facilitate comparison of the {\em rates} of luminosity
  density evolution among the three different bands.
 }
\label{figldBRU}
\end{figure}

However, although the {\em rate} $P+Q$ of luminosity density evolution for
a particular galaxy type is similar in different bands, the {\em
  normalization} $\rho_L(z=0)$ is in general different. 
This makes the overall
evolution of $\rho_L$ somewhat different for the different bands, and
Figure~\ref{figldBRU} compares $\rho_L(z)$ for the three bands,
including subdivision by galaxy type. 
For $R_c$ and $U$, we have first applied $AB$ corrections $R_{cAB} =
R_c + 0.169$ and $U_{AB} = U + 0.69$ (\cite{fsi95}) before applying
the same conversion we used for $B_{AB}$ to convert to
$h$ W Hz$^{-1}$ Mpc$^{-3}$ units (see end of \S~\ref{computeLF}).
The relative contribution of
late-type galaxies is strongest for $U$ and weakest for $R_c$, and
vice versa for early-type galaxies, as expected.
This causes the total $\rho_L$ to increase with redshift at a
somewhat faster rate in $U$ than 
in $R_c$, but Figure~\ref{figldBRU} (upper left panel) shows
that the difference is not very strong, at least over the CNOC2
redshift range.

\subsection{SED Type Distributions, Number Counts, and Color Distributions}
\label{counts}

\begin{figure}
\plotone{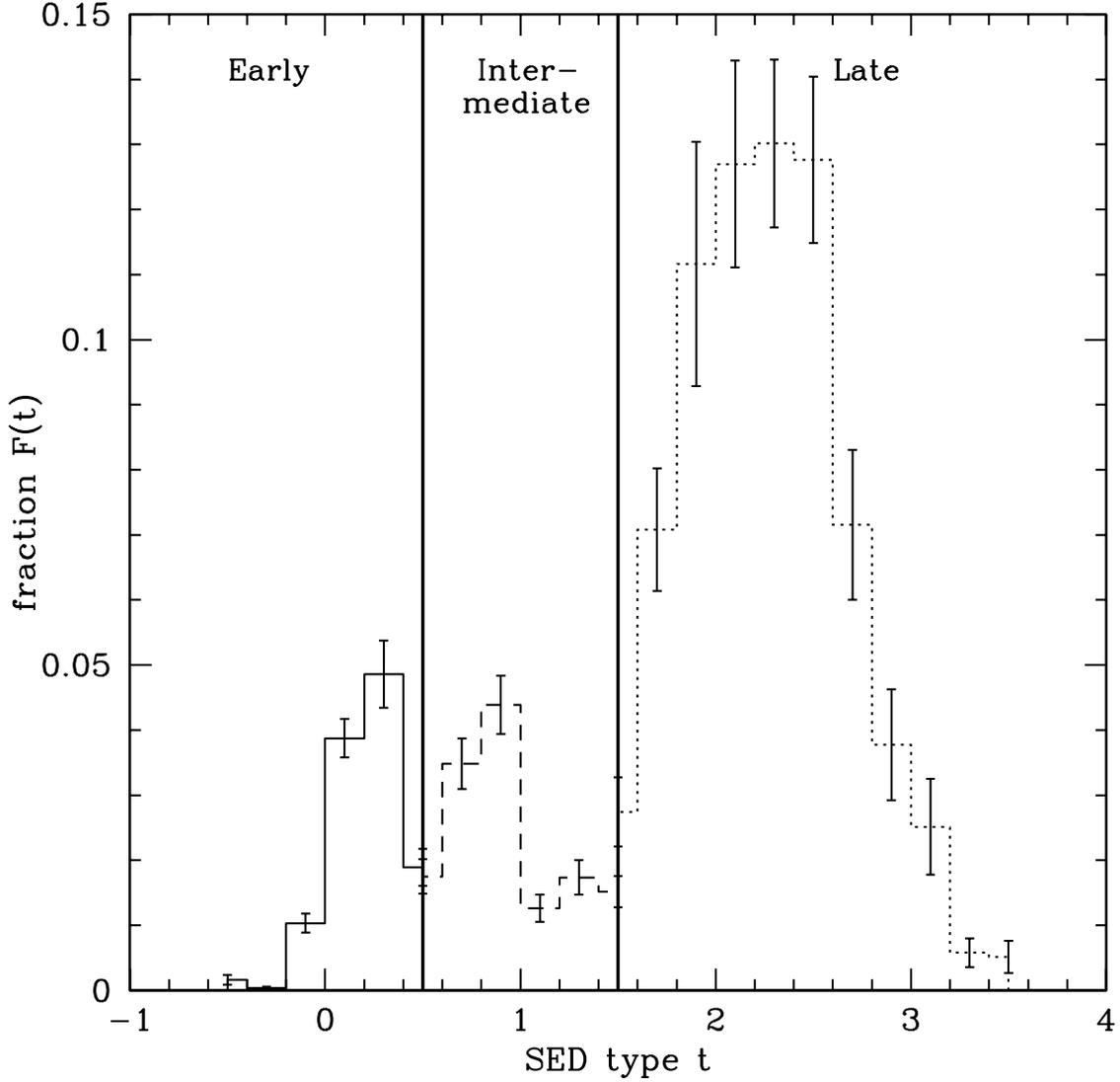}
\caption{
 Fractional distribution $F(t)$ of SED types $t$, calculated for
  the redshift range $0.12 < z < 0.55$ using equation (\ref{eqFt}) as
  described in the text.
 All uncertainties are computed assuming simple $\protect\sqrt{N}$
  errors.
 Note that we have corrected $F(t)$ so that it is appropriate for a
  {\em volume-limited} sample with $-22 < M_{B_{AB}} - 5 \log h <
  -16$; see text for details.
 }
\label{fighstp}
\end{figure}

So far we have not explicitly needed the distribution of SED types,
but in order to obtain the LF-computed galaxy number counts and color 
distributions, the SED type distribution will be helpful.
Otherwise we
will have to make some ad hoc assumptions, e.g., a uniform type distribution
within each of the three galaxy categories, or delta functions at a number
of characteristic SED types.
We first define an overall fractional type distribution $F$ in bins of SED
type $t$ by
\begin{equation} \label{eqFt}
F(t_1 < t < t_2) \equiv 
      \sum_{t_1 < t_i < t_2} W_i / S_j(z_i)
    \left/ 
      \sum_{\rm all \ galaxies} W_i / S_j(z_i)
    \right.
\end{equation}
Here we are weighting by the usual statistical weights $W_i$, as well
as by the appropriate selection function $S_j(z_i)$ from
equation~(\ref{eqselfunc}), where $j = $ early, intermediate, or late
indicates the category to which galaxy $i$ belongs.
The inverse selection function weighting corrects $F(t)$ to what
one would obtain for a volume-limited sample with 
$-22 < M_{B_{AB}} - 5 \log h < -16$,
and there is also an implicit assumption that the LF is independent of
SED type $t$ {\em within} each of the three galaxy categories.
Also, because the LF evolves, $F(t)$ will change with redshift, but
for illustrative purposes we will neglect this complication
and simply plot in Figure~\ref{fighstp} 
the $F(t)$ computed over the full redshift completeness
range $0.12 < z < 0.55$.
In our number count and color distribution calculations we actually only need
the fractional distribution $G_j(t)$ {\em within} each of the three galaxy
categories $j$:
\begin{equation}
G_j(t_1 < t < t_2) \equiv 
      \sum_{t_1 < t_i < t_2 \ 
           \cap {\rm \ galaxy \ } i {\rm \ in \ category \ } j}
          W_i / S_j(z_i)
    \left/ 
      \sum_{{\rm all \ galaxies \ in \ category \ } j} W_i / S_j(z_i)
    \right. \ .
\end{equation}
We have explicitly checked that $G_j$ changes only weakly with
redshift, so that it is a good approximation to adopt the $G_j$
computed for the full redshift range $0.12 < z < 0.55$ in our
subsequent calculations.
Note that the appearance of the histogram in Figure~\ref{fighstp} (the
presence of peaks and valleys, how smooth or not it appears) may
depend to some extent upon the choice of the set of original SED's used 
to define the classification scheme.

\begin{figure}
\plotone{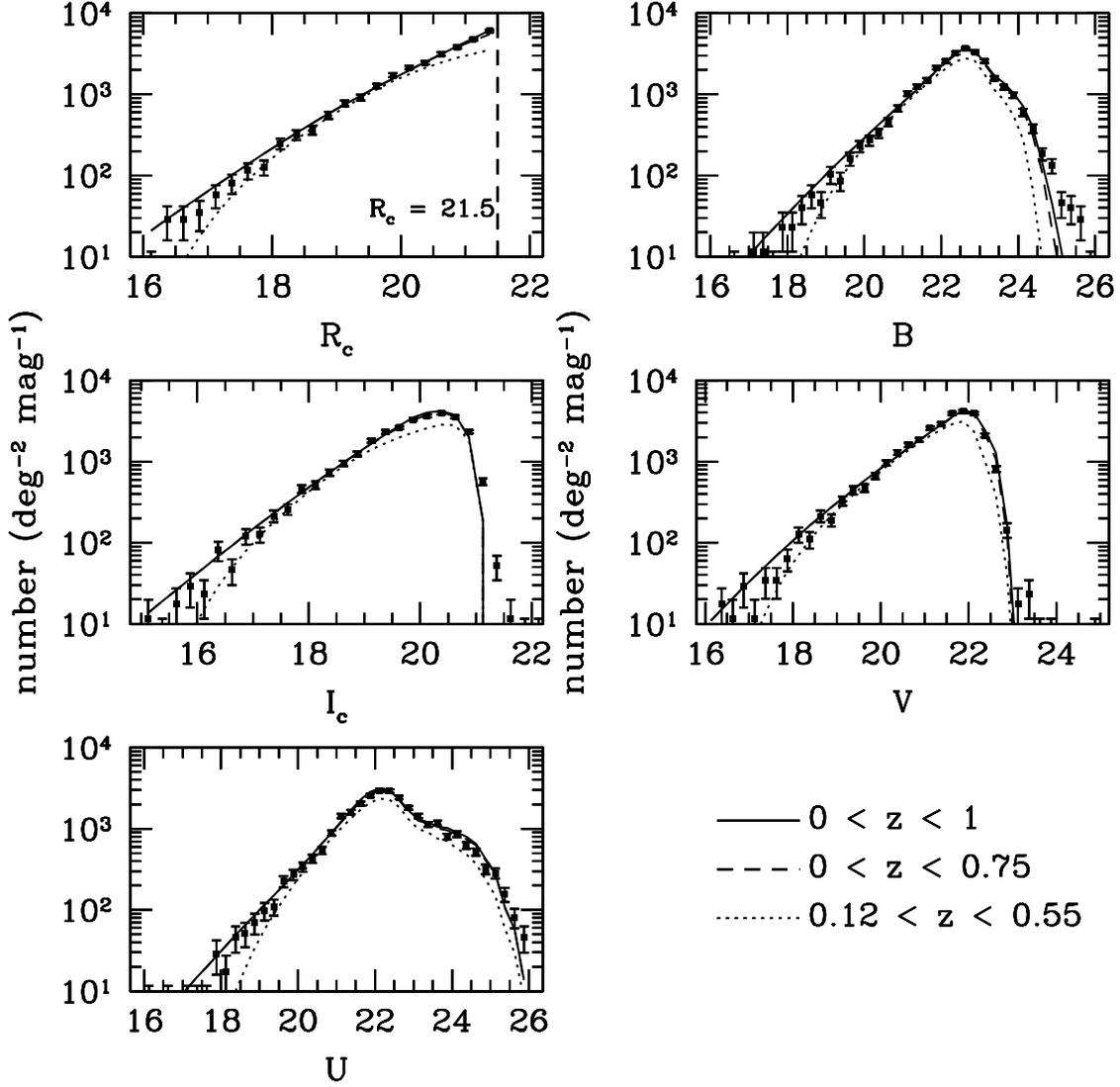}
\caption{
 The differential galaxy number counts ({\em points}) for the 
  CNOC2 $U \! B V \! R_c I_c$ bands, restricted to those galaxies with
  $R_c < 21.5$.
 Uncertainties are computed assuming simple $\protect\sqrt{N}$ errors, 
  and thus do not account for fluctuations due to galaxy clustering. 
 Also shown are counts computed from our best-fit evolving $B_{AB}$ LF
  model, using equation (\ref{eqct}) as described in the text.
 The various curves show the contributions to the counts from galaxies
  with $0.12 < z < 0.55$ ({\em dotted curves}), 
  $0 < z < 0.75$ ({\em dashed curves}), and 
  $0 < z < 1$ ({\em solid curves}).
 }
\label{figct}
\end{figure}

We now consider the CNOC2 photometric sample with $R_c < 21.5$, the
nominal spectroscopic limit.
The galaxy number counts in the $U \! B V \! R_c I_c$ bands for this
sample are plotted in Figure~\ref{figct}.
Note the turnover at faint magnitudes in the $U \! B V I_c$ 
bands is due to our explicit $R_c < 21.5$ cut and is {\em not} a
result of incompleteness in the photometry in these bands.
We also plot the number counts computed using the evolving $B_{AB}$-band 
LF's $\phi_j(M,z)$ and the fractional distributions $G_j(t)$ derived
previously. 
In particular, galaxies with redshifts $z_1 < z < z_2$ will contribute
to the number counts $N$ in an apparent magnitude interval
$m_1 < m < m_2$ of a particular band according to
\begin{equation} \label{eqct}
N(m_1 < m < m_2; z_1 < z < z_2) = 
  \sum_j \sum_t G_j(t) 
  \int_{z_1}^{z_2} \left( \frac{dV}{dz} \right) dz 
  \int_{M_{\rm min}(z,m_1,t)}^{M_{\rm max}(z,m_2,t)}
       \phi_j(M,z) dM \ ,
\end{equation}
where $M_{\rm min}(z,m_1,t)$ and $M_{\rm max}(z,m_2,t)$ are the
absolute magnitude limits observable at redshift $z$, given the
apparent magnitude limits $m_1$ and $m_2$ and 
the $k$-corrections connecting absolute $B_{AB}$
magnitudes to the apparent magnitudes for the band in question.
These $k$-corrections depend on the 
galaxy type $t$ and are calculated using our usual CWW SED's.
Also, in the second sum above, $G_j$ is evaluated using bins of width 
$\Delta t = 0.2$ (as in Figure~\ref{fighstp}).
We calculate $N$ first considering only the contribution of galaxies with 
$0.12 < z < 0.55$, the nominal redshift completeness
range adopted for the LF analysis.
We can clearly see the shortfall compared to the actual counts at both
bright and faint magnitudes, resulting from neglect of low- and and high-$z$
galaxies, respectively.
The match between the observed and LF-computed counts
is much improved by extending the redshift range to $0 < z < 0.75$, 
and further extension to $0 < z < 1$ makes little difference.
The good agreement seen is {\em not} a circular result, since the
LF is fit only for galaxies within $0.12 < z < 0.55$, so that
including the
LF-{\em extrapolated} contribution from galaxies outside that redshift range
serves as an independent 
check on the validity of our LF and evolution models.

\begin{figure}
\plotone{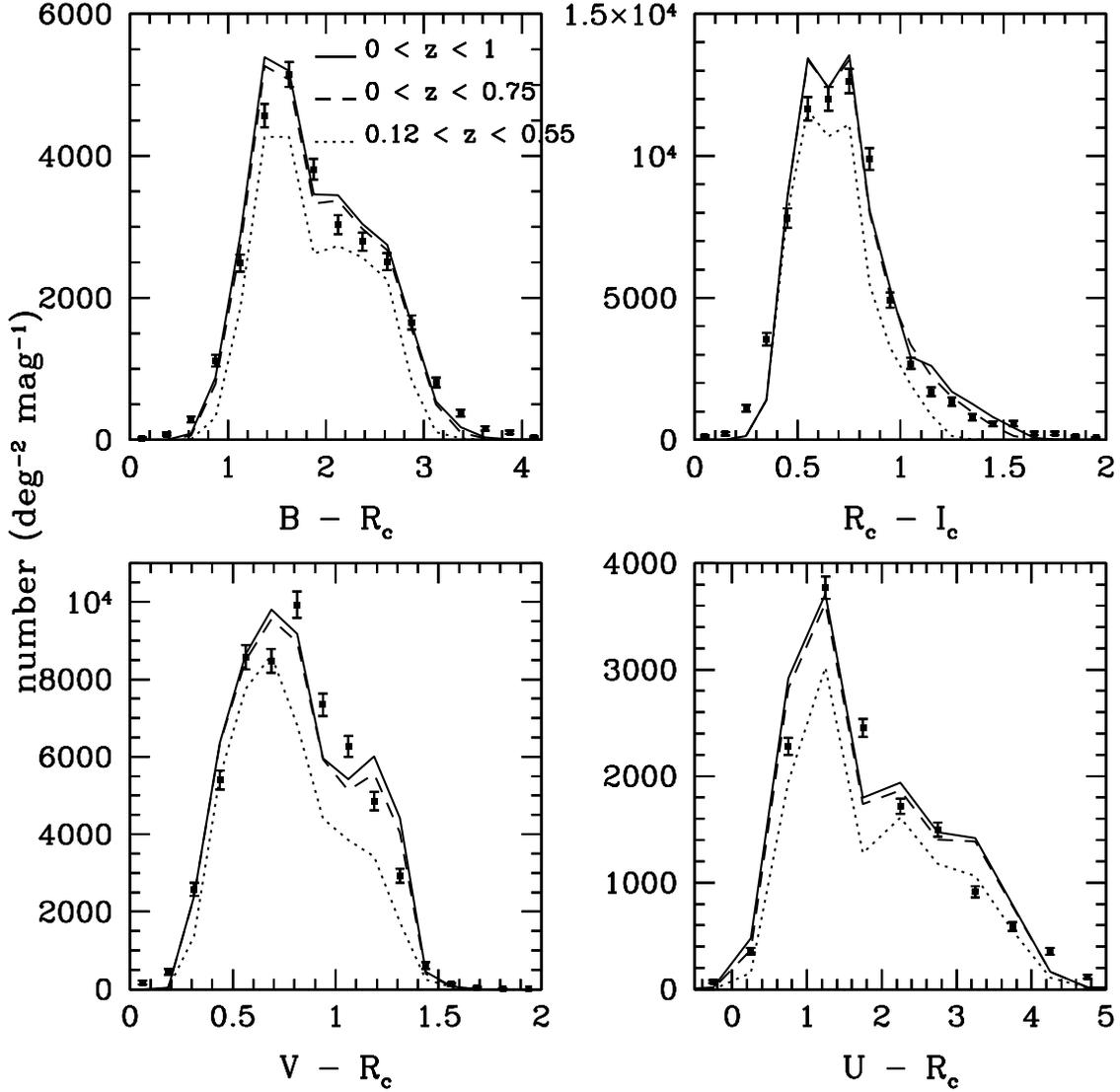}
\caption{
 The color distributions ({\em points}) in $B-R_c$, $R_c-I_c$, $V-R_c$,
  and $U-R_c$, calculated for those CNOC2 galaxies with
  $R_c < 21.5$.
 Uncertainties are computed assuming simple $\protect\sqrt{N}$ errors, 
  and thus do not account for fluctuations due to galaxy clustering. 
 Also shown are color distributions computed from our best-fit 
  evolving $B_{AB}$ LF model, 
  for the contributions of galaxies 
  with $0.12 < z < 0.55$ ({\em dotted curves}), 
  $0 < z < 0.75$ ({\em dashed curves}), and 
  $0 < z < 1$ ({\em solid curves}).
 }
\label{figcd}
\end{figure}

We then repeat the same exercise but using various color
distributions, as shown in Figure~\ref{figcd}.
The LF-computed color distributions are calculated using an expression
analogous to equation~(\ref{eqct}), but augmented with the appropriate 
limits in the observed colors.
The LF-computed results again converge by $z = 0.75$ and the match to
the observed color distributions is good for all four colors shown:
$B-R_c$, $R_c-I_c$, $V-R_c$, and $U-R_c$.
(We should recall here that we did adjust the CWW SED $I_c$
magnitudes to improve the match to the $R_c-I_c$ distribution, 
as mentioned in \S~\ref{classification}).
The overall reasonable agreement between the $0 < z < 1$ LF-computed
color distributions and the
observed distributions provides further validation of our evolving LF model
and of our magnitude- and color-dependent weighting scheme defined
back in \S~\ref{weights}.

\subsection{Potential Systematic Effects}

Here we will consider a number of potentially important
systematic sources of error which may affect our LF and evolution
fits, specifically: (1) differences between the 0223 and 0920 patches;
(2) random photometric errors; (3) potential redshift incompleteness;
and (4) potential apparent magnitude incompleteness.
We will find that typically our LF and evolution parameters
are biased at less than the $1\sigma$ level.

\subsubsection{Patch-to-Patch Variation}
 
Since there are large-scale density fluctuations in our survey
(Figure~\ref{fighist}), we should check how well our results for the
0223 and 0920 patches agree.
We do so in some detail, comparing the $M^*$-$\alpha$ (Figure~\ref{figmaB}) 
and $P$-$Q$ error contours,
as well as the trend of $\rho_L(z)$ by galaxy type.
Encouragingly, the LF parameters $M^*$, $\alpha$, $P$, and $Q$ for the
two patches are all
consistent within their respective $2\sigma$ error contours.
Examination of the $\rho_L(z)$ comparison shows excellent
agreement of the LF-computed luminosity density evolution trends for all
three galaxy types, despite the obvious density fluctuations seen in
both patches.

\subsubsection{Random Photometric Errors} \label{randomerr}

\begin{figure}
\plotone{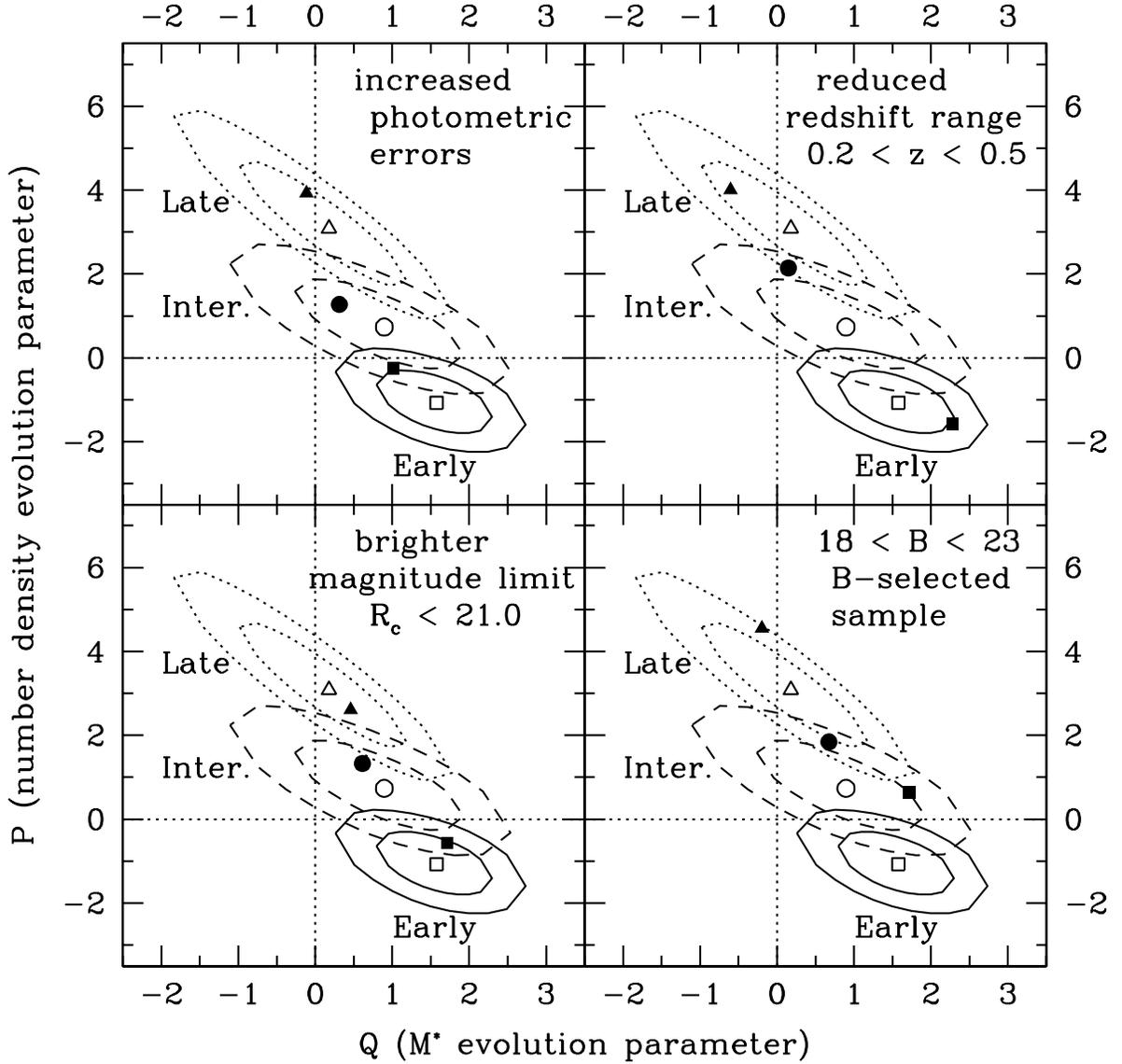}
\caption{
 The impact of various systematic effects on the best-fit values
  of $P$ and $Q$ for the $B_{AB}$ LF.
 The original $P$-$Q$ values ({\em open points)} 
  plus $1\sigma$ and $2\sigma$ 
  contours from Figure~\protect\ref{figpqB} are
  reproduced here. 
 The {\em solid points} show the modified $P$-$Q$ values resulting
  from use of an ``error-boosted'' sample ({\em top left}; see text for
  details), 
  from a reduced redshift range $0.2 < z < 0.5$ ({\em top right}), 
  from adoption of a brighter magnitude limit $R_c < 21.0$ ({\em
    bottom left}),
  and from use of a $B$-selected $18 < B < 23$ sample ({\em bottom right}).
 }
\label{figpqBsys}
\end{figure}

Random photometric errors will in general cause an Eddington-type effect
on the LF, such that $M^*$ is biased brighter and $\alpha$ is biased 
steeper (see EEP). 
This effect is appreciable for photographic-plate-based surveys
with magnitude errors $\sigma_m \sim 0.3$ mag (e.g.,
\cite{lov92}; \cite{mar94b}), but is essentially negligible for
CCD-based surveys with magnitude errors $\sigma_m \sim 0.1$ mag (e.g.,
\cite{lin96lum}), though one can correct for it nonetheless by taking
the LF to be a Schechter function {\em convolved} with a Gaussian 
magnitude error distribution with dispersion $\sigma_m$ (see EEP).
For CNOC2, $\sigma_{R_c} < 0.1$ mag at the nominal $R_c = 21.5$
spectroscopic limit and the consequent effects on $M^*$ and $\alpha$
should be small.
However, we should also confirm that the impact of photometric errors
are likewise negligible for the $P$ and $Q$ evolution parameters.
Moreover, since our galaxy classifications (and consequent derivation
of $k$-corrections and absolute magnitudes) also make use of the 
$U \! B V \! I_c$ magnitudes apart from just $R_c$, the photometric error
distributions in these various bands will affect our derivation of the LF
and evolution parameters in a complicated way.
The median magnitude errors for our $R_c < 21.5$ galaxies are
0.04 mag for $R_c$ and $I_c$, 0.05 mag for $V$, 0.08 mag for $B$, and 0.16
mag for $U$.
One could estimate the potential biases by fitting the LF's of Monte Carlo
mock CNOC2 galaxy catalogs, generated
using the best-fit type-dependent LF and evolution parameters of the
real sample,
combined with the appropriate photometric error distributions in each
of the CNOC2 bands. 
We will however use a less complicated procedure, and simply see what
happens if we artificially {\em increase} the photometric errors of
the real CNOC2 sample.
Specifically, for each photometric band of each galaxy, we modify the
observed magnitude by adding a random magnitude error.
The random magnitude error is drawn from a Gaussian distribution with zero
mean, and with dispersion equal to the PPP-estimated magnitude error for that
particular band and galaxy.
The photometric errors appropriate for the modified magnitudes are thus 
$\sqrt{2}$ times the original estimated errors.
We repeat this procedure five independent times for each galaxy in the
full CNOC2 {\em photometric} sample, so we end up with a five-fold
larger ``error-boosted'' sample.
Galaxy classifications and statistical weights are then computed using the
same procedure as for the original sample, but now based on the
modified magnitudes, and we then re-fit for the LF and evolution parameters.
The biases in the fitted parameters of the ``error-boosted'' sample relative
to the original CNOC2 sample give us an estimate of the biases
inherent in the CNOC2
sample relative to a hypothetical sample with no photometric errors.
Our ``error-boosted'' $P$ and $Q$ values are plotted as solid points in 
Figure~\ref{figpqBsys} (upper left panel), and are within the
1$\sigma$ error contours of the original $P$-$Q$ values. 
Likewise, the ``error-boosted'' $M^*$ and $\alpha$ values (not
plotted) are also within about 1$\sigma$ of the original values
(but systematically biased bright or steep, as expected).
We thus conclude that the existing photometric errors of our sample do
not significantly bias our LF and evolution parameter fits.

\subsubsection{Redshift Incompleteness} \label{zincomplete}

As described in \S~\ref{data}, our nominal redshift limits $z = 0.12$
and 0.55 are set by the observability of important
absorption and emission features over the 4400-6300\AA \ spectroscopic range. 
Examination of Figure~\ref{fighist}
shows that, as expected, the observed (weighted) redshift distribution
outside the $0.12 < z < 0.55$ range 
tends to lie low compared to the distribution from the best-fit 
LF model, although the effect is primarily seen for early and
intermediate types at higher redshifts (compare also
Figures~\ref{figsed} and \ref{figcol4}).
However, in the highest-$z$ bin $0.5 < z < 0.55$
{\em within} our nominal redshift range, 
there is already a noticeable dip in the redshift distributions for early and
intermediate galaxies.
This perhaps indicates some unaccounted residual redshift incompleteness
in that bin, and we
should check what happens if we exclude that bin from our analysis.
Also, we note that the $0.1 < z < 0.2$ bins may suffer from
incompleteness in late-type galaxies, if H$\beta$ and [OIII] 
$\lambda\lambda$5007,4959 do not 
adequately pick up for the unobservable [OII] $\lambda$3727 line.

Thus we redo our fits for the more redshift-complete range $0.2 < z <
0.5$, over which the most important redshift-identification features, 
Ca~II H+K and [OII] $\lambda$3727, are always observable.
We in fact find $M^*$ and $\alpha$ values to be in good agreement with the
original ones, and as shown in Figure~\ref{figpqBsys} (top right
panel), the $P$ and $Q$ parameters agree within 
$\lesssim 1.5\sigma$ of the original values.
It thus appears that our LF parameters are not significantly biased by
any residual redshift incompleteness effects, even though there are
some possible hints
of incompleteness in the $0.5 < z < 0.55$ bin for early- and
intermediate-type galaxies.

\subsubsection{Apparent Magnitude Incompleteness}

As shown in Figure~\ref{figsf} and discussed
in \S~\ref{weights}, the (uncorrected) raw
differential redshift success rate 
at the nominal spectroscopic limit $R_c = 21.5$ is 0.5, and the
redshift sampling rate is 0.2 so that
the typical galaxy weight is about 5.
Here we check if using a 0.5 mag brighter limit of $R_c = 21.0$, where
there is an improved raw redshift success rate of 0.6 and a smaller
typical galaxy weight of about 2, will make a
significant difference in the LF evolution results.
Figure~\ref{figpqBsys} (bottom left) shows that the $P$ and $Q$ values
for the $17.0 < R_c < 21.0$ sample are always within the original 1$\sigma$
contours.
Likewise, the $M^*$ and $\alpha$ values are within the original
2$\sigma$ contours.
We thus conclude that potential unaccounted incompleteness
over the $21.0 < R_c < 21.5$ magnitude range does not 
make a significant difference to our results.

\subsection{$B$-band Selection} \label{Bselect}

Here we examine the effects of using a $B$-selected CNOC2 sample
compared to our usual $R_c$-selected sample.
We do this partly in anticipation of our later comparison with the
$B$-selected Autofib Redshift Survey in \S~\ref{compautofib}.
We define a $18 < B < 23$ CNOC2 sample ($N = 1936$) and compute
new weights, using the bound $\vert B_i - B_j \vert \leq 0.25$ in place of
the corresponding $R_c$ bound in equation~(\ref{eqwtbds}).
The $B$-selected $P$-$Q$ results are shown in Figure~\ref{figpqBsys} (bottom right
panel), where we find agreement within 2$\sigma$ with the
$R_c$-selected results, except for the early-type galaxies, which now show
weak {\em positive} density evolution $P = 0.6$.
In general the $B$-selected sample shows more positive density
evolution compared to the $R_c$-selected sample, but the luminosity
evolution parameters $Q$ are very similar.
The corresponding $M^*$ and $\alpha$ values agree well within 1$\sigma$ for
the early and intermediate types, and are within 2$\sigma$ for the late
types.
We thus conclude that the $B$- and $R_c$-selected samples do give LF
evolution results that are generally in good agreement, with the sole
exception of the $P$ value for the early types.

\section{Comparisons with Previous Surveys} \label{comparisons}

In this section we compare our LF evolution results with those
obtained from three previous intermediate-$z$ redshift surveys.
We first briefly compare against the field galaxy sample from the
CNOC1 Cluster Redshift Survey, the immediate predecessor of CNOC2.
We then continue with the two next largest intermediate-$z$ 
redshift survey samples, specifically the Canada-France Redshift
Survey and the composite Autofib Redshift Survey.

\subsection{CNOC1 Cluster Redshift Survey --- Field Sample} \label{compcnoc1}

The CNOC1 Cluster Redshift Survey (\cite{car96}; \cite{yee96})
included observations of both cluster and field galaxies in the fields
of 16 high X-ray luminosity clusters.
The observational techniques used in the CNOC1 survey
are very similar to those used in CNOC2, but CNOC1 galaxies only have Gunn $r$
and $g$ photometry available.
Lin et al.\ (1997) examine the LF for a sample of
389 CNOC1 field galaxies, with redshifts $0.2 < z < 0.6$ and apparent
magnitudes $18 < r < 22$.
Non-evolving luminosity functions in $B_{AB}$ and Gunn $r$ are
computed, for the whole CNOC1 field sample, as well as for blue and red
subsets divided by the observed $g - r$ color of a CWW Sbc galaxy.
Consistent with the CNOC2 results, the CNOC1 LF's show the same trend of
a steeper faint-end slope for blue galaxies relative to red ones.
The CNOC1 sample is too small for the LF evolution analysis of the
present paper, but Lin et al.\ (1997) have computed luminosity densities
and have shown that the CNOC1 blue galaxy $\rho_L(z)$ increases
strongly with redshift, while the CNOC1 red galaxy $\rho_L(z)$ is
essentially constant with $z$.
These results are consistent with those found for the
CNOC2 sample.

We have also computed non-evolving $B_{AB}$ and $r$ LF's for CNOC2
galaxies, using basically the same CWW Sbc cut applied to CNOC1.
For each of the all, blue, and red samples and for both rest-frame bandpasses,
we confirm that the CNOC1 and CNOC2 results are indeed in 
good quantitative agreement in $M^*, \alpha$, and normalization.
The errors of the CNOC1 LF's are fairly large, however, primarily because of its
much smaller sample size, so that unfortunately little improvement in
the LF constraints is gained by adding the CNOC1 field data into the 
CNOC2 sample.
The present CNOC2 LF results essentially supercede those obtained
earlier from CNOC1.

\subsection{Canada-France Redshift Survey (CFRS)} \label{compcfrs}

The Canada-France Redshift Survey (CFRS; \cite{lil95a}) consists of
591 galaxy redshifts up to $z \sim 1$. 
The sample is
selected in the $I$-band with $17.5 \leq I_{AB} \leq 22.5$, and is 
distributed over five widely separated fields, totaling 125 arcmin$^2$
on the sky. 
Lilly et al.\ (1995b) examine the evolution of the CFRS
LF for $0 \lesssim z \lesssim 1$. 
They divide their sample by
observed $(V-I)_{AB}$ color, also using CWW SED's, and find rapid
evolution in the LF of galaxies bluer than a CWW Sbc galaxy,
contrasted with little change in the LF of redder-than-Sbc galaxies.

\begin{figure}
\plotone{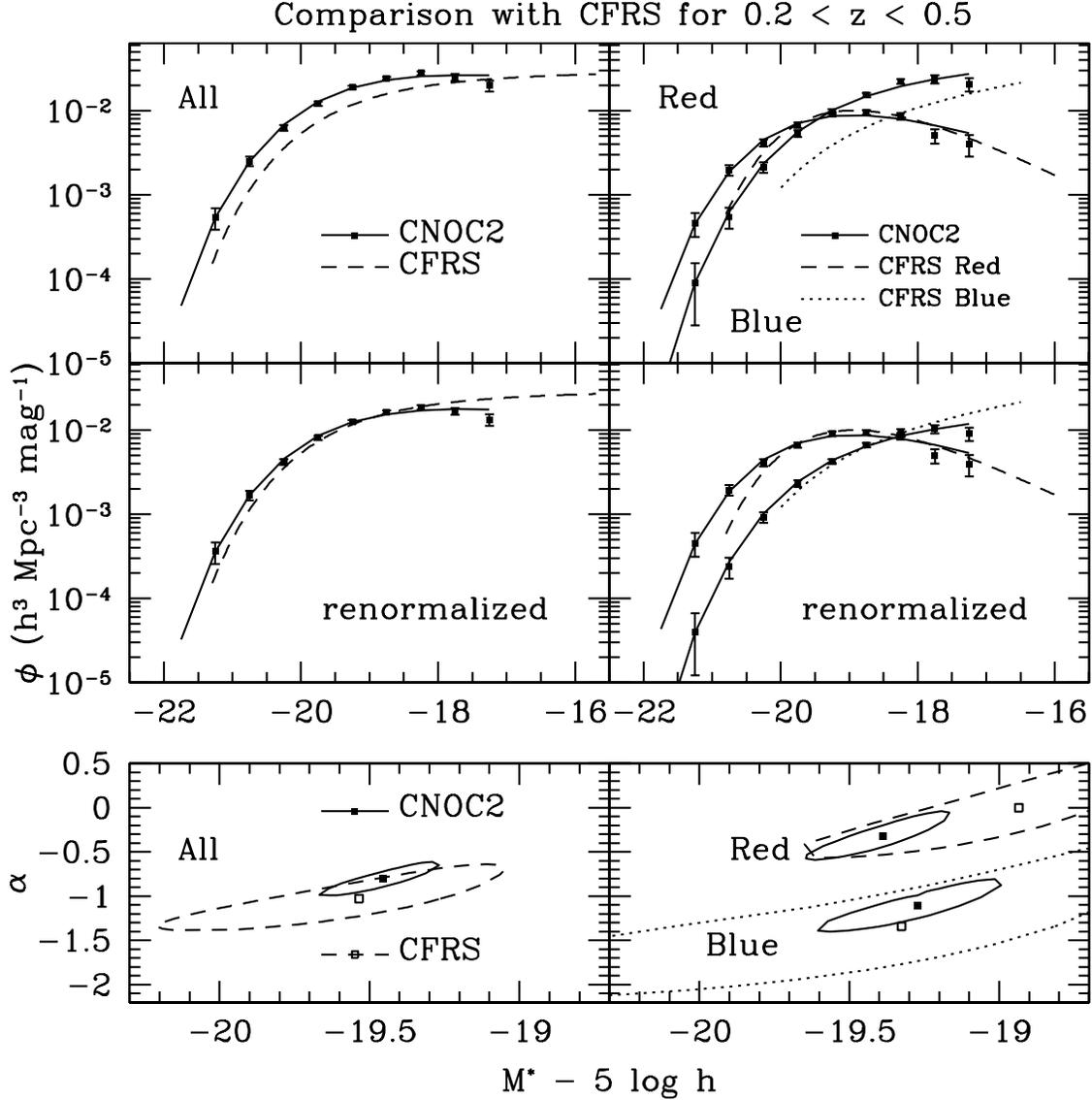}
\caption{
 ({\em Top panels}) Comparison of non-evolving $B_{AB}$ LF's for the 
  CNOC2 ({\em solid curves and filled points}) and 
  CFRS (\protect\cite{lil95a}; {\em dashed and dotted curves}) 
  samples in the overlapping redshift range $0.2 < z < 0.5$.
 Comparisons are shown for the full galaxy samples
  ({\em top left}), and for red and blue galaxy subsamples split at the
  color/SED of a CWW Sbc galaxy ({\em top right}).
 ({\em Middle panels}) Same as the corresponding top 
  panels except that the CNOC2 LF's have been renormalized to match
  the CFRS LF's using an analog of equation~(\ref{eqnorm}).
 ({\em Bottom panels}) $2\sigma$ $M^*$-$\alpha$ error
  contours for the all, red, and blue CNOC2 and CFRS samples.
 }
\label{figlfcfrs}
\end{figure}

Lilly et al.\ (1995b) also split their sample into several redshift bins,
including 208 galaxies in a $0.2 < z < 0.5$ bin which overlaps most 
with the CNOC2 redshift limits. 
In Figure~\ref{figlfcfrs}, we compare
the CFRS $B_{AB}$ LF results (the ``best'' estimates of \cite{lil95b})
against those for a nearly ten times larger sample of 1842 CNOC2 
galaxies with $0.2 < z < 0.5$. 
We also use the CWW Sbc cut
to divide our sample into red and blue subsets; we initially do {\em
  not} include evolution, as the CFRS results are fit using non-evolving
Schechter functions. 
The bottom panels in the figure show that the $M^*$-$\alpha$ values
for the two surveys are in good agreement (the CFRS error contours
have been calculated by us using CFRS redshift catalog data kindly
supplied by Simon Lilly).
This is also demonstrated 
in the middle panels, where we have renormalized the CNOC2 LF's
to match the normalizations of the CFRS LF's, using an equation
analogous to equation~(\ref{eqnorm}), in order to focus on comparing 
the LF shapes.
The renormalizations affect the red-galaxy LF's very
little, as the CNOC2 and CFRS results agree well in the first place.
However, as shown in the top panels, there is a noticeable difference
in the blue-galaxy LF's, where CNOC2 shows a higher number
density than CFRS.

\begin{figure}
\plotone{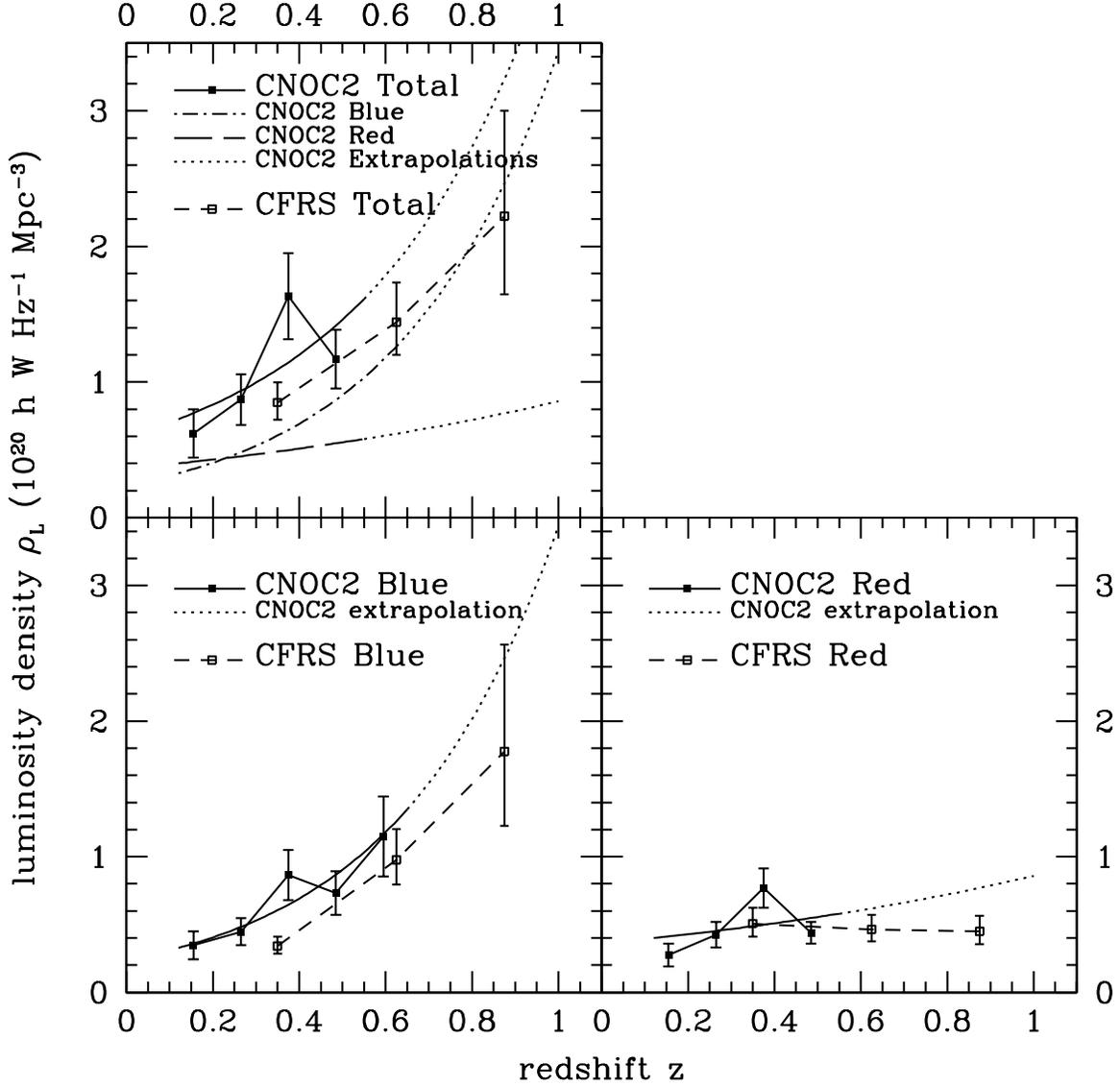}
\caption{
 Comparison of $B_{AB}$ luminosity densities for the CNOC2 
  ({\em solid curves and filled points}) and 
  CFRS (\protect\cite{lil96}; {\em short dashed curves and open points}) samples.
 The CNOC2 results are obtained using our usual 5-parameter LF
  evolution model, but now fit for blue ({\em bottom left}) and red
  ({\em bottom right}) CNOC2 galaxy samples split at the color/SED of a CWW
  Sbc galaxy.
 Note that the blue CNOC2 sample is defined for an extended redshift
  completeness range $0.12 < z < 0.65$; the red galaxy sample covers the
  usual $0.12 < z < 0.55$ CNOC2 redshift range.
 The {\em dotted curves} in the bottom panels show extrapolations of
  the CNOC2 LF fits beyond the redshift ranges indicated above.
 The {\em top left panel} compares the sum of the red and blue 
  luminosity densities for the CNOC2 and CFRS data sets, and the
  individual CNOC2 blue and red galaxy $\rho_L(z)$ fits and extrapolations are
  also shown as indicated in the panel legend.
 }
\label{figldcfrs}
\end{figure}

We next add evolution into our CNOC2 LF fits, using our usual
5-parameter method, in order to extrapolate our luminosity density results
into the $0.5 < z < 1$ redshift range probed by CFRS, as shown in
Figure~\ref{figldcfrs}.
We have extended the upper redshift limit to $z = 0.65$ for the
CNOC2 blue sample, in order have an additional data point to show.
Notice from Figure~\ref{figsed} that there does not
appear to be any obvious incompleteness for bluer-than-Sbc CNOC2
galaxies for $0.55 < z < 0.65$ (as there is for redder-than-Sbc
galaxies), and also note that
there is no obvious dip due to incompleteness in the last CNOC2 blue-galaxy 
$\rho_L$ point at $z \approx 0.6$ in Figure~\ref{figldcfrs}.
The CFRS $\rho_L$ results are taken from Lilly et al.\ (1996; their
``LF-estimated'' 4400\AA \ values),
and the two surveys do appear consistent within the errors.
The CNOC2 blue $\rho_L(z)$ and extrapolation more or less parallel 
the CFRS results, but are about 50\% higher overall.
The CNOC2 red $\rho_L(z)$ and extrapolation 
agree well with CFRS at $z \lesssim 0.7$, but appears to
overshoot CFRS in the highest redshift bin, $z \sim 0.9$.

We note that the difference in the CNOC2 and CFRS blue-galaxy luminosity 
densities may be consistent with the galaxy density fluctuations expected
for these two surveys.
We estimate (using the procedure described at the end of \S~\ref{computeLF})
that the density fluctuations $\delta\rho / \rho$ are approximately $12\%$ and
$13\%$ for the $0.2 < z < 0.5$ volumes in CNOC2 and CFRS, respectively.
The ratio of roughly 1.5 in the blue-galaxy $\rho_L$ for the two surveys would
then have a 1$\sigma$ uncertainty (assuming Gaussian galaxy density fluctuations)
of about $\pm 0.27$, so the luminosity densities differ at the $< 2\sigma$ 
level, and even less so if we include the remaining sampling and LF-fit
contributions to the total error on $\rho_L$.
On the other hand, it is unclear why we do not see any differences in the 
respective red galaxy populations, which should show stronger 
density fluctuations than the blue galaxies (e.g., Figure~\ref{fighist}).
Thus the blue-galaxy differences may be caused instead by some systematic 
differences in, e.g., galaxy classification and/or photometry for blue
galaxies in the two surveys, although one might then have expected to see
a more significant difference in the shapes of the CNOC2 and CFRS
luminosity functions.

Recently, galaxy evolution results have also been reported for a sample of 341
galaxies drawn from the CFRS and the Autofib/Low Dispersion Survey
Spectrograph (LDSS) data sets, which have morphologies classified from
Hubble Space Telescope images (\cite{bri98}; \cite{lil98}).
Though there is clearly correlation between the early, intermediate,
and late SED/color classifications adopted in this paper and the 
``elliptical,'' ``spiral,'' and ``peculiar'' morphological categories,
respectively, defined by Brinchmann et al.\ (1998), the correlations
are broad enough to preclude a detailed quantitative comparison.
We will defer this for a future paper on morphological classifications
of CNOC2 galaxies.
Here we will simply mention two LF-related results from Brinchmann et
al.\ (1998) which are qualitatively consistent with our results: 
(1) the LF of the spiral CFRS/LDSS sample indicates about 1 magnitude
of luminosity evolution in $B_{AB}$ by $z \simeq 1$, similar to the $Q
= 0.9$ value we find for the CNOC2 intermediate-type galaxies, which
should be dominated by spirals;
and (2) the peculiar/irregular CFRS/LDSS galaxies appear to be primarily
responsible for the rapid rise with redshift of the blue galaxy
luminosity density, a result consistent with our 
observation that late-type CNOC2 galaxies cause the strong observed
increase in the overall $\rho_L$ with redshift.

\subsection{Autofib Redshift Survey} \label{compautofib}

The Autofib Redshift Survey is a composite of various galaxy survey samples
(\cite{ell96} and references therein) and contains over 1700 redshifts with
$0 < z \lesssim 0.75$. 
The survey is selected in the blue, with apparent magnitudes in the
range $11.5 < b_J < 24.0$. 
Ellis et al.\ (1996) find an overall steepening of the LF
at higher redshifts, and similar to CNOC2 and CFRS, the LF evolution is
dominated by late-type galaxies, which also show increased [OII] $\lambda$3727
emission and thus stronger star formation at higher redshifts.

\begin{figure}
\plotone{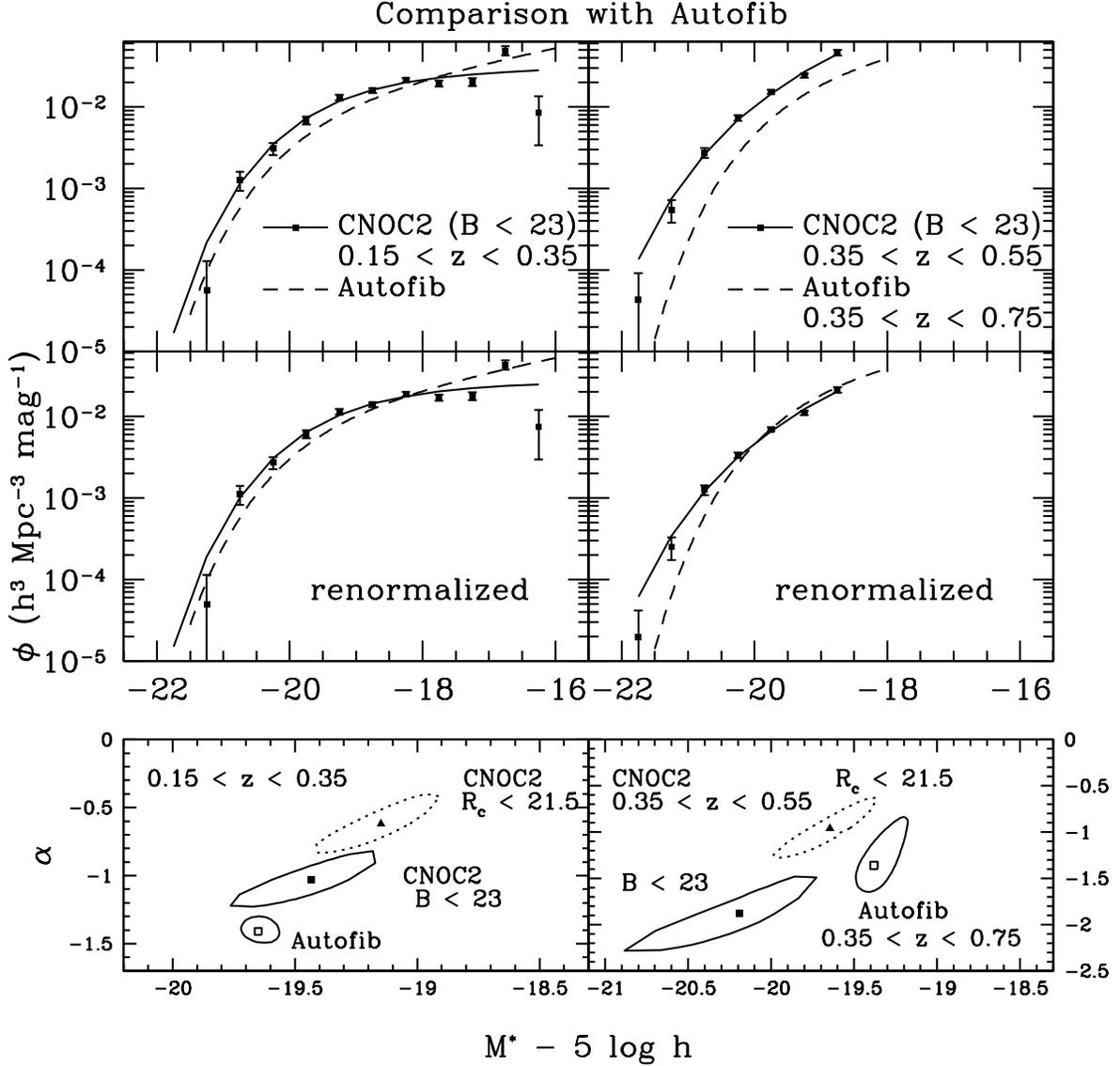}
\caption{
 ({\em Top panels}) Comparison of non-evolving $B$ LF's for the 
  CNOC2 ({\em solid and curves and filled points}) and 
  composite Autofib (\protect\cite{ell96}; {\em dashed curves}) 
  samples in the redshift ranges  $0.15 < z < 0.35$ ({\em left panels}) 
  and $0.35 < z < 0.75$ ({\em right panels}).
 Unlike previous plots, the CNOC2 LF's have been computed using a 
  $18 < B < 23$ sample to better match the $B$-selected Autofib sample (see text).
 ({\em Middle panels}) Same as the corresponding top 
  panels except that the CNOC2 LF's have been renormalized to match
  the Autofib LF's using an analog of equation~(\ref{eqnorm}).
 ({\em Bottom panels}) $2\sigma$ $M^*$-$\alpha$ error
  contours for the CNOC2 and Autofib ({\em solid contours and open squares}) LF's.
 Note that CNOC2 results for both our $18 < B < 23$ ({\em solid contours and 
  filled squares}) and our standard $17 < R_c < 21.5$ ({\em dotted contours
  and filled triangles}) samples are shown.
 }
\label{figlfautofib}
\end{figure}

Ellis et al.\ (1996) give overall $b_J$ LF's in several redshift intervals,
including $0.15 < z < 0.35$ and $0.35 < z < 0.75$ bins which overlap with CNOC2.
In Figure~\ref{figlfautofib}, we make the same redshift cuts (but with
a $z = 0.55$ upper limit), and compute non-evolving LF's for
comparison.
We show results both for the standard CNOC2 $17 < R_c < 21.5$ sample
($N = 2076$; {\em filled triangles})
and for a blue-selected $18 < B < 23$ CNOC2 sample 
($N = 1830$; {\em filled squares}).
There is a significant
systematic difference between the two CNOC2 samples, where the
$B$-selected sample shows a steeper $\alpha$ and brighter $M^*$, because
of the increased contribution (due to $k$-correction effects)
of bluer late-type galaxies in the $B$-selected sample.
However, this systematic offset between the $R_c$- and $B$-selected
samples is an artifact of trying to force fit a single LF to the full galaxy
population, and does not occur if we subdivide into three populations
as we did before (\S~\ref{Bselect}).
Using a $B$-selected CNOC2 sample
significantly improves the agreement with the $0.15 < z < 0.35$
Autofib results, 
although the 2$\sigma$ $M^*$-$\alpha$ error contours 
still do not quite overlap, as
Autofib shows a steeper $\alpha$ and a brighter $M^*$. 
(Note that although the CNOC2 and Autofib samples here are similar in
size, the Autofib error contours are smaller because of Autofib's much
wider {\em apparent} magnitude limits compared to CNOC2.)
Nonetheless, a visual comparison of the two lower-$z$ LF's (top left panel
of Figure~\ref{figlfautofib}) does show reasonable agreement.
However, in the higher-$z$ bin there is a
noticeable mismatch in $M^*$ and/or normalization between the 
CNOC2 and Autofib results.

The causes of these discrepancies are not known, but can include
sampling fluctuations, as well as unaccounted systematic differences
in sample selection, galaxy classifications and $k$-corrections,
photometry, and the 
like (see \cite{lin97} for additional discussion).
Note that sample size may be an important consideration for the comparison 
in the higher-redshift bin.
Though the overall Autofib sample contains some 1700 redshifts, the
relevant sample sizes here are smaller (\cite{ell97}, Figure~6b):
$N = 665$ for $0.15 < z < 0.35$ and only $N = 152$ for $0.35 < z < 0.75$.
The corresponding CNOC2 ($B$-selected) sample sizes are $N = 940$ and
$N = 890$, so that the high-$z$ CNOC2 sample is nearly 6 times larger
than the corresponding Autofib data set.
Also, large-scale galaxy density fluctuations may play a role.
The values of $\delta\rho / \rho$ are estimated to be $16\%$ and $13\%$ for 
the low- and high-$z$ CNOC2 volumes, respectively, and are presumably 
somewhat larger for the corresponding Autofib volumes (although we
have not done the exact calculations as we lack certain needed Autofib
sample details). 
As we saw earlier in our CFRS comparison, such values of $\delta\rho / \rho$ 
do not preclude a factor of 1.5 in the relative LF normalizations,
which would significantly reduce the discrepancy in the high-$z$ bin.
We have also checked whether random $k$-correction errors and photometric 
errors in the Autofib sample might be responsible for the $M^*$ and $\alpha$
differences.
The Autofib $k$-corrections are assigned primarily on the basis of
galaxy spectral classifications, rather than more directly via multicolor
photometry as we do. 
Ellis et al.\ (1996) estimate that their $k$-correction 
errors due to spectral misclassifications have a redshift dependence
$\sigma_B \sim 0.5 z$ mag (our interpretation of their Figure~6).
Also, the Autofib photometry is based mainly on photographic plate 
data with errors typically 0.1-0.2 mag, in contrast to
CNOC2 CCD photometry with errors $< 0.1$ mag.
Both these effects will tend to bias the Autofib results to brighter
$M^*$ and steeper $\alpha$ (see \S~\ref{randomerr}), so we have checked 
the effect of adding such $k$-correction errors and photometric errors 
($0.2$ mag Gaussian) to our $B$ magnitudes.
In agreement with Ellis et al.\ (1996) and Heyl et al.\ (1997), we find that 
the differences are small, with $\vert \Delta M^* \vert \lesssim 0.2$
and $\vert \Delta \alpha \vert \lesssim 0.1$, not enough to significantly
improve the agreement of the $M^*$-$\alpha$ contours in the low-$z$ bin,
and of the wrong sign for the high-$z$ bin.
We have also tried computing $b_J$ absolute magnitudes and 
LF's for CNOC2 galaxies using the $b_J$ response function (instead of
our usual Johnson $B$) but it makes negligible difference to our results.
Additional exploration of Autofib vs.\ CNOC2 photometry systematics
likely requires us to apply our photometry codes to the original Autofib
data.
Such a detailed comparison may not be warranted given that the 
main CNOC2/Autofib differences lie in the high-$z$ bin, where the main culprits
may very well be galaxy density fluctuations and the small Autofib sample size
there.

Heyl et al.\ (1997) have classified Autofib galaxies into six
types based on cross-correlation against local galaxy spectral
templates, and examined the evolution of the LF divided by galaxy
spectral type.
Note that in discussing LF evolution, Heyl et al.\ typically use the three 
broader categories ``early-type E/S0,'' ``early-type spirals,'' and
``late-type spirals'' (each including two of their original six
types), which have obvious but broad correlations relative to our
early, intermediate, and late types, respectively.
Also, their LF evolution model is similar but not identical to ours, and involves
six parameters compared to our five, with the additional parameter
characterizing the rate of evolution of $\alpha$, which we have taken
as fixed with redshift.
In addition, unlike our analysis, Heyl et al.\ do not plot error contours,
like our $P$-$Q$ diagrams, to show the correlations among their LF 
evolution parameters.
Because both their classification and analysis methods are different
from ours, and because we have already noted some discrepancies
between the CNOC2 and Autofib results above, 
we will not attempt a detailed quantitative comparison here.
We will note however, that generally speaking the Heyl et al.\ (1997) results 
are qualitatively consistent with ours.
Specifically, they find: (1) no significant evolution of the E/S0 LF
out to at least $z \sim 0.5$; 
(2) modest evolution in the LF of early-type
spirals, characterized by steepening of $\alpha$ at higher redshifts
rather than by changes in $M^*$ or $\phi^*$; 
and (3) strong evolution in the LF of late-type spirals, described by
steepening $\alpha$, brightening $M^*$, and increasing $\phi^*$ at
higher $z$.
The main difference compared to CNOC2 lies in the steepening
$\alpha$ observed in Autofib, contrasted with the good match of 
our $\alpha(z) =$ constant models to the CNOC2 data (see
Figure~\ref{figlfsBa} in particular).
Also, the $Q \approx 1$ luminosity evolution we find in our early- and
intermediate-type LF's is somewhat different from the trends seen in
the Autofib E/S0 and early-spiral LF's.
It is not clear at present what is responsible for these detailed 
CNOC2/Autofib evolution differences, but we note in particular that the different
galaxy classification schemes involved may play an important role.
We will return to this comparison again in a future paper on application of 
spectral classifications to the full CNOC2 sample.

\section{Conclusions} \label{conclusions}

In this paper we have examined the evolution of the luminosity
function for a sample of over 2000 field galaxies, with $0.12 < z <
0.55$ and $17.0 < R_c < 21.5$, drawn from two different sky patches of the
CNOC2 Field Galaxy Redshift Survey.
Although this sample comprises only half the ultimate CNOC2 data set,
it is nonetheless the largest intermediate-redshift 
galaxy survey sample at present.
The availability of $U \! B V \! R_c I_c$ photometry for our sample 
allows galaxy classifications by SED type, as well as computation of
LF's in different rest-frame bandpasses.
In addition, the multicolor photometry permits us to examine sample
selection effects in detail, and allows us to construct galaxy weights to 
account for our redshift success rates as functions of galaxy
magnitude and color.

In particular, we have calculated LF parameters in the $B_{AB}$,
$R_c$, and $U$ bands for early-, intermediate-, and late-type galaxies,
classified using $U \! B V \! R_c I_c$ 
colors derived from the non-evolving galaxy spectral energy distributions
of Coleman, Wu, \& Weedman (1980).
We present a description of the LF evolution in terms of a
five-parameter model involving the usual three Schechter function
parameters, plus two additional parameters $P$ and $Q$ describing
number density and luminosity evolution rates, respectively
(eq.\ \ref{eqmodel}).
The best-fit parameters of our LF evolution models are given in
Tables~\ref{tablfB} and \ref{tablfRU}.
We find that the faint-end slope of the LF is steeper for later-type
galaxies relative to earlier-type objects, consistent with previous
LF studies at both intermediate and low redshifts.

The principal results of this paper are the quantitative separation of
luminosity and density evolution in the LF's of different galaxy
populations, and the finding that the character of the LF evolution is
strongly type dependent, varying from primarily luminosity
evolution for early-type galaxies to predominantly density evolution
for late-type objects.
We quantify the rates of luminosity function evolution 
using our $P$ and $Q$ parameters.
Specifically, we see that (for $q_0 = 0.5$): 
(1) the late-type galaxy LF is best fit
by strong positive density evolution ($P = 3.1$), with
nearly no luminosity evolution ($Q = 0.2$);
(2) the intermediate-type LF shows 
positive luminosity evolution ($Q = 0.9$) plus weak positive density
evolution ($P = 0.7$), resulting in mild positive evolution in the
luminosity density $\rho_L$;
and (3) the early-type LF shows positive luminosity evolution 
($Q = 1.6$) which is nearly compensated by negative density evolution 
($P = -1.1$), resulting in a very weak positive evolution in $\rho_L$.
However, we should note that the $P$ and $Q$ parameters are
strongly correlated for late-type galaxies, and ``no-evolution'' for
early- and intermediate-type objects is ruled out at only about the
$2\sigma$ confidence level.
Nonetheless, it is a robust result that the LF's of late and early+intermediate
galaxies are evolving differently and occupy different regions of
$P$-$Q$ parameter space.
Moreover, there is a distinct
contrast between the sharply rising luminosity density of late-type galaxies
and the relatively constant $\rho_L$ of early- and intermediate-type
objects.
(This is probably not too surprising given that one expects the
rapidly evolving population to consist of those galaxies actively
forming stars in the past, which are essentially the late types.)
These general conclusions are little changed by adopting a different
value of $q_0 = 0.1$.

The rates of luminosity evolution ($Q \approx 1$) for our early and
intermediate types are in the range expected from models of galaxy 
evolution (e.g., \cite{bc96}). 
At face value, the strong density evolution observed for late types
suggests that mergers play an important role in the evolution of these
galaxies.
However, other processes, particularly those affecting star formation
properties, may mimic the effect of mergers and cause
similar changes in number density (e.g., a starbursting 
sub-population among the late types at high-$z$ may be responsible for
the density evolution).
In Paper II, we will test various physical galaxy evolution models in detail,
including the effects of different star formation histories, ages, 
stellar initial mass functions, dust content, and the like. 
Nonetheless, whatever the responsible physical mechanisms are, they
will need to explain the strong correlation between galaxy type and the
character of the LF evolution, in particular the strong increase in the 
apparent density evolution as one proceeds to later galaxy types.
The relevant underlying physical variables controlling the evolution
should thus be closely correlated with the galaxy SED type.
On the other hand, within each of our galaxy categories, those
physical variables are probably not strongly
correlated with galaxy luminosity, since the 
data are well-fit by our fixed-$\alpha$ evolution models (so that the
evolution does not vary much with luminosity within each galaxy
category).
It may be a challenge for physical models to explain this combination
of strong type-dependence in the LF evolution, coupled with relatively
little luminosity-dependence of the evolution within each galaxy type.

We also compute SED type distributions, $U \! B V \! R_c I_c$ galaxy
number counts, and various color distributions for CNOC2 galaxies.
In particular, we find that extrapolations of our LF evolution models
to $z \approx 0.75$ yield good matches to the observed number counts and color
distributions, thus providing an additional check on the validity of our LF 
evolution results.
In addition, we have verified that various systematic effects, specifically
patch-to-patch variations, photometric errors, surface brightness
selection, redshift incompleteness, and apparent magnitude
incompleteness, do not significantly affect our results
($\lesssim 1\sigma$ difference typically).

Finally, we note that our LF results are generally consistent with those
found in previous intermediate-$z$ redshift surveys, as verified in
specific comparisons against results from the next two
largest samples, CFRS and Autofib.
However, there are still some unresolved detailed discrepancies, particularly 
with respect to the $B$-selected Autofib survey, which may be due to
differences in galaxy classification or sample selection methods.

In this paper, we have simply presented a {\em description} of the evolution
of the LF's of different intermediate-redshift galaxies, without
delving into the possible underlying physical processes.
As mentioned earlier, in our second LF paper
we will actually confront the CNOC2 observations against various 
galaxy evolution models, in order to better understand and constrain
those physical mechanisms.
Subsequent papers on galaxy population evolution in CNOC2 
will also make use of the morphological and spectral
information that will become available for CNOC2 galaxies once the
appropriate data are fully reduced.
Ultimately the doubled size of the full CNOC2 sample over
the present interim sample will
significantly improve upon the LF evolution constraints that 
we have presented here.
We are also in the process of deriving properly calibrated 
photometric redshifts, which should provide another factor of two
increase in useful sample size for $R_c < 21.5$ galaxies.
Future papers will re-examine the question of 
LF evolution using these even larger CNOC2 galaxy data sets.

\acknowledgments

The CNOC project is supported in part by a Collaborative Program grant
from the Natural Sciences and Engineering Research Council of Canada (NSERC).
HL acknowledges support provided by NASA through Hubble Fellowship grant
\#HF-01110.01-98A awarded by the Space Telescope Science Institute, which 
is operated by the Association of Universities for Research in Astronomy, 
Inc., for NASA under contract NAS 5-26555.
HY and RC wish to acknowledge support from NSERC operating grants.
Thanks also to Pat Hall for a careful reading of the manuscript and 
for providing useful comments and suggestions.
Thanks in addition to Simon Lilly and Matthew Colless for kindly
providing CFRS and Autofib data, respectively.
Finally, we wish to thank CTAC of CFHT for the generous allotment of 
telescope time, as well as to acknowledge the fine support received from CFHT 
staff.

\clearpage

\begin{deluxetable}{lrrrrrr}
\tablewidth{0pt}
\tablecaption{$B_{AB}$ LF Parameters \tablenotemark{a}}
\tablehead{
   \colhead{Sample} 
 & \colhead{$N$ \tablenotemark{b}} 
 & \colhead{$M^*(z=0.3)$ \tablenotemark{c}}
 & \colhead{$\alpha$}
 & \colhead{$\phi^*(z=0)$ \tablenotemark{d}}
 & \colhead{$P$}
 & \colhead{$Q$}
 }
\startdata
 & & & $q_0 = 0.5$ & & \nl
Early        &  611 &  $ -19.06 \pm   0.12 $  &  $   0.08 \pm   0.14 $  &  $   0.0203 \pm   0.0036 $ 
                    &  $  -1.07 \pm   0.49 $  &  $   1.58 \pm   0.49 $  \nl 
Intermediate &  518 &  $ -19.38 \pm   0.16 $  &  $  -0.53 \pm   0.15 $  &  $   0.0090 \pm   0.0023 $ 
                    &  $   0.73 \pm   0.70 $  &  $   0.90 \pm   0.72 $  \nl 
Early+Inter. & 1129 &  $ -19.19 \pm   0.10 $  &  $  -0.20 \pm   0.10 $  &  $   0.0291 \pm   0.0049 $ 
                    &  $  -0.27 \pm   0.40 $  &  $   1.29 \pm   0.41 $  \nl 
Late         & 1016 &  $ -19.26 \pm   0.16 $  &  $  -1.23 \pm   0.12 $  &  $   0.0072 \pm   0.0033 $ 
                    &  $   3.08 \pm   0.99 $  &  $   0.18 \pm   0.71 $  \nl 
 & & & & & \nl
 & & & $q_0 = 0.1$ & & \nl
Early        &  611 &  $ -19.19 \pm   0.12 $  &  $   0.08 \pm   0.14 $  &  $   0.0197 \pm   0.0036 $ 
                    &  $  -1.79 \pm   0.49 $  &  $   2.00 \pm   0.49 $  \nl 
Intermediate &  518 &  $ -19.51 \pm   0.17 $  &  $  -0.53 \pm   0.15 $  &  $   0.0087 \pm   0.0023 $ 
                    &  $   0.00 \pm   0.71 $  &  $   1.32 \pm   0.72 $  \nl 
Early+Inter. & 1129 &  $ -19.32 \pm   0.10 $  &  $  -0.20 \pm   0.10 $  &  $   0.0284 \pm   0.0048 $ 
                    &  $  -1.00 \pm   0.40 $  &  $   1.72 \pm   0.41 $  \nl 
Late         & 1016 &  $ -19.38 \pm   0.16 $  &  $  -1.23 \pm   0.12 $  &  $   0.0071 \pm   0.0034 $ 
                    &  $   2.34 \pm   0.98 $  &  $   0.61 \pm   0.71 $  \nl 
\enddata
\tablenotetext{a}{All tabulated errors are 
  $1\sigma$ {\em one}-parameter errors. 
  See Figures~\ref{figmaB} and \ref{figpqB} for the joint
  {\em two}-parameter $M^*$-$\alpha$ and $P$-$Q$ error contours, respectively.}
\tablenotetext{b}{We apply apparent magnitude limits
  $17.0 < R_c < 21.5$, absolute magnitude limits
  $-22.0 < M_{B_{AB}} - 5 \log h < -16.0$, and redshift limits
  $0.12 < z < 0.55$ in defining our samples.}
\tablenotetext{c}{We take Hubble constant $h = 1$.}
\tablenotetext{d}{Units are $h^3$~Mpc$^{-3}$~mag$^{-1}$.}
\label{tablfB}
\end{deluxetable}

\clearpage

\begin{deluxetable}{lrrrrrr}
\tablewidth{0pt}
\tablecaption{$R_c$ and $U$ LF Parameters \tablenotemark{a}}
\tablehead{
   \colhead{Sample} 
 & \colhead{$N$ \tablenotemark{b}} 
 & \colhead{$M^*(z=0.3)$ \tablenotemark{c}}
 & \colhead{$\alpha$}
 & \colhead{$\phi^*(z=0)$ \tablenotemark{d}}
 & \colhead{$P$}
 & \colhead{$Q$}
 }
\startdata
 & & $R_c$ & $q_0 = 0.5$ & & \nl
Early        &  611 &  $ -20.50 \pm   0.12 $  &  $  -0.07 \pm   0.14 $  &  $   0.0185 \pm   0.0037 $ 
                    &  $  -0.88 \pm   0.52 $  &  $   1.24 \pm   0.53 $  \nl 
Intermediate &  517 &  $ -20.47 \pm   0.17 $  &  $  -0.61 \pm   0.15 $  &  $   0.0080 \pm   0.0023 $ 
                    &  $   0.89 \pm   0.74 $  &  $   0.69 \pm   0.76 $  \nl 
Early+Inter. & 1128 &  $ -20.61 \pm   0.11 $  &  $  -0.44 \pm   0.10 $  &  $   0.0230 \pm   0.0046 $ 
                    &  $   0.08 \pm   0.45 $  &  $   0.70 \pm   0.48 $  \nl 
Late         & 1012 &  $ -20.11 \pm   0.18 $  &  $  -1.34 \pm   0.12 $  &  $   0.0056 \pm   0.0030 $ 
                    &  $   3.17 \pm   1.03 $  &  $   0.11 \pm   0.74 $  \nl 
 & & & & & \nl
 & & $R_c$ & $q_0 = 0.1$ & & \nl
Early        &  609 &  $ -20.59 \pm   0.12 $  &  $  -0.03 \pm   0.14 $  &  $   0.0179 \pm   0.0036 $ 
                    &  $  -1.54 \pm   0.52 $  &  $   1.51 \pm   0.53 $  \nl 
Intermediate &  518 &  $ -20.62 \pm   0.18 $  &  $  -0.63 \pm   0.15 $  &  $   0.0077 \pm   0.0023 $ 
                    &  $   0.15 \pm   0.75 $  &  $   1.11 \pm   0.78 $  \nl 
Early+Inter. & 1127 &  $ -20.73 \pm   0.11 $  &  $  -0.43 \pm   0.10 $  &  $   0.0223 \pm   0.0046 $ 
                    &  $  -0.59 \pm   0.46 $  &  $   1.02 \pm   0.48 $  \nl 
Late         & 1012 &  $ -20.20 \pm   0.17 $  &  $  -1.30 \pm   0.12 $  &  $   0.0053 \pm   0.0029 $ 
                    &  $   2.84 \pm   1.06 $  &  $   0.22 \pm   0.76 $  \nl 
 & & & & & \nl
 & & $U$ & $q_0 = 0.5$ & & \nl
Early        &  611 &  $ -18.54 \pm   0.11 $  &  $   0.14 \pm   0.15 $  &  $   0.0213 \pm   0.0036 $ 
                    &  $  -1.19 \pm   0.48 $  &  $   1.85 \pm   0.48 $  \nl 
Intermediate &  518 &  $ -19.27 \pm   0.16 $  &  $  -0.51 \pm   0.15 $  &  $   0.0092 \pm   0.0026 $ 
                    &  $   0.68 \pm   0.69 $  &  $   0.97 \pm   0.70 $  \nl 
Early+Inter. & 1129 &  $ -18.92 \pm   0.10 $  &  $  -0.22 \pm   0.10 $  &  $   0.0302 \pm   0.0051 $ 
                    &  $  -0.35 \pm   0.40 $  &  $   1.40 \pm   0.41 $  \nl 
Late         & 1017 &  $ -19.32 \pm   0.15 $  &  $  -1.14 \pm   0.13 $  &  $   0.0095 \pm   0.0039 $ 
                    &  $   2.67 \pm   0.92 $  &  $   0.51 \pm   0.66 $  \nl 
 & & & & & \nl
 & & $U$ & $q_0 = 0.1$ & & \nl
Early        &  611 &  $ -18.67 \pm   0.11 $  &  $   0.15 \pm   0.15 $  &  $   0.0209 \pm   0.0037 $ 
                    &  $  -1.93 \pm   0.48 $  &  $   2.27 \pm   0.48 $  \nl 
Intermediate &  518 &  $ -19.40 \pm   0.16 $  &  $  -0.51 \pm   0.15 $  &  $   0.0090 \pm   0.0023 $ 
                    &  $  -0.05 \pm   0.69 $  &  $   1.39 \pm   0.71 $  \nl 
Early+Inter. & 1129 &  $ -19.05 \pm   0.10 $  &  $  -0.22 \pm   0.10 $  &  $   0.0294 \pm   0.0049 $ 
                    &  $  -1.06 \pm   0.40 $  &  $   1.82 \pm   0.41 $  \nl 
Late         & 1016 &  $ -19.44 \pm   0.16 $  &  $  -1.13 \pm   0.12 $  &  $   0.0087 \pm   0.0038 $ 
                    &  $   2.29 \pm   0.96 $  &  $   0.64 \pm   0.68 $  \nl 
\enddata
\tablenotetext{a}{All tabulated errors are 
  $1\sigma$ {\em one}-parameter errors.}
\tablenotetext{b}{We apply apparent magnitude limits
  $17.0 < R_c < 21.5$, absolute magnitude limits
  $-23.0 < M_{R_c} - 5 \log h < -17.0$ or
  $-22.0 < M_U - 5 \log h < -16.0$, and redshift limits
  $0.12 < z < 0.55$ in defining our samples.}
\tablenotetext{c}{We take Hubble constant $h = 1$.}
\tablenotetext{d}{Units are $h^3$~Mpc$^{-3}$~mag$^{-1}$.}
\label{tablfRU}
\end{deluxetable}

\clearpage

\begin{deluxetable}{lrrrr}
\tablewidth{0pt}
\tablecaption{Luminosity Density Values \tablenotemark{a}}
\tablehead{
   \colhead{Sample} 
 & \colhead{$\rho_L ({\rm fit}, z=0)$}
 & \colhead{$\rho_L(0.12 < z < 0.25)$}
 & \colhead{$\rho_L(0.25 < z < 0.40)$}
 & \colhead{$\rho_L(0.40 < z < 0.55)$}
 }
\startdata
 & $B$ & $q_0 = 0.5$ & & \nl
Early         &  $  0.258 \pm  0.042 $ 
              &  $  0.301 \pm  0.074 $   &  $  0.387 \pm  0.071 $   &  $  0.265 \pm  0.045 $   \nl 
Intermediate  &  $  0.159 \pm  0.028 $ 
              &  $  0.217 \pm  0.055 $   &  $  0.351 \pm  0.063 $   &  $  0.259 \pm  0.045 $   \nl 
Late          &  $  0.189 \pm  0.030 $ 
              &  $  0.390 \pm  0.094 $   &  $  0.580 \pm  0.109 $   &  $  0.720 \pm  0.123 $   \nl 
Total         &  $  0.606 \pm  0.078 $ 
              &  $  0.907 \pm  0.213 $   &  $  1.318 \pm  0.230 $   &  $  1.244 \pm  0.191 $   \nl 
 & & & & \nl
 & $B$ & $q_0 = 0.1$ & & \nl
Early         &  $  0.252 \pm  0.042 $ 
              &  $  0.282 \pm  0.072 $   &  $  0.360 \pm  0.069 $   &  $  0.223 \pm  0.040 $   \nl 
Intermediate  &  $  0.155 \pm  0.030 $ 
              &  $  0.203 \pm  0.053 $   &  $  0.317 \pm  0.061 $   &  $  0.224 \pm  0.040 $   \nl 
Late          &  $  0.183 \pm  0.030 $ 
              &  $  0.364 \pm  0.092 $   &  $  0.521 \pm  0.106 $   &  $  0.615 \pm  0.114 $   \nl 
Total         &  $  0.591 \pm  0.076 $ 
              &  $  0.849 \pm  0.208 $   &  $  1.197 \pm  0.225 $   &  $  1.063 \pm  0.177 $   \nl 
 & & & & \nl
 & $R_c$ & $q_0 = 0.5$ & & \nl
Early         &  $  0.785 \pm  0.130 $ 
              &  $  0.896 \pm  0.221 $   &  $  1.115 \pm  0.202 $   &  $  0.750 \pm  0.126 $   \nl 
Intermediate  &  $  0.351 \pm  0.066 $ 
              &  $  0.472 \pm  0.118 $   &  $  0.764 \pm  0.139 $   &  $  0.565 \pm  0.096 $   \nl 
Late          &  $  0.320 \pm  0.050 $ 
              &  $  0.657 \pm  0.158 $   &  $  1.012 \pm  0.199 $   &  $  1.225 \pm  0.215 $   \nl 
Total         &  $  1.455 \pm  0.186 $ 
              &  $  2.024 \pm  0.475 $   &  $  2.891 \pm  0.510 $   &  $  2.539 \pm  0.386 $   \nl 
 & & & & \nl
 & $R_c$ & $q_0 = 0.1$ & & \nl
Early         &  $  0.778 \pm  0.134 $ 
              &  $  0.841 \pm  0.217 $   &  $  1.006 \pm  0.194 $   &  $  0.620 \pm  0.112 $   \nl 
Intermediate  &  $  0.346 \pm  0.066 $ 
              &  $  0.444 \pm  0.116 $   &  $  0.692 \pm  0.133 $   &  $  0.486 \pm  0.088 $   \nl 
Late          &  $  0.302 \pm  0.050 $ 
              &  $  0.604 \pm  0.152 $   &  $  0.901 \pm  0.187 $   &  $  1.055 \pm  0.198 $   \nl 
Total         &  $  1.426 \pm  0.189 $ 
              &  $  1.889 \pm  0.463 $   &  $  2.599 \pm  0.489 $   &  $  2.160 \pm  0.356 $   \nl 
 & & & & \nl
 & $U$ & $q_0 = 0.5$ & & \nl
Early         &  $  0.086 \pm  0.014 $ 
              &  $  0.102 \pm  0.025 $   &  $  0.136 \pm  0.024 $   &  $  0.092 \pm  0.016 $   \nl 
Intermediate  &  $  0.077 \pm  0.014 $ 
              &  $  0.106 \pm  0.026 $   &  $  0.171 \pm  0.031 $   &  $  0.127 \pm  0.022 $   \nl 
Late          &  $  0.118 \pm  0.019 $ 
              &  $  0.241 \pm  0.057 $   &  $  0.344 \pm  0.064 $   &  $  0.438 \pm  0.076 $   \nl 
Total         &  $  0.281 \pm  0.036 $ 
              &  $  0.449 \pm  0.104 $   &  $  0.651 \pm  0.114 $   &  $  0.657 \pm  0.104 $   \nl 
 & & & & \nl
 & $U$ & $q_0 = 0.1$ & & \nl
Early         &  $  0.084 \pm  0.014 $ 
              &  $  0.096 \pm  0.025 $   &  $  0.123 \pm  0.024 $   &  $  0.079 \pm  0.014 $   \nl 
Intermediate  &  $  0.075 \pm  0.014 $ 
              &  $  0.099 \pm  0.026 $   &  $  0.154 \pm  0.030 $   &  $  0.109 \pm  0.020 $   \nl 
Late          &  $  0.113 \pm  0.019 $ 
              &  $  0.224 \pm  0.057 $   &  $  0.310 \pm  0.062 $   &  $  0.381 \pm  0.070 $   \nl 
Total         &  $  0.273 \pm  0.035 $ 
              &  $  0.419 \pm  0.103 $   &  $  0.587 \pm  0.110 $   &  $  0.569 \pm  0.096 $   \nl 
\enddata
\tablenotetext{a}{Units are $10^{20} \ h$~W Hz$^{-1}$ Mpc$^{-3}$.
  As discussed in the text, the tabulated 1$\sigma$ errors include both bootstrap resampling 
   errors (accounting for uncertainties in the LF fits and in galaxy sampling) and estimated
   uncertainties due to large-scale galaxy density fluctuations.}
\label{tabrhoL}
\end{deluxetable}

\clearpage

\clearpage

\clearpage

\clearpage

\clearpage

\clearpage

\clearpage

\clearpage

\clearpage

\clearpage

\clearpage

\clearpage

\clearpage

\clearpage

\clearpage

\clearpage

\clearpage

\clearpage

\clearpage

\clearpage

\clearpage

\end{document}